\documentclass[preprint,12pt]{elsarticle}

\usepackage{amssymb}

\usepackage{lineno}

\journal{Ocean Modelling}

\usepackage{graphicx}
\usepackage{color}
\usepackage{gensymb}  
\usepackage{textcomp}  
\usepackage{ulem}  

\renewcommand{\vec}{\mathbf}
\def\NEU#1{{\textcolor{blue}{\bf #1}}} 

\def\NEU#1{#1} 

\begin{document}

\begin{frontmatter}


\title{{A framework for the evaluation of turbulence closures used in
mesoscale ocean large-eddy simulations}}
\author{\corref{cor1}Jonathan Pietarila Graham, Todd Ringler\corref{cor2}}
\ead{jpietarilagraham@mailaps.org, ringler@lanl.gov}
\cortext[cor1]{+1-(505)-665-1830, FAX: +1-(505)-665-2659 }
\cortext[cor2]{+1-(505)-667-7744}
\address{Solid Mechanics and Fluid Dynamics
  (T-3), Los Alamos National
  Laboratory, MS-B258, Los Alamos, NM 87545, USA}




\begin{abstract}

{We present a methodology to determine the best turbulence closure for
  an {eddy-permitting} ocean model {through} measurement of the
  error-landscape of the closure's subgrid spectral transfers and
  flux.  \NEU{We apply this method to 6 different closures for
    forced-dissipative simulations of the barotropic vorticity
    equation on a f-plane (2D Navier-Stokes equation).}  Using a
  high-resolution benchmark, we compare each closure's model of energy
  and enstrophy transfer to the actual transfer observed in the
  benchmark run.  The error-landscape norm enable\NEU{s} us to both
  make objective comparisons between the closures and to optimize each
  closure's free parameter for a fair comparison.  The hyper-viscous
  closure most closely reproduces the enstrophy cascade{,} especially
  at larger scales due to the concentration of its {dissipative}
  effects to the very smallest scales.  The viscous and Leith closures
  perform nearly as well{,} especially at smaller scales where all
  three models were dissipative. The Smagorinsky closure dissipates
  enstrophy at the wrong scales.  The anticipated potential vorticity
  closure was the only model to reproduce the {upscale transfer} of
  kinetic energy from the unresolved scales{,} but would require
  high-order Laplacian corrections {in order} to concentrate
  dissipation {at the smallest scales}.  The Lagrangian-averaged
  $\alpha-$model closure did not perform successfully for {forced 2D
    isotropic Navier-Stokes:} small-scale filamentation is {only
    slightly reduced} by the model {while} small-scale {roll-up is
    prevented.  {Together, this reduces} the effects of diffusion.}}

\end{abstract}

\date{\today}

\end{frontmatter}



\section{\label{sec:intro}Introduction}

Turbulence closure models are required in the dynamical cores of
global ocean-climate simulations.  While grand challenge coupled
climate simulations can use an ocean resolution of 0.1\textdegree
($\sim10\,$km) to simulate timescales of decades, resolving the
turbulent cascade for submesocale, $O(1\,$km), eddies remains
computationally unachievable.  For this reason, mesoscale ocean
large-eddy simulations (MOLES; \cite{FoKeMe2008}) are employed.  {The
  goal of a MOLES is to anticipate $1\,$km results at a much coarser
  resolution.  While such closures are sometimes compared subjectively
  by visualizing the simulation results,} what is needed is a
{prescription} to objectively {and rigorously} compare {between} the
various proposed MOLES closures.  Such a method is presented here:
\NEU{the computation of fluxes and comparison via the error-landscape measured against a high resolution benchmark.
Our application is to an idealized system, but the framework can be generalized}
for the
evaluation and development of closures applicable to World Ocean
simulations.  \NEU{In \ref{sec:appendix}, we present the details for generalization to a 3D baroclinic zonally-reentrant channel.}

Often, the closure approach taken is {to set the 
  dissipation scale equal to the grid scale.  This is equivalent to
  setting the {appropriately-averaged} grid-scale Reynolds number to unity and is accomplished by simply
  using a} constant viscosity, $\nu$, that is much larger than the
physical value ($\sim10^{-6}\,$m$^2$s$^{-1}$) so that a {numerically} resolved
simulation results.  These large viscosities, however, also
result in unphysical damping of the large scales.  To reduce this
effect while remaining in the paradigm of a linear dissipative model,
the order of the Laplacian, $\Delta=\nabla^2$, can be increased to
$\Delta^2=\nabla^4$ or higher.  Such hyper-viscous models are more
scale-selective, applying dissipation concentrated near the grid scale
{(a new dissipation scale is derived from dimensional analysis of
  the $\Delta^n$ dissipation and this scale is set equal to the grid scale).}
Turbulence is {far} more than a dissipative phenomenon, however, and purely
dissipative models cannot reproduce up-scale energy transfers due to
interactions between scales \NEU{(nor can they reproduce ``backscatter'' in the 3D case \cite{2000AnRFM..32....1M}).}

Another approach is to {use what is known about turbulent cascades and} apply dissipation only where it is required
with a spatio-temporally varying viscosity, e.g., the
{Smagorinsky \citep{Sm1963} and}
 Leith
\citep{Le1996} models.  In the {Smagorinsky} model, the global average {energy}
dissipation (due to a spatially uniform viscosity) is equated to the
local dissipation at the grid scale because the turbulence is assumed
homogeneous. The expression for $\nu_\ast(x,t)$ then follows from the
{3D turbulence} spectrum and dimensional analysis. 
{For Leith, enstrophy dissipation and the 2D turbulence spectrum in
an enstrophy cascade are used {to derive the appropriate $\nu_\ast(x,t)$}.  However,}
 the assumption of
homogeneity is controversial \citep{LeMeCo2005} and there are also
issues with vorticity dissipation at the boundaries
\citep{FoKePe2004,FoKe2005}.  Yet, the Leith model has been successful
in improving numerical stability in global eddy-permitting models
\citep{FoKeMe2008}.

{In 2D turbulent systems where enstrophy is clearly the quantity cascading to
unresolved scales, methods to} 
{dissipate potential enstrophy while conserving energy} {have merit}.  
{This {can be} accomplished by modifying the Coriolis force in the momentum equation such that the transport of potential vorticity is appropriately diffusive while still being energetically neutral  \cite{SaBa1985}. The anticipated potential vorticity method (APVM)}
 reproduces both the physical
{transfer of energy to larger scales} and the dissipation of small-scale
enstrophy \citep{VaHu1988}.  APVM has also been extended to
variable-resolution grids \cite{ChGuRi2011}, {and it has been generalized to 3D rotating Boussinesq flows \cite{2012arXiv1206.2607G}.}  However, it requires a
high-order Laplacian correction to concentrate the eddy viscosity to
the smallest scales \citep{VaHu1988}.

A {more} recent approach is to use a mathematical regularization {of the
  underlying equations,} which ensures smooth {(hence, computable)}
solutions, as the closure model: e.g., the Lagrangian-averaged
$\alpha-$model
{\citep{HoMaRa1998,1998PhRvL..81.5338C,1999PhyD..133...49C,1999PhyD..133...66C,1999PhFl...11.2343C,2001PhyD..152..505F}.}
It is dispersive rather than dissipative: the transport is by a
spatially-smoothed velocity field (filter width $\sim\alpha$).  For
three-dimensional (3D) incompressible, non-rotating, and
non-stratified flows the $\alpha-$model does not produce sizeable
computational gains because it {unphysically} develops rigid
bodies in the flow \citep{PiGrHoMi+2007}.  This limitation {disappears when modelling systems that include} a body force.  It has
been used successfully where there is a Lorentz force, in electrically
conducting fluids \citep{PiGrMiPo2009,PiGrMiPo2011}, and where there
is a Coriolis force, in rotating fluids, e.g., the two-dimensional
(2D) barotropic vorticity equation (BVE) on a $\beta-$plane
\citep{NaMa2001,HoNa2003}, the shallow water equations \citep{Wi2004},
a two-layer quasigeostrophic (QG) model \citep{HoWi2005}, {and the primitive equations \cite{2008JCoPh.227.5691H,2008JPhA...41H4009H}.}

For 2D
flows, relevant to this paper, the $\alpha-$model enhances the inverse
cascade of energy \citep{NaSh2001} and in the enstrophy cascade
regime, the rough kinetic energy and enstrophy spectra remain
unchanged ($k^{-3}$ and $k^{-1}$, respectively) in the limit
$\alpha\rightarrow\infty$ \citep{LuKuTa+2007}.  With forcing {applied in the wavenumber shells $2<k<4$ with an amplitude proportional to $\alpha^2$,}
 \cite{LuKuTa+2007} found that increasing $\alpha$ {led} to
increasing the amount of fine structure and, consequently, to the need
for increased resolution.  They posited that with forcing unscaled,
computational gains (instead of losses) might be realized.  We will
test whether or not this is so.

{The challenge in 
evaluating the effectiveness of LES closures for MOLES should already be 
clear. Not only do many possible closures exist, but these closures often 
differ at the conceptual level of how unresolved turbulent motion should be
modeled. As such, we expect that the various possible closures will each
excel in some plausible evaluation metric. The challenge is then to determine
an approach, i.e {\it an evaluation framework}, that is both unbiased and fairly
measures the effectiveness of the various closures in mimicking the influence
of unresolved scales. The goal of this contribution is to do exactly that. }

Our approach here is to begin with the simplest system that we believe
might be applicable to MOLES, with the understanding that the results
obtained in such idealized systems will have to be reevaluated as the
system complexity and realism increases. With this caveat in mind, we
solve the 2D barotropic vorticity equation (2D BVE) in a
doubly-periodic domain. The motivation for using the 2D BVE is to
exploit the similarity of the the QG vorticity equation to the 2D BVE.
{(MOLES will be applied at grid resolutions near $5-10\,$km.)}  The QG
vorticity equation has a potential enstrophy cascade of QG eddies
below the scale of the baroclinic instability.  Similarly, the 2D BVE
has an enstrophy cascade below the forcing scale, which serves here as
an analog to the scale of the baroclinic instability. Furthermore, the
robust analysis of spectral fluxes of energy and enstrophy in 3D
systems, needed for more complex, realistic flows \NEU{(see
  \ref{sec:appendix}),} is sufficiently complicated to warrant starting at a
lower spatial dimension. Since the 2D BVE system lacks the process of
baroclinic instability to initiate the turbulent mixing, we use
large-scale, slowly varying in time, wind stress to activate the
turbulence. As used in Ocean General Circulation Models (OGCMs),
quadratic bottom-drag is used to obtain realistic equilibrium
solutions.

Details of the enstrophy cascade process can be measured using {spectral}
enstrophy transfer analysis \citep{Kr1971,MaVa1993}.  {The goal of
  any LES is to anticipate higher resolution results.  This is
  accomplished by accurately modeling the interactions with the missing
  scales.  The statistics of these interactions, on a wavenumber
  basis, are measured with spectral transfer analysis.  If this
  analysis shows an accurate reproduction, we can be sure we are
  getting the right answer for the right reason.} The error-landscape
of enstrophy flux {is likely, then, the best measure of MOLES performance.  We use it to} quantify the performance of the {six} popular
MOLES closures discussed above (the two linear dissipative models and
the {four} nonlinear models derived from hypotheses about turbulence)
employing a single, exponentially convergent, numerical model, the
Geophysical High Order Suite for Turbulence (GHOST;
\cite{MiRoRe+2011}).

To compare the models, we start by computing a fully-resolved
numerical solution of a flow with a fixed, physical viscosity as the
benchmark.  This eliminates the possibility of any bias between the
parameterizations that could result from using any single MOLES at
higher resolution as the benchmark.  {It also serves as our best
  hope for the MOLES simulations: that they reproduce the benchmark.}  In Section
\ref{sec:technique}, {spectral} enstrophy transfer analysis is reviewed: its
application to MOLES and how this will be combined with the
error-landscape is given.  In Section \ref{sec:results}, the details
of the parameterizations are introduced and each parameterization is
optimized with the error-landscape technique in order to make a fair
{and objective}
comparison.

\section{\label{sec:technique}{Theory}}

\subsection{{2D turbulence}}

{For scales much smaller than the deformation radius, the quasi-geostrophic
potential vorticity equation reduces to the 2D-BVE (see, e.g., \cite{Va2006}).}
The 2D-BVE are 
\begin{eqnarray}
\partial_t\zeta + \{\psi,\zeta\} = F+\nu\nabla^2\zeta-\frac{C_D}{h}\hat{\vec{z}}\cdot\nabla\times(|u|\vec{u})\nonumber\\
\zeta = \nabla^2 \psi\nonumber\\
\vec{u}=-\nabla\times(\psi\hat{\vec{z}})\,,
\label{eq:bve}
\end{eqnarray}
where $\zeta$ is the vorticity, $\psi$ the stream function, $\vec{u}$
the 2D velocity, $F$ an external time-varying forcing to mimic
wind stress, $\nu$ the viscosity,
$\hat{\vec{z}}$ the out-of-plane unit vector,
 and $C_D/h$ the coefficient of
quadratic bottom drag. 
As a constant Coriolis parameter has no effect
on 2D motion, Eqs. (\ref{eq:bve}) also describe the 2D-BVE on a
$f-$plane.  

\begin{figure}[htbp]
\includegraphics[width=10cm]{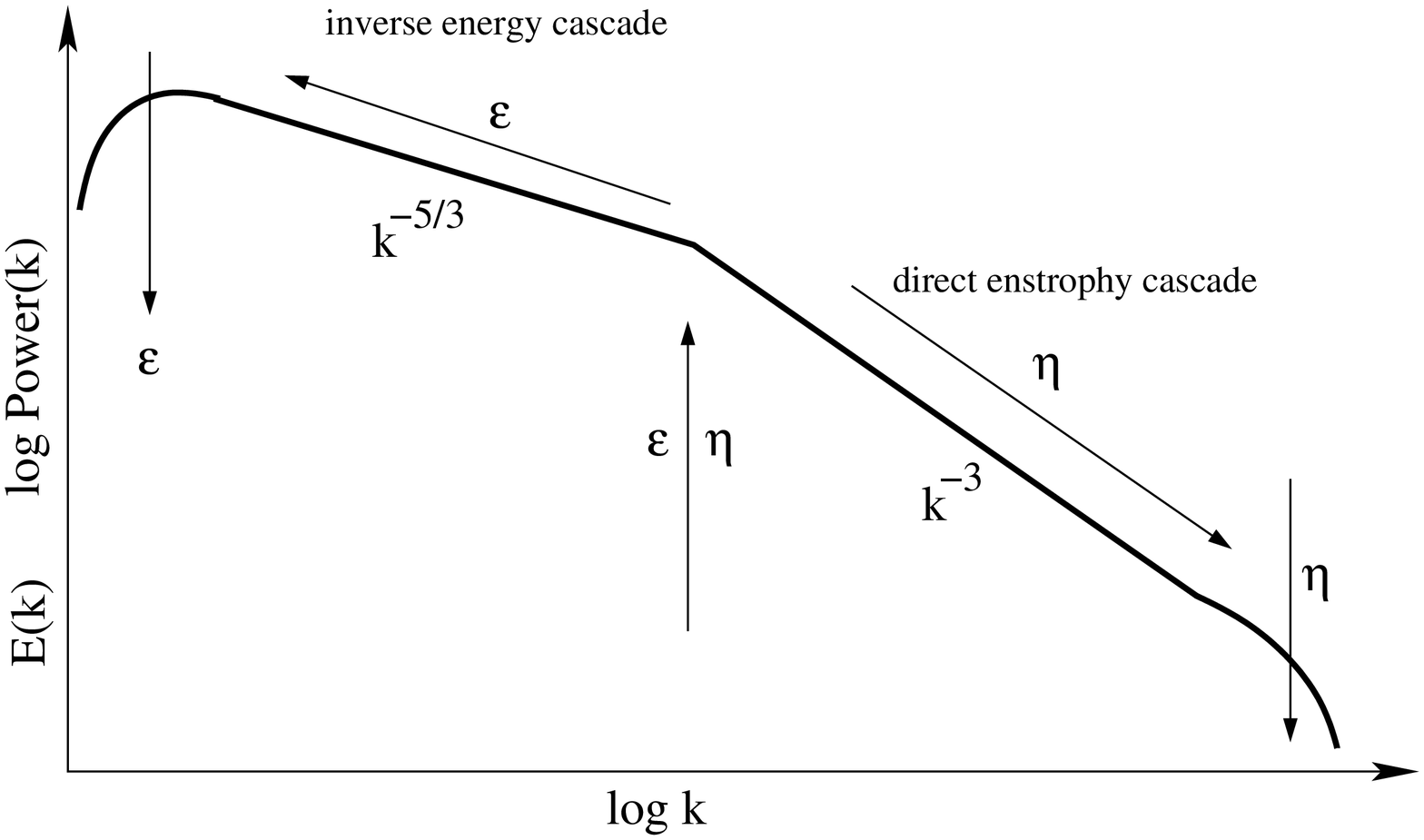}\\
\includegraphics[width=10cm]{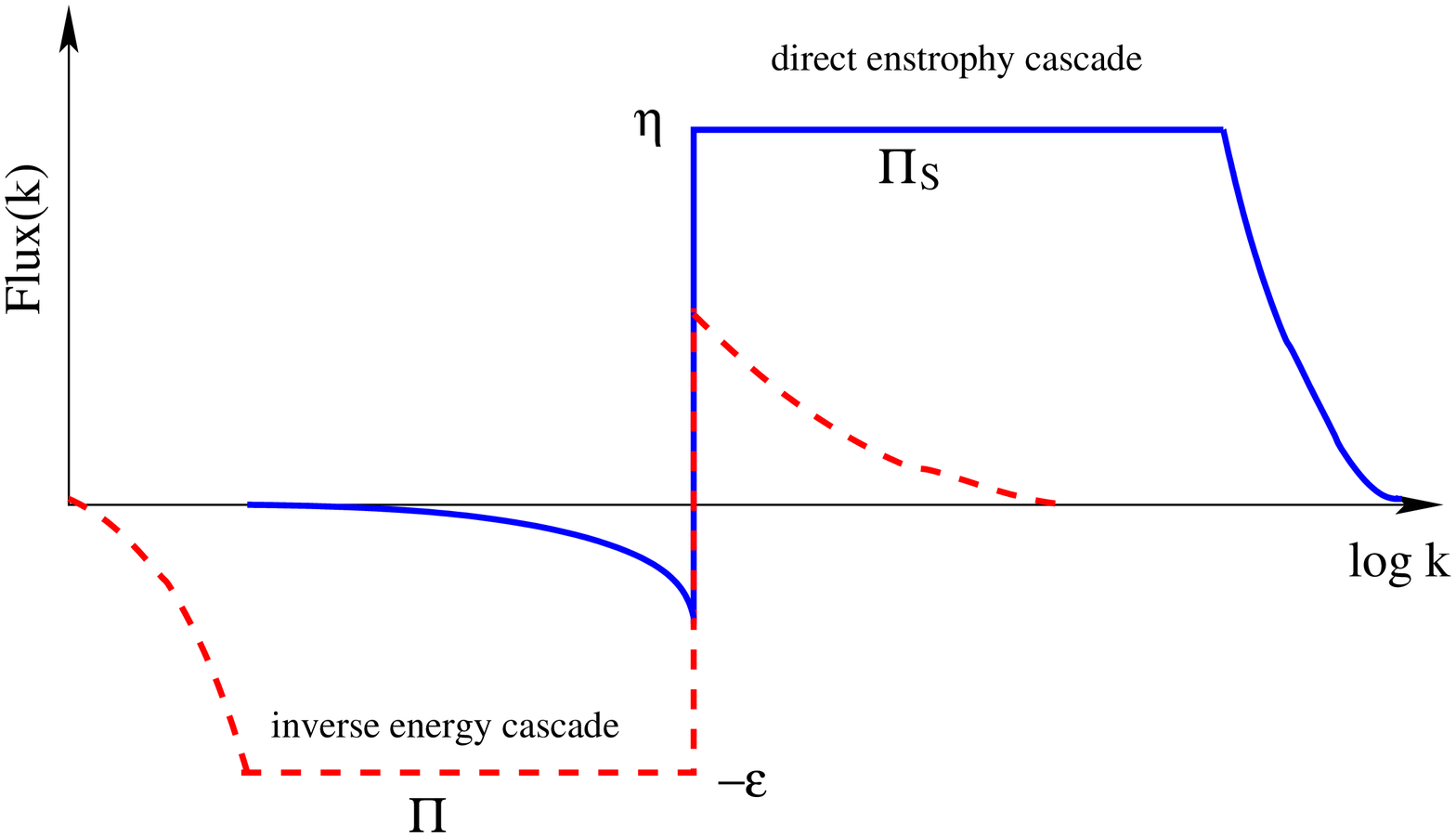}
\caption{{Cartoon depicting 2D turbulence theory: kinetic energy
    spectrum ($E(k)$, top panel) and fluxes (bottom panel) of enstrophy ($\Pi_S(k)$,  blue
    solid line) and energy ($\Pi(k)$,  red dashed line). Kinetic energy
    undergoes an inverse cascade to large scales (negative flux) at
    the kinetic energy injection rate, $\varepsilon$.  Enstrophy undergoes a
    direct cascade to small scales at the enstrophy injection rate,
    $\eta$.}}
\label{fig:2dtheory}
\end{figure}

{A general overview of 2D turbulence theory (see, e.g., \cite{Va2006})
  is presented {in} Fig. \ref{fig:2dtheory}.}  Kinetic energy,
$|\vec{u}|^2/2$, and hence enstrophy, $|\zeta|^2/2$,
are injected into the fluid.  Because both are quadratic ideal
invariants (conserved in the absence of forcing and viscosity) and
$\zeta=\hat{\vec{z}}\cdot\nabla\times\vec{u}$, enstrophy cascades to
smaller scales and energy undergoes an inverse cascade to larger
scales (Fjortoft's theorem).  {(The central point \NEU{in deriving}
  Fjortoft's theorem is \NEU{to realize that} energy, $E(k)$, and
  enstrophy, $Z(k)$, spectra are related by $k^2E(k)=Z(k)$.)}  Under
the assumption of spectral locality, forcing and dissipation
\NEU{cannot} affect the flow \NEU{except} over a finite range of
scales \NEU{near where they are prescribed: far from these ranges,} both cascades must \NEU{therefore} have a
constant flux (Fig. \ref{fig:2dtheory}, lower panel).  The constant
flux cascade regimes are called inertial ranges because only the
inertial terms, $\vec{u}\cdot\nabla\vec{u}$ for energy and
$\vec{u}\cdot\nabla\zeta=\{\psi,\zeta\}$ for enstrophy, are
non-negligible.  Dimensional analysis after equating constant fluxes
to the inertial terms for energy and enstrophy yields a $k^{-5/3}$
energy spectrum in the inverse energy cascade and a $k^{-3}$ energy
spectrum in the enstrophy cascade (Fig. \ref{fig:2dtheory}, upper
panel).  {Fine theoretical details such as the logarithmic correction
  to the $k^{-3}$ spectrum \citep{Kr1971} and arguments about locality
  \citep{XiWaCh+2008} have here been omitted.  }

\subsection{Transfer analysis}

{The $\{\psi,\zeta\}$ term is the only non-negligible term
in the enstrophy cascade regime.  It will also be shown in Section
\ref{sec:LES} to be the term whose small-scale interactions we need to
parameterize.   It is thus the focus of our comparison methodology.
Other terms in the analysis will heretofore be abbreviated as $\mathcal{F}$
for forcing, $\mathcal{D}$ for dissipation, and $\mathcal{Q}$ for large-scale
drag (where, e.g., $\mathcal{F}\equiv\zeta F$).}
The time evolution of enstrophy at any physical position is given by the
enstrophy-balance equation,
\begin{eqnarray}
\partial_t\frac{1}{2}\zeta^2=\zeta\partial_t\zeta=
-\zeta\{\psi,\zeta\}+\mathcal{F}+\mathcal{D}+\mathcal{Q}\,.
\end{eqnarray}  The time evolution of the
enstrophy spectrum at wavenumber $k$, $Z(k)$, is similarly,
\begin{equation}
\partial_tZ(k) = \hat{\zeta}^\ast\partial_t\hat{\zeta}= S(k)
+\mathcal{F}(k) +\mathcal{D}(k) +\mathcal{Q}(k) \,,
\end{equation}
where $S(k)$ is the enstrophy transfer function (i.e., net enstrophy
received by wavenumber $k$ from all other wavenumbers),
\begin{equation}
S(k)=-\hat{\zeta}^\ast(k)\widehat{\{\psi,\zeta\}}(k)\,,
\end{equation}
and where the Fourier transform is represented by
$\hat{\cdot}$ and complex-conjugation by
$\cdot^\ast$.
The flux of enstrophy through wavenumber $k$, i.e., the sum
of {the rate of change of} enstrophy {leaving} all wavenumbers $\leq k$ {and going} to wavenumbers $>k$ {(i.e., moving to smaller scales)}, is
given by
\begin{equation}
\Pi_S(k) = -\int_\NEU{0}^k S(k') dk'\,,
\end{equation}
that is, the total {rate} of enstrophy flowing past wavenumber $k$ {to larger wavenumbers.}
The divergence of the flux is the transfer, {$S(k)$.} 
{Because of the relation between energy and enstrophy spectra, the
  transfer of energy is $T(k)=S(k)/k^2$.}

\subsection{\label{sec:LES}MOLES}

To reduce computational cost, MOLES solve only the largest scales of a
flow.  The remaining {\sl unresolved} scales {from the anticipated higher-resolution simulation} are filtered out.  The
filtering operation is indicated by $\bar{\cdot}$ and the resulting
equations are
\begin{equation}
\partial_t\bar{\zeta} + \{\bar{\psi},\bar{\zeta}\} = 
\sigma + \bar{F}+\nu\nabla^2\bar{\zeta}
-\overline{\frac{C_D}{h}{\hat{\vec{z}}\cdot}\nabla\times(|u|\vec{u})}\,,
\label{eq:LES}
\end{equation}
where we have defined the
subgrid term $\sigma\equiv
-\overline{\{\psi,\zeta\}}+\{\bar{\psi},\bar{\zeta}\}$.  The subgrid
term is the effects on the resolved scales by unresolved fluid
motions. {How well it is modeled is the measure of the success of the MOLES.}  (Note that $\sigma = {\hat{\vec{z}}\cdot}\nabla\times\nabla\cdot\tau$ where
$\tau = -\overline{\vec{u}\vec{u}}+\bar{\vec{u}}\bar{\vec{u}}$, the
\NEU{momentum-equation} LES subgrid stress tensor.)  The time evolution of the
enstrophy spectrum is now given by
\begin{equation}
\partial_tZ(k) =  \bar{S}(k)
+ L(k)+{\mathcal{F}}(k) +\bar{\mathcal{D}}(k) +\mathcal{Q}(k)\,,
\end{equation}
where $\bar{S}(k)$ is {the rate of} enstrophy received by wavenumber $k$ from all other {\sl resolved} wavenumbers,
\begin{equation}
\bar{S}(k)=-\hat{\bar{\zeta}}^\ast\widehat{\{\bar{\psi},\bar{\zeta}\}}\,,
\end{equation}
and $L(k)$ is the {rate of} enstrophy received from all {\sl unresolved} wavenumbers,
\begin{equation}
L(k) = \hat{\bar{\zeta}}^\ast\hat{\sigma}\,.
\end{equation}
{Note that the rate of energy received from unresolved wavenumbers is $L(k)/k^2$.} 
{How closely the sum of enstrophy transfer functions, from resolved and unresolved wavenumbers, approximate} the enstrophy transfer function from a
fully resolved system,
\begin{equation}
\bar{S}(k)+L(k) \approx S(k)\,,
\end{equation}
(for all wavenumbers smaller than the filter wavenumber) {is the spectral measure of the success of the model.}
\NEU{For $k$ in the inertial range, $S(k)=0$ and a successful model will produce $L(k)\approx-\bar{S}(k)$.}
 The flux
of enstrophy through wavenumber $k$ due to resolved and modeled
interactions is given by
\begin{equation}
\Pi_T(k) = -\int_\NEU{0}^k {\big{[}}\bar{S}(k') + L(k'){\big{]}} dk'\,.
\end{equation}

\subsection{Objective method: error-landscape of enstrophy flux}

To objectively compare parameterizations, we make use of the
error-landscape assessment \citep{MeGeBa2003,MeSaGe2006,MeGeSa2007,Me2011}
{on the enstrophy flux.}
{We modify the method of \cite{Me2011} and employ $L^1$ instead of $L^2$
error norms,
\begin{equation}
D_p = \frac{\int_{1}^{k_{max}}\big{|}\Pi_S(k)-\Pi_T(k)\big{|}k^pdk}{\int_{1}^{k_{max}}\big{|}\Pi_S(k)\big{|}k^pdk}\,,
\end{equation}
where $k_{max}$ is determined from the MOLES resolution (see below).
We chose $p=0$ to obtain a good balance between the smaller resolved
scales and the largest, less model-sensitive, scales.  The optimal
parameter value for each method is the point where this error norm
is minimized.  (The term landscape is intuitive for two-parameter
models.)  Inter-model comparisons are also made using the $D_0$ norm.}

\subsection{{Design of numerical experiments}}

We employ a well-tested parallelized pseudo-spectral code
\citep{MiRoRe+2011}. The computational box has size $[2\pi]^2$, and
wave numbers vary from $k_{min}=1$ to $k_{max}=N/3$ using a standard
2/3 de-aliasing rule, where $N$ is the number of grid points per
direction.  To cast our results in meaningful units, the results are
dimensionalized by $l=l_0l'$, $t=t_0t'$ where $\cdot'$ indicates
non-dimensionalized pseudo-spectral result and $l_0 = 504 \times 10^4/
\pi\,$m and $t_0=1.2\times10^6$s.  {To spin up our runs we begin with
a $1008^2$ simulation (dimensionalized grid spacing $\NEU{\Delta x=}10\,$km) initialized
with a few large-scale Fourier modes.  The forcing is designed to mimic
wind-stress at $k=4$: 
\begin{equation}
F = A(t)\bigg{[}\cos{(4y+\phi_y)} - \cos{(4x+\phi_x)}\bigg{]}\,,
\end{equation}
where $\phi_x=\pi\sin(1.2\times10^{-6}\,$s$^{-1} t)$
and	       $\phi_y=\pi\sin(1.2\times10^{-6} \pi\,$s$^{-1} t/3)$
so that the wind varies with a period of about 60 days.
The coefficient $A$ is dynamically controlled to hold a stead{y}  enstrophy
injection rate of $1.75\times10^{-18}\,$s$^{-3}$ to reduce the amount of required statistics to measure a constant flux cascade, i.e.,
\begin{equation}
\frac{\int \zeta F dA}{ \int dA} = 1.75\times10^{-18}\,s^{-3}\,.
\end{equation}
Time step is $600\,$s, $\nu=88\,$m$^2$s$^{-1}$, and $C_D/h=1.25\times10^{-8}\,$m$^{-1}$.
The resulting root-mean-squared velocity is $v_{rms}=2.6\,$ms$^{-1}$
and the forcing scale ($k=4$) is $L_F=2520\,$km.   
The corresponding forcing-scale turnover time is $11\,$days and the
Reynolds number is $Re\equiv v_{rms}L_F/\nu\approx75,000$.
{Simulations are integrated} for over $1300\,$days.  The
final turbulent state of this run is used as initial conditions for
the benchmark and MOLES runs at $\nu=1.375\,$m$^2$s$^{-1}$.  }

\section{\label{sec:results}Analysis of parameterizations}

The goal of MOLES is to anticipate higher resolution results at an
affordable resolution by representing the effects of the unresolved
eddies.  To avoid any bias between the parameterizations, we use as
the benchmark a fully resolved direct numerical solution (DNS) at a
resolution of $8192^2$ of a flow with $\nu=1.375\,$m$^2$s$^{-1}$.
Each MOLES is then run at a resolution of $1008^2$ and tested for its
ability to reproduce the benchmark. This allows us to test the models'
representations against a known solution: a DNS flow.  Accordingly,
the MOLES simulations also must use $\nu=1.375\,$m$^2$s$^{-1}$ in
addition to the subgrid term or they should be compared, instead, to a
$\nu=0$ benchmark which cannot be produced.

\begin{figure}[htbp]
\includegraphics[width=8cm]{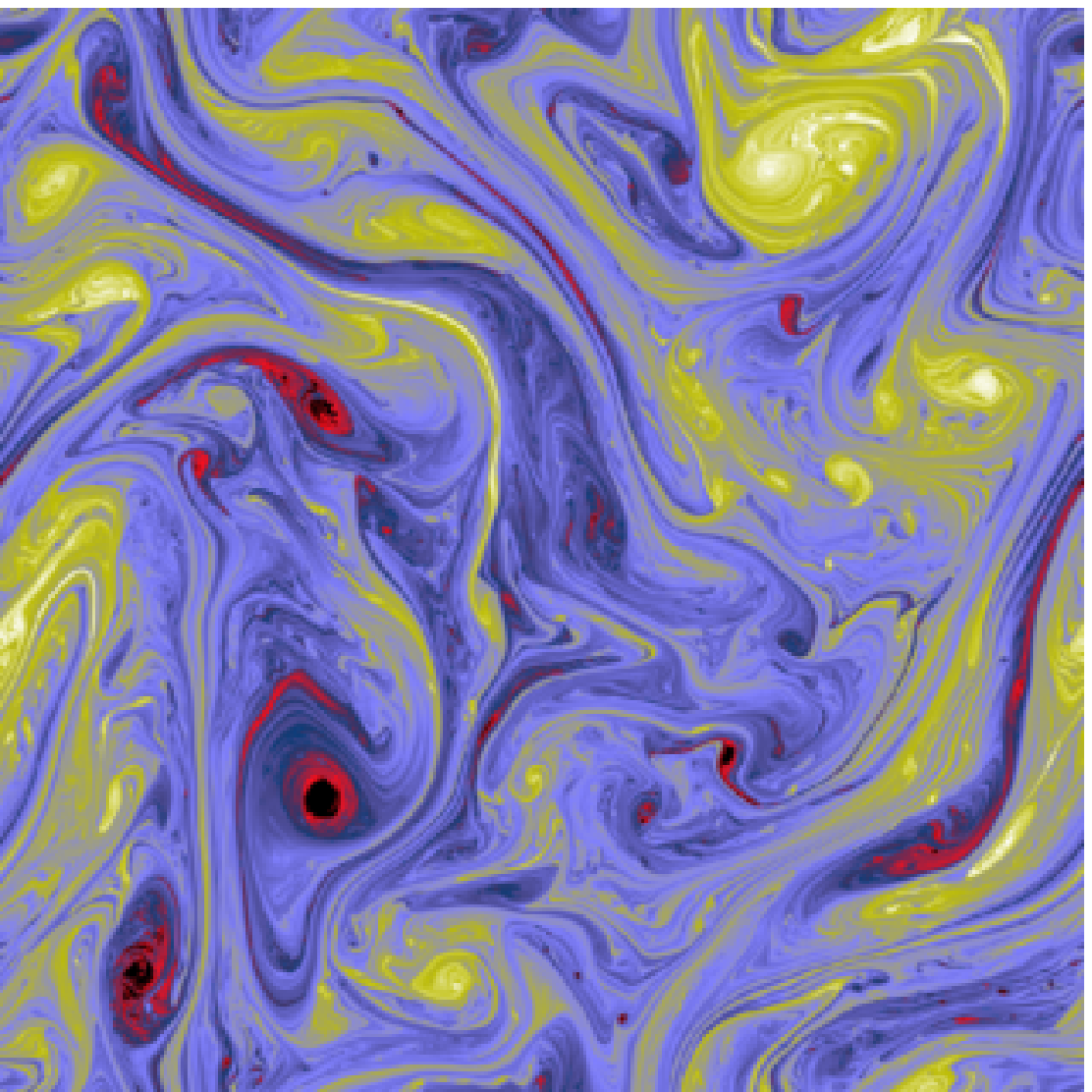}
\includegraphics[width=8cm]{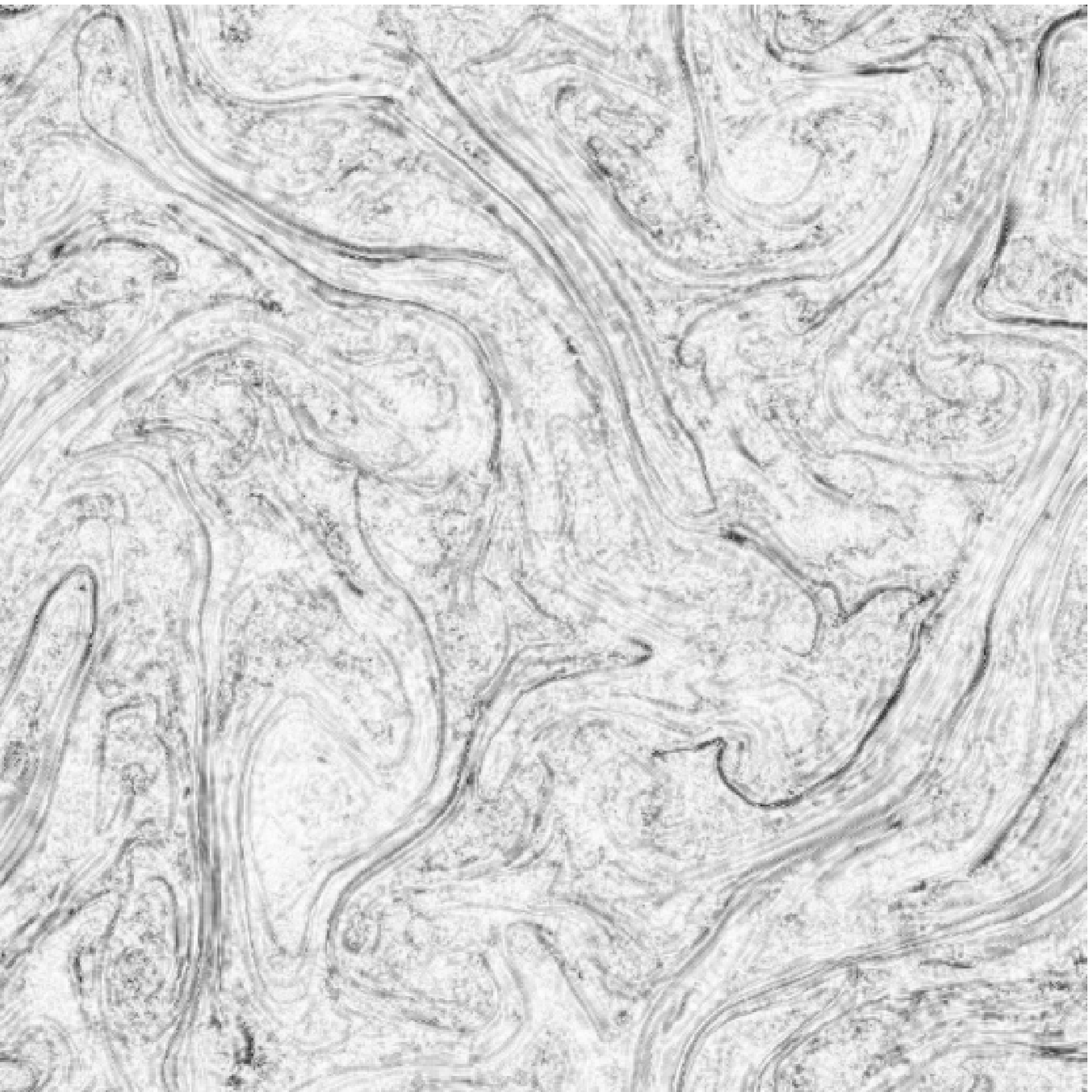}\\
\hspace{0.01cm}\\ 
\includegraphics[width=8cm]{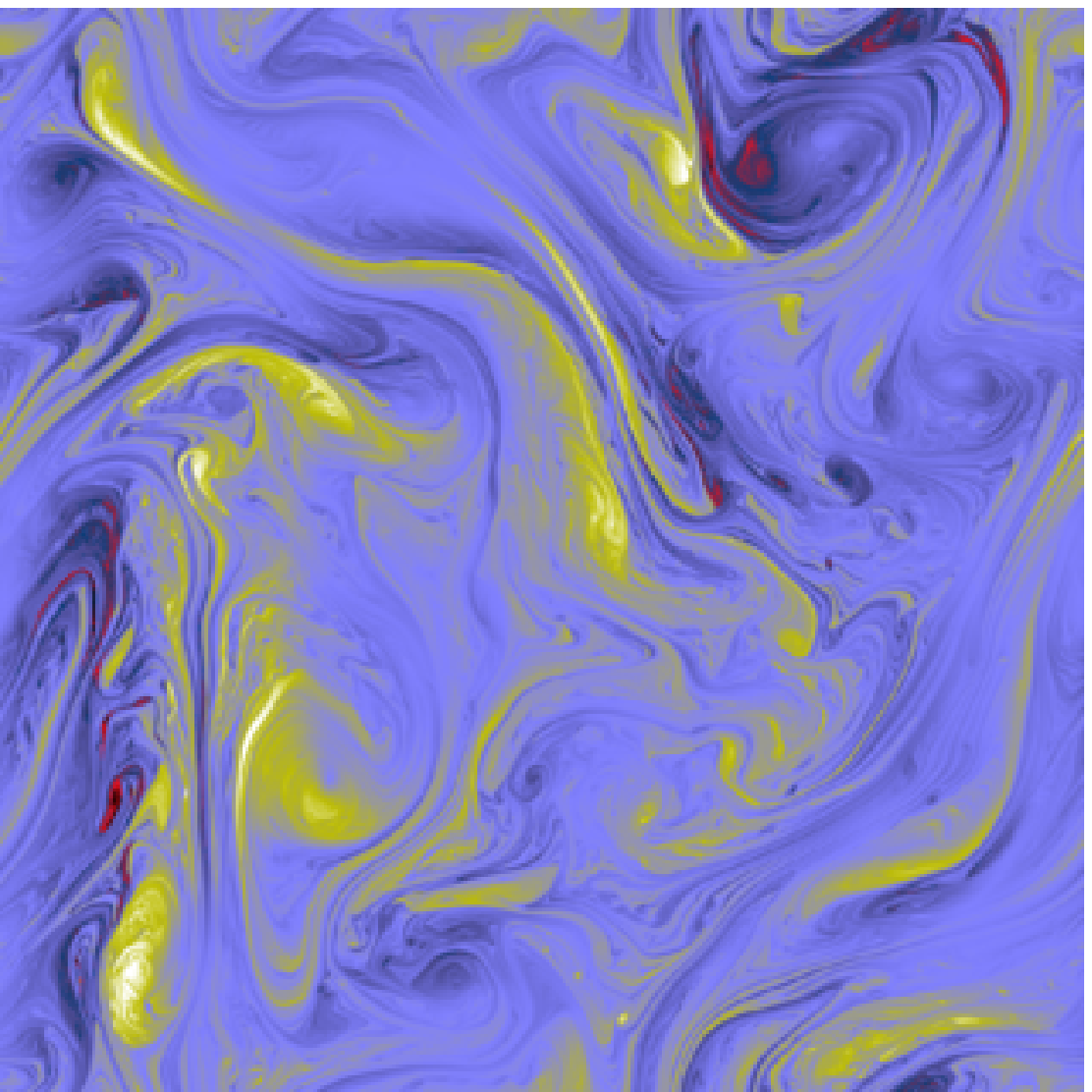}
\includegraphics[width=8cm]{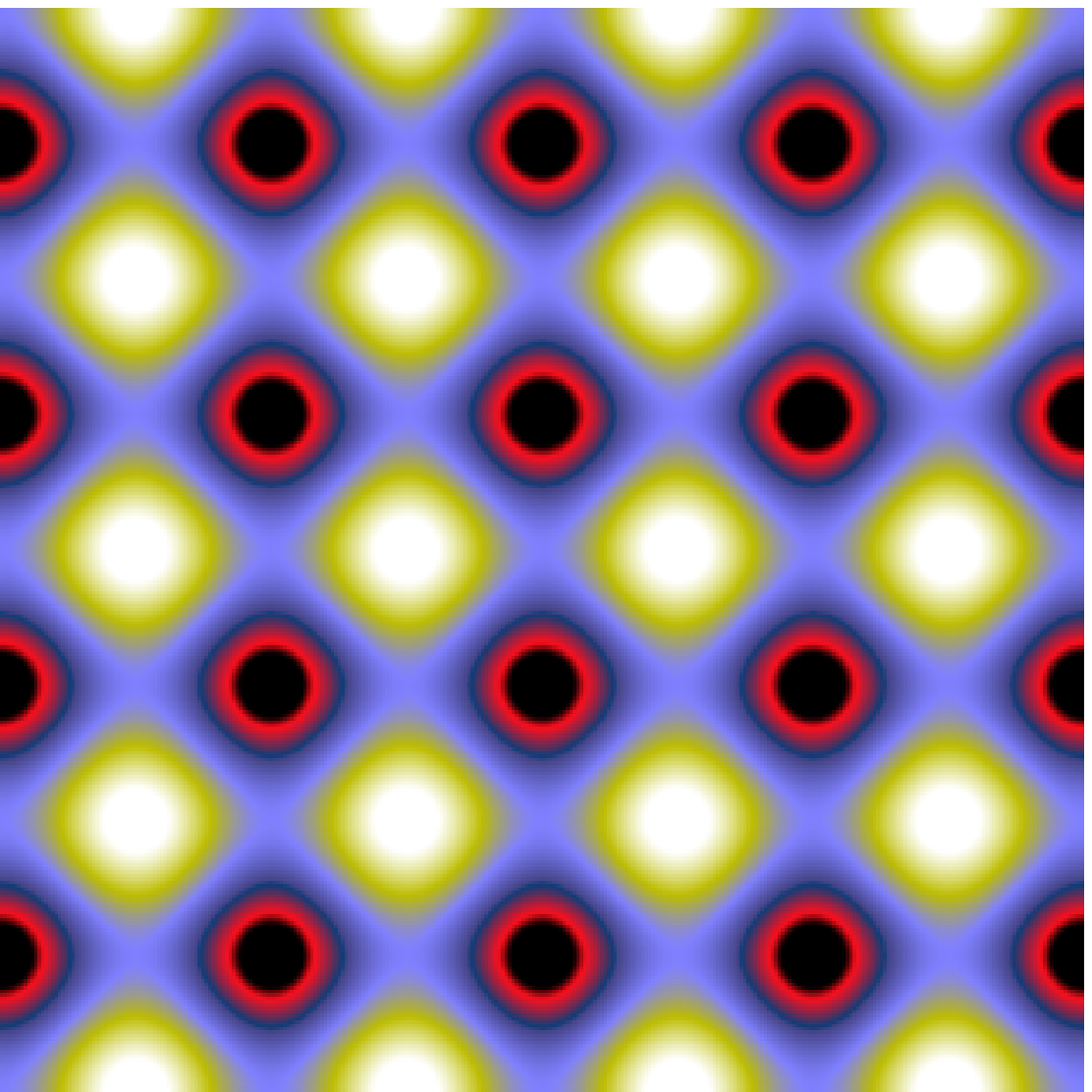}\\ 
\caption{{$8192^2$
    benchmark; snapshot at 390 days for (Upper Left) vorticity,
    $\zeta$, with thresholds $\pm1.5\times10^{-5}\,$s$^{-1}$
    (counter-clockwise vorticity is shown in yellow; clockwise in
    red); (Upper Right) absolute value of vorticity tendency due to dissipation,
    $\nu\nabla^2\zeta$, black pixels are
    $2.25\times10^{-7}\,$s$^{-2}$; (Lower Left) vorticity tendency
    due to quadratic drag, $-\frac{C_D}{h}\nabla\times(|u|\vec{u})$,
    with thresholds $\pm1.38\times10^{-6}\,$s$^{-2}$; (Lower Right)
    vorticity tendency due to forcing, $F$, with thresholds
    $\pm1\times10^{-4}\,$s$^{-2}$.  }}
\label{fig:benchreal}
\end{figure}

{The benchmark {is} run for 390 days,  $v_{rms}=2.6\,$ms$^{-1}$ and
the corresponding forcing-scale turnover time is $11\,$days. The
Reynolds number is $\approx4.8\times10^6$.
A snapshot of the vorticity of the benchmark run is shown in the
  Upper Left panel of Fig. \ref{fig:benchreal}.  There are several
  large vortices of both signs.  Over time, vortices stretch and fold
  vortex filaments into the fine-scale features as seen.  This is the
  enstrophy cascade process.  This simulation is completely resolved
  and this cascade is arrested at the smallest scales by dissipation
  (Upper Right panel in Fig. \ref{fig:benchreal}).  Energy is injected
  by the forcing term (Lower Right panel in Fig. \ref{fig:benchreal})
  at a constant injection rate: an inverse cascade of energy and
  direct cascade of enstrophy result.  The quadratic drag term serves
  to arrest the inverse cascade of kinetic energy and primarily
  removes energy (and enstrophy) at the largest scales.  Though, it
  does remove both from a wide range of scales (Lower Left panel in
  Fig. \ref{fig:benchreal}).  }

\begin{figure}[htbp]
\includegraphics[width=10cm]{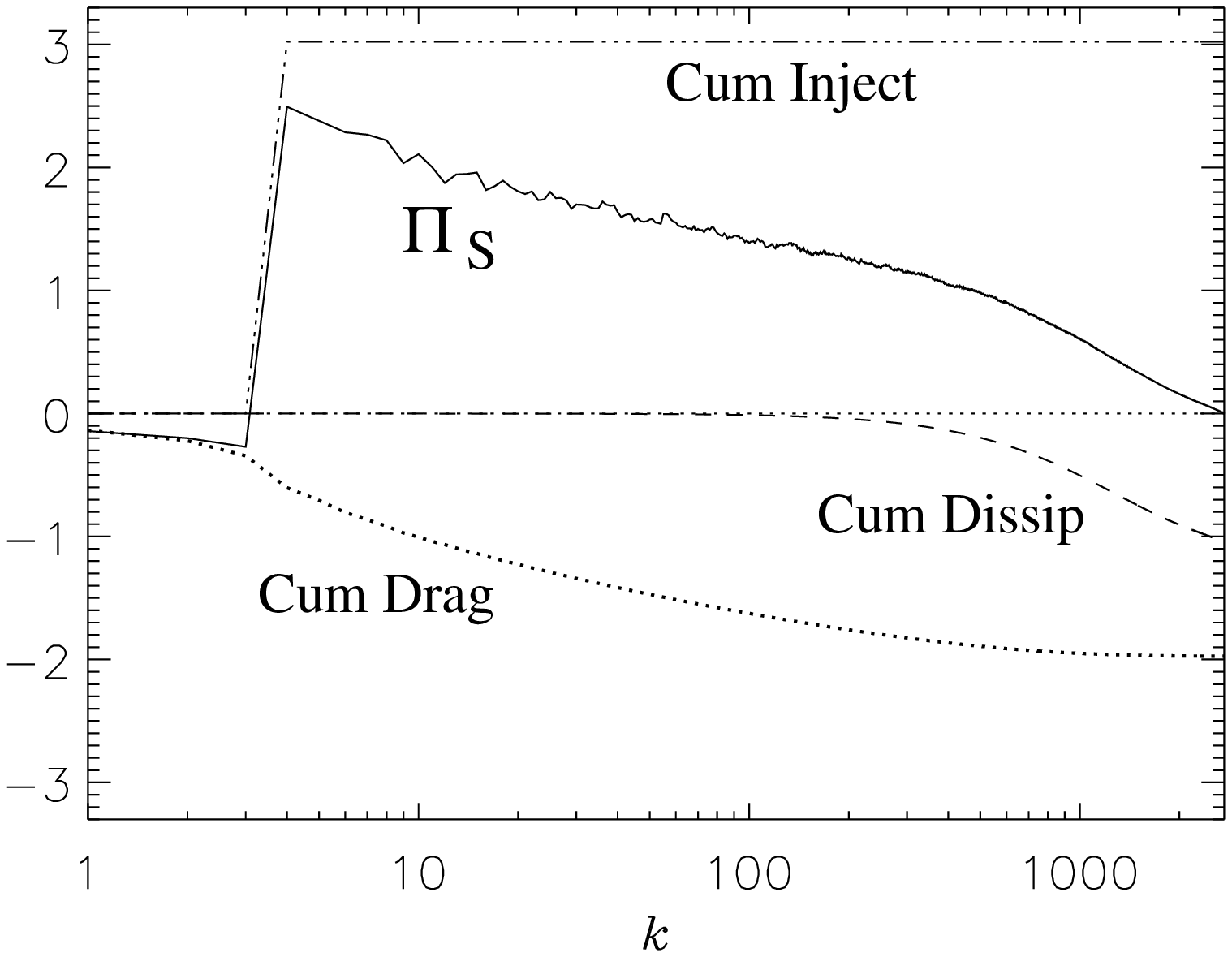}\\
\includegraphics[width=10cm]{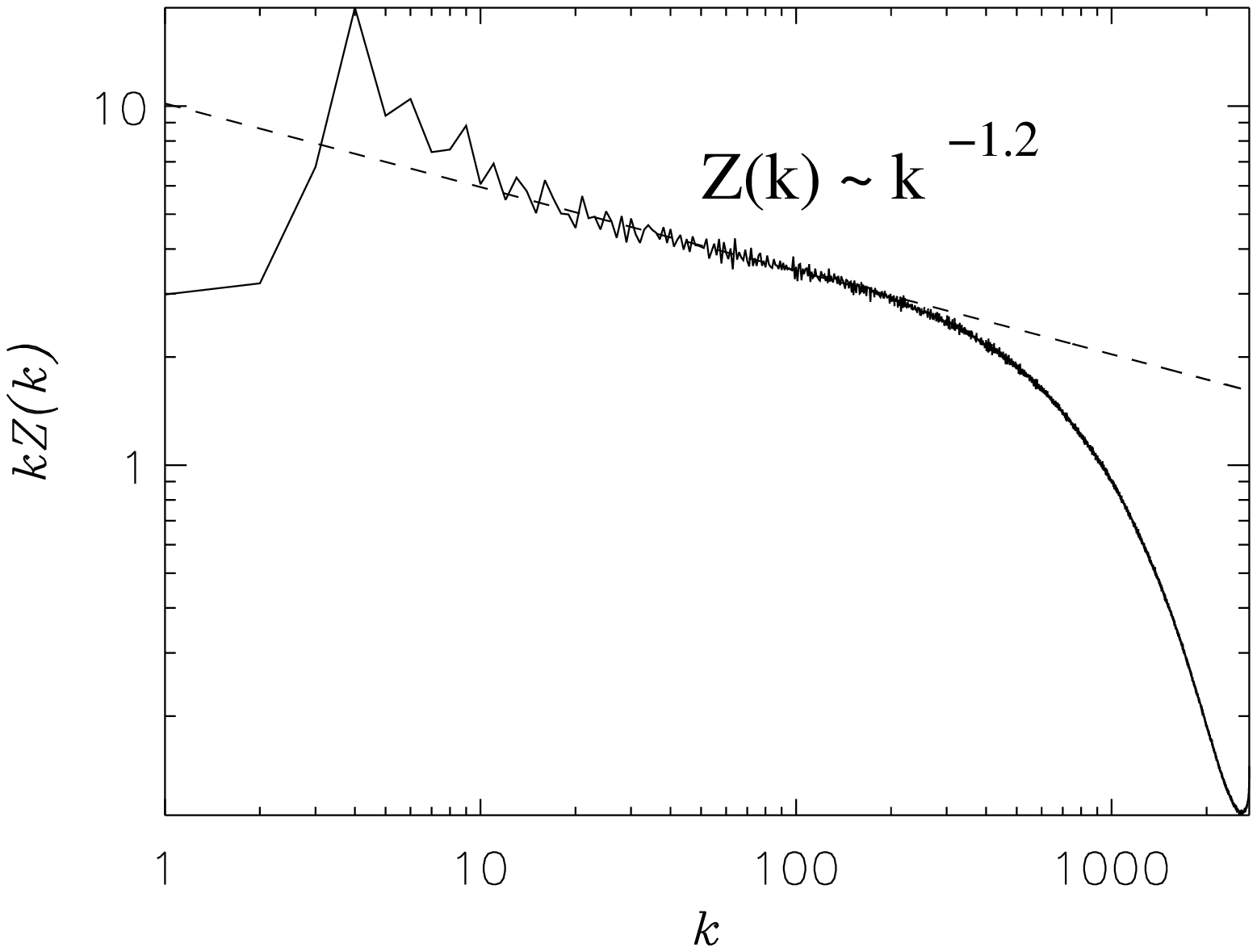}
\caption{Benchmark run: (Top) Enstrophy flux
  ($\Pi_S(k)$, solid) and cumulative enstrophy injection
  (
dash-triple-dotted), dissipation (
dashed), and quadratic drag
  (
dotted).  As quadratic drag operates at all but the dissipative
  scales, a constant enstrophy flux range is not seen. (Bottom) Compensated enstrophy spectrum,
  $kZ(k)$, versus wavenumber, $k$, for $8192^2$ BVE benchmark.    Quadratic drag acts at all scales
  and precludes a pure $Z(k)\sim k^{-1}$ spectrum.   }
\label{fig:bench}
\end{figure}

 The flux and resulting enstrophy spectrum for the benchmark are shown
 in Fig. \ref{fig:bench}.  A power-law spectrum, $Z(k)\sim k^{-1.2}$,
 is observed in the enstrophy cascade inertial range.  It is steeper
 than the predicted $k^{-1}$ spectrum due to the quadratic drag which
 acts at all scales of the flow: the difference between the enstrophy
 flux (solid line) and a constant flux is exactly the cumulative drag
 (dotted line).  This steeper spectrum is similar to the result for
 linear drag \citep{DaGu2001}.  {Note that dissipation is not
   significant for wavenumbers, $k<300$.  Reproducing this flow at a
   resolution of $1008^2$ ($k_{max}=336$) will thus be a {onerous test
   for} the parameterizations.}

\begin{figure}[htbp]
\includegraphics[width=10cm]{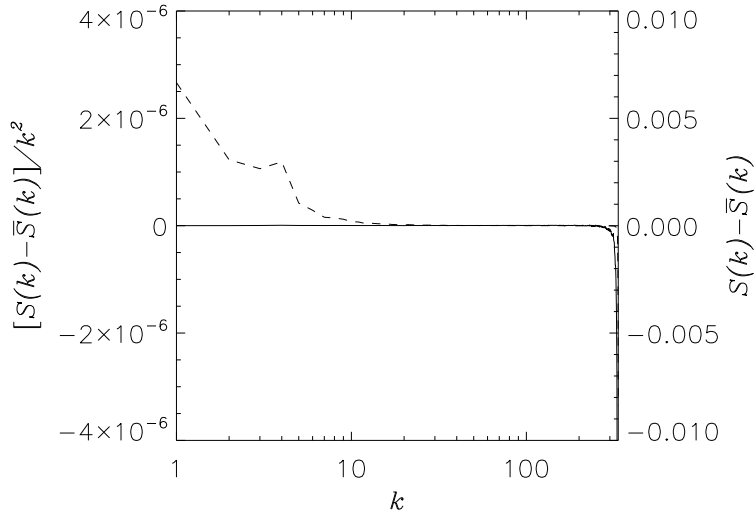}
\caption{\NEU{Benchmark run:
Transfers with what will be unresolved scales for MOLES simulations
for enstrophy, $S(k)-\bar{S}(k)$ (solid line),
and energy,  $[S(k)-\bar{S}(k)]/k^2$ (dashed line). An ideal MOLES would
exactly reproduce these transfers with $L(k)=S(k)-\bar{S}(k)$.}}
\label{fig:bench2}
\end{figure}

\NEU{The benchmark run contains all scales of motion at this $Re$.  It
  can be used to calculate the true transfers with scales that will be
  unresolved at MOLES resolution by {spectral cut-off filtering}
  the benchmark run down to a resolution of $1008^2$.  These subgrid
  transfers for energy and enstrophy are plotted in
  Fig. \ref{fig:bench2}.  The effects of the subgrid scales are to
  remove enstrophy from a narrow band of wavenumbers near the
  resolution limit and to generate a small amount of energy at the
  very largest scales.  These transfers can also been seen in Fig. 7
  of \cite{VaHu1988}.  The upscale energy transfer is a strong function
  of the resolution, $\Delta x$: as can be seen by comparison with
  \cite{VaHu1988}, the smaller $\Delta x$ is, the smaller in magnitude
  is the upscale energy transfer.  In fact, in the limit as $\Delta x$
  approaches $\nu^{1/2}$ times some constant, both subgrid transfers
  will tend to zero \cite{LuKuTa+2007}.  However, at fixed $\Delta x$ both transfers will
  tend to a non-zero function of $k$ that remains the same in the
  limit of zero viscosity.  This is due to spectral locality: only
  those scales nearest $\Delta x$ will contribute to the transfers.
  As $\nu$ decreases, and more and more scales are added, they will
  contribute less and less to the transfers for $k<1/\Delta x$.}

\NEU{Given that an ideal MOLES will have $L(k)$ that exactly reproduces
  Fig. \ref{fig:bench2}, we can anticipate the performance of the
  proposed closures.  None of the purely dissipative models (viscous,
  hyper-viscous, Leith, or Smagorinsky) will be able to reproduce the
  upscale transfer of energy.  The hyper-viscous model should better
  confine enstrophy dissipation to large wave numbers as its subgrid
  term contains fourth-order derivatives compared to second-order for
  the viscous model and first-order derivatives of the product of
  first-order derivatives for Leith.  Smagorinsky is derived for 3D
  flow and is not expected to perform well in 2D.  It has been
  previously shown that AVM can produce the correct forms of the
  transfers if high enough order viscosities and small enough
  anticipation times are employed \cite{VaHu1988}. The $\alpha-$model
  is non-dissipative, but could {\sl potentially} transport energy in
  the correct direction \cite{NaSh2001}.  }

\subsection{\label{sec:viscous}Linear viscous parameterizations and their performance}

The simplest parameterization is to assume the main
effect of subgrid turbulence is dissipative.  Accordingly, the
viscosity is often increased until a numerically resolved solution is
possible.  The subgrid term, $\sigma$, in the MOLES equation,
Eq. (\ref{eq:LES}) is then
\begin{equation}
\sigma = \big{(}\nu'-\nu\big{)}\nabla^2\zeta\,,
\end{equation}
with $\nu' \gg \nu$.  A slightly more sophisticated 
approach is to add higher-order dissipation, hyper-viscosity, e.g.
\begin{equation}
\sigma = \nu_4\nabla^4\zeta\,,
\end{equation}
or even higher order.  We focus on $\nabla^2$ and $\nabla^4$
parameterizations here.

\begin{figure}[htbp]
\includegraphics[width=10cm]{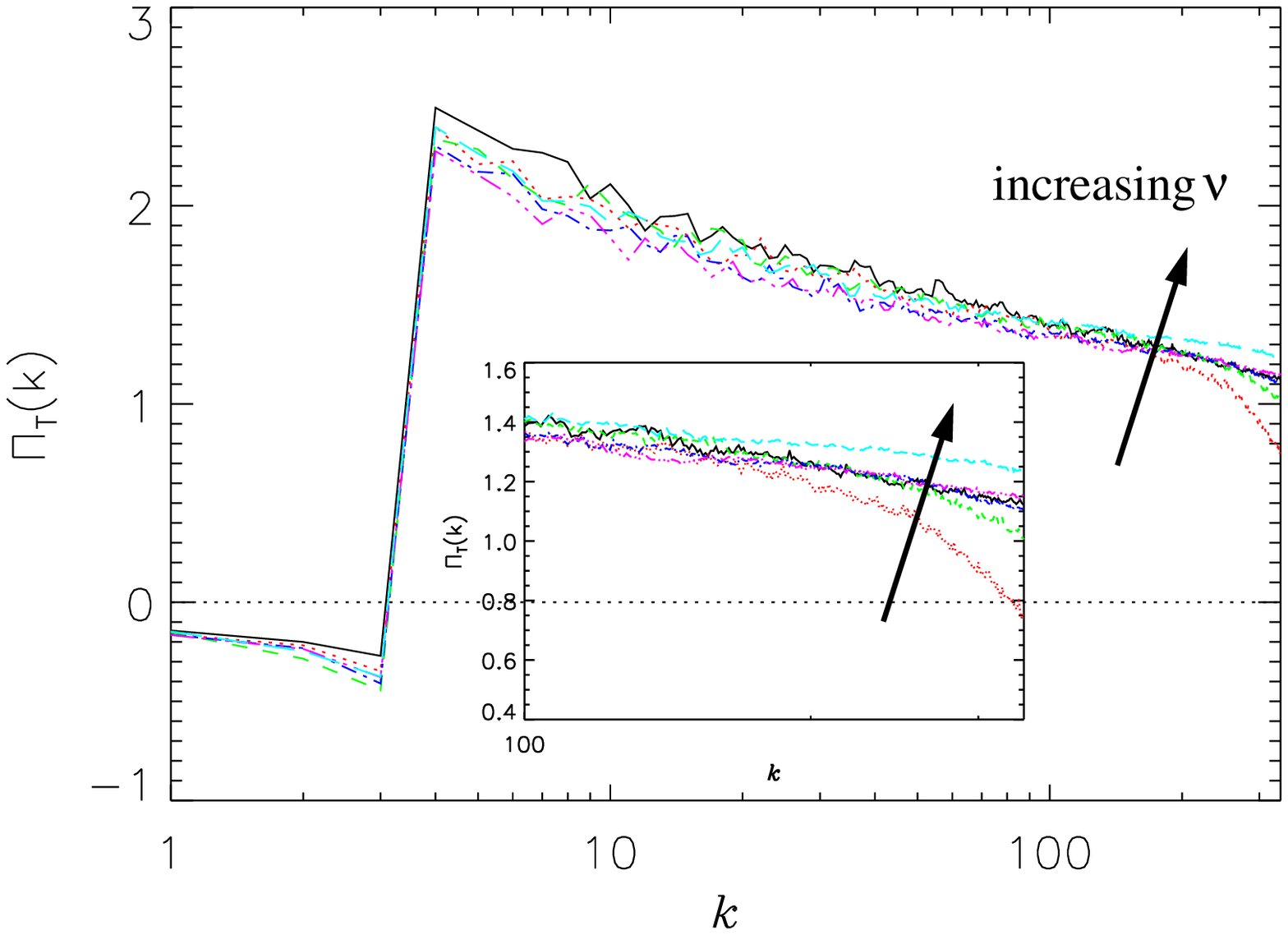}\\
\includegraphics[width=10cm]{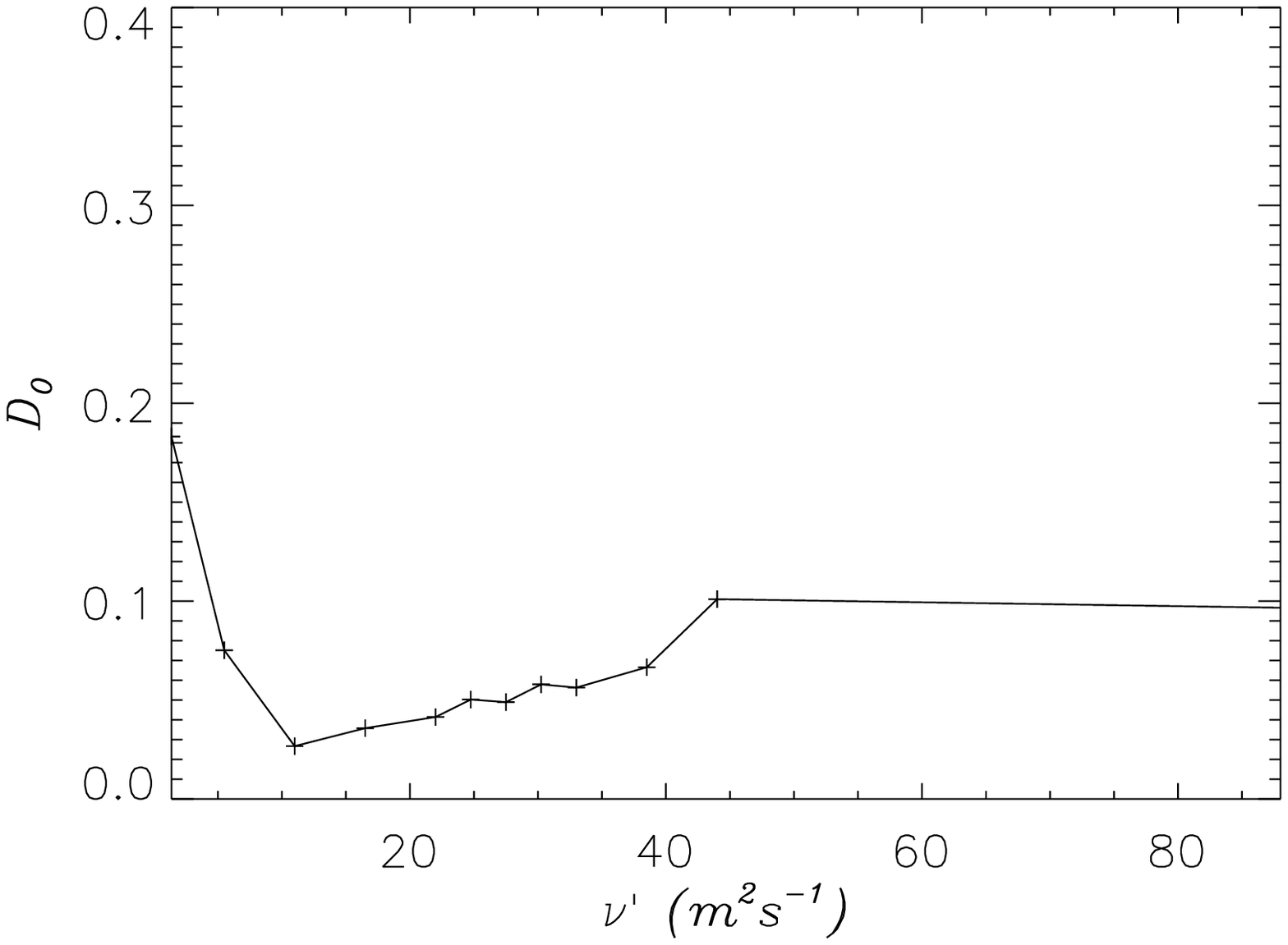}
\caption{Viscous model: (Top) Modeled flux, $\Pi_T(k)$, for
  ${\nu'}=5.5\,$m$^2$s$^{-1}$ (red dotted), $11\,$m$^2$s$^{-1}$
  (green dashed), $16.5\,$m$^2$s$^{-1}$ (blue dash-dotted),
  $22\,$m$^2$s$^{-1}$ (pink dash-triple-dotted), and
  $24.75\,$m$^2$s$^{-1}$ (cyan long-dashed) and $\Pi_S(k)$ for
  $8192^2$ BVE benchmark (solid black).  (Bottom) Flux error-landscape
  {norm $D_0$.}  The optimal value is
  ${\nu'}=11\,$m$^2$s$^{-1}$.}
\label{fig:viscoustrue}
\end{figure}

We apply the error-landscape of enstrophy flux technique to
optimize the viscous model.  The modeled flux, $\Pi_T(k)$, for the
viscous model is shown in Fig.  \ref{fig:viscoustrue}.  Note that as
the viscosity is varied, the modeled flux brackets both sides of the
benchmark flux.  This suggests an optimal $\nu'$ for the model should
be indicated by the enstrophy flux error-landscape.  Indeed, {$D_0$} has its minimum for
${\nu'}=11\,$m$^2$s$^{-1}$.  This is the optimal viscous model which we will
compare to the other parameterizations.

\begin{figure}[htbp]
\includegraphics[width=10cm]{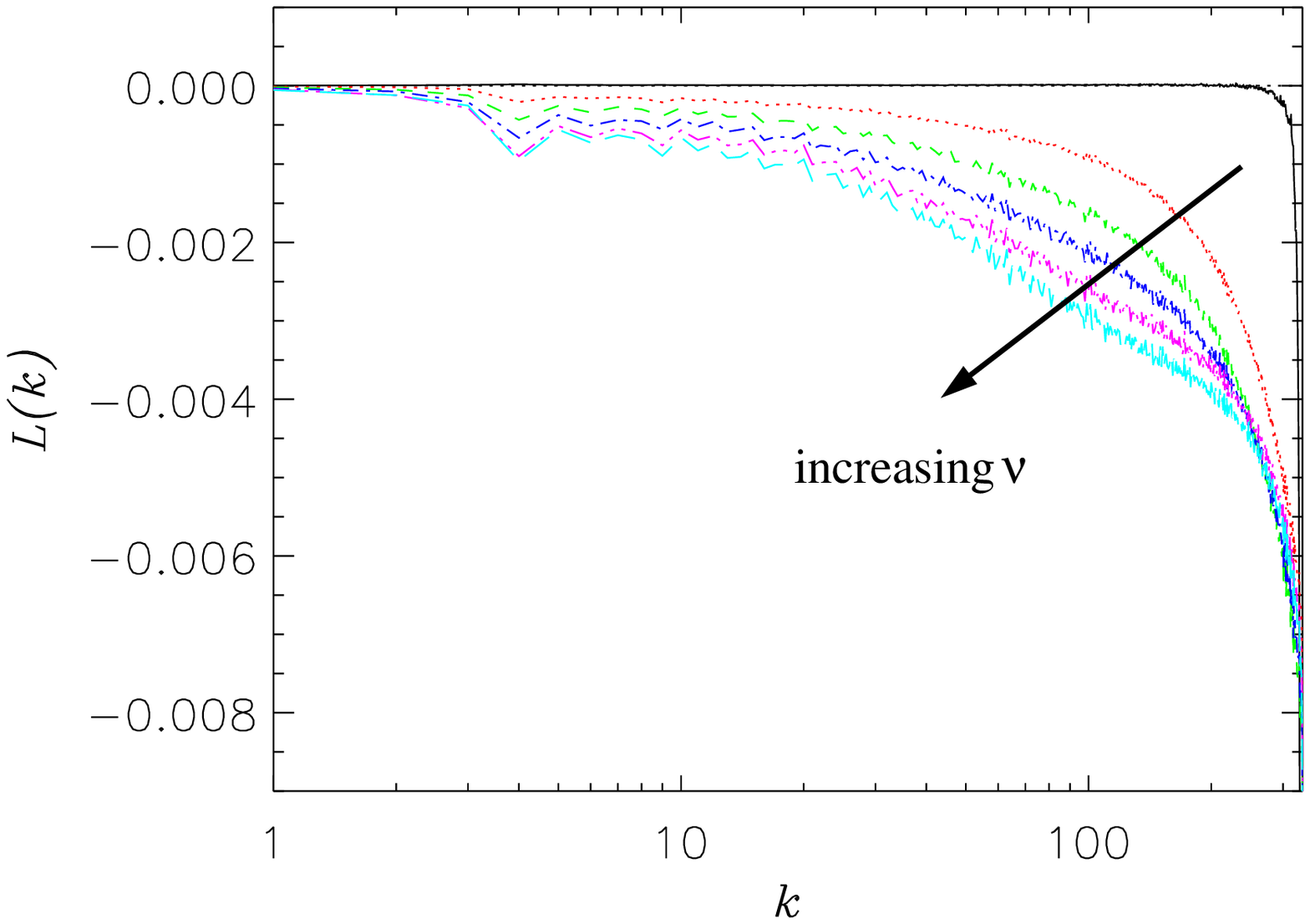}\\
\includegraphics[width=10cm]{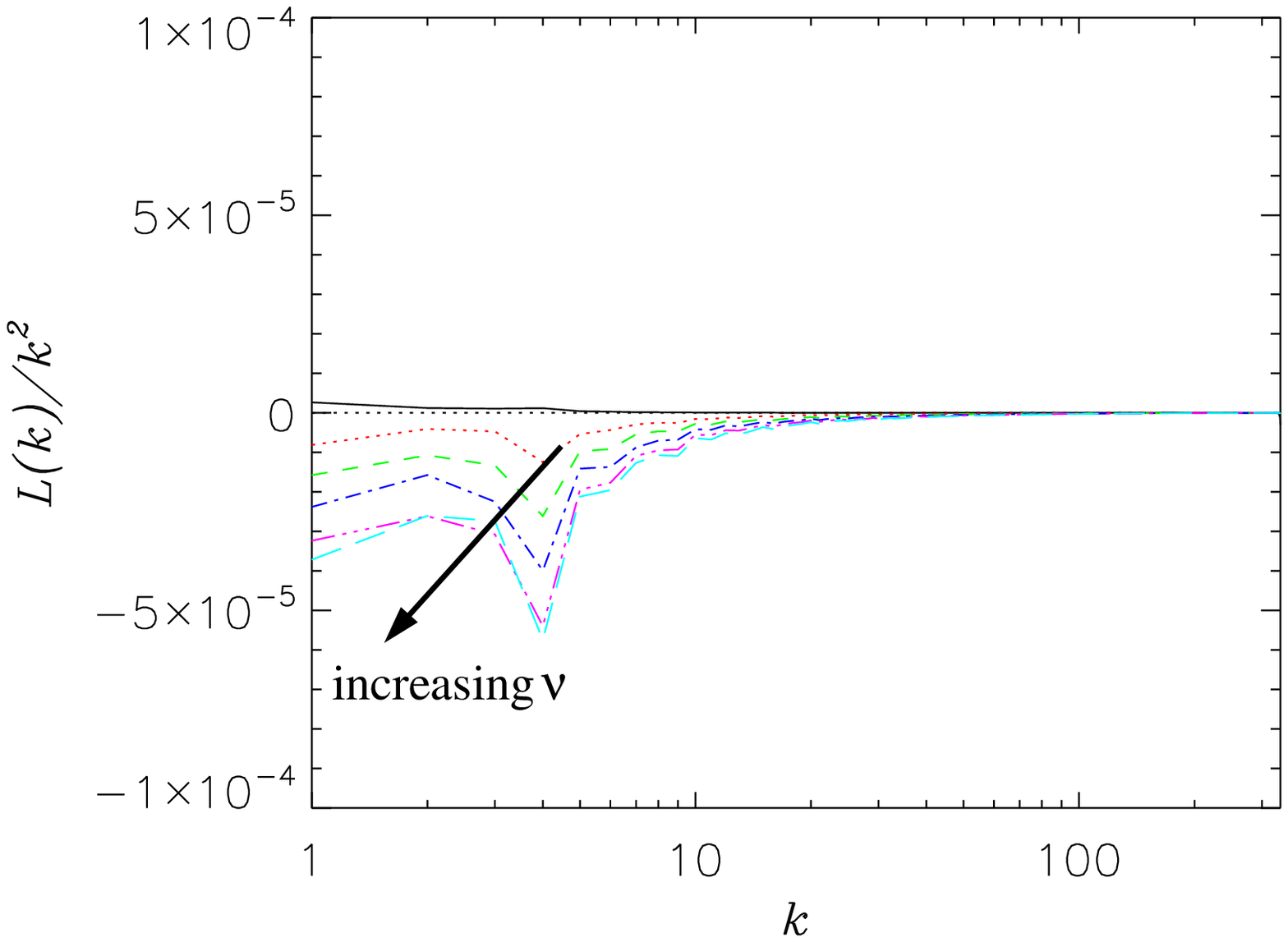}
\caption{Viscous model: subgrid transfers for enstrophy ($L(k)$, Top)
  and energy ($L(k)/k^2$, Bottom)  {and
  $S(k)-\bar{S}(k)$ for benchmark (solid black).}  The model is solely dissipative of
  enstrophy and energy.  {Exact viscosities are denoted} in
  Fig. \ref{fig:viscoustrue}.}
\label{fig:viscexp}
\end{figure}

\begin{figure}[htbp]
\includegraphics[width=10cm]{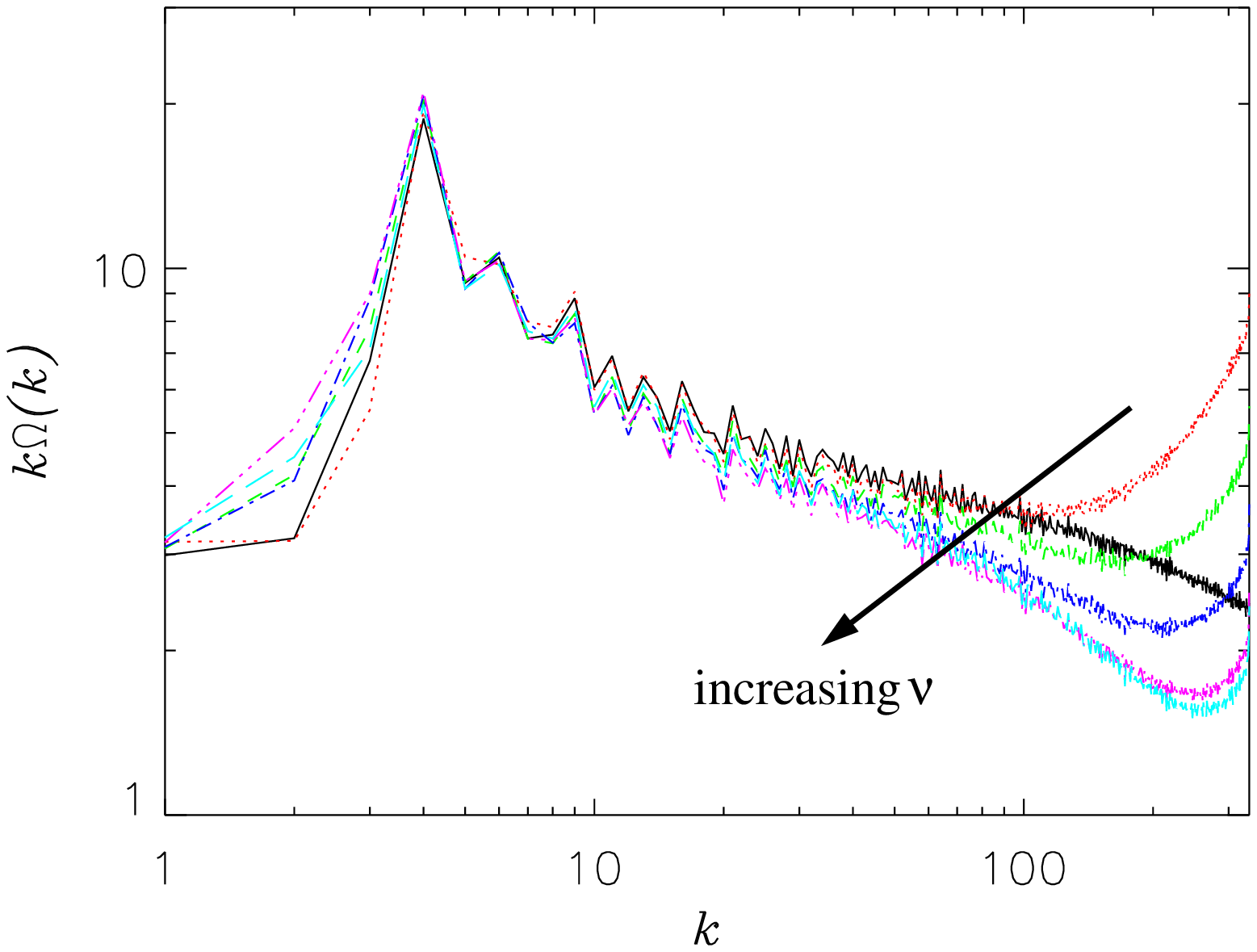}
\caption{Viscous model:  Compensated enstrophy spectrum; 
 {exact viscosities are denoted} in
  Fig. \ref{fig:viscoustrue}.}
\label{fig:viscous}
\end{figure}

The approximate reproduction of the benchmark flux is accomplished by
the action of the subgrid enstrophy transfer $L(k)$
(Fig. \ref{fig:viscexp}).  As expected, the action of the viscous
model is solely dissipative.  The solid black line indicates what the
true transfer with the unresolved scales should be, $S(k)-\bar{S}(k)$ \NEU{(see Fig. \ref{fig:bench2}).}  The viscous model dissipates enstrophy over a much larger
range of scales.  Moreover, since energy is dissipated as
$\sim\nu'Z(k)\sim k^{-1.2}$, \NEU{eddy viscosity} is unphysically \NEU{positive} at
large scales.  What the unresolved scales should be doing is
contributing to the {upscale transfer} of energy as shown by the solid,
black benchmark line.  The enstrophy spectra are shown in
Fig. \ref{fig:viscous}.  The result of too little dissipation is the
piling of small-scale thermal noise in the spectrum
\citep{CiBoDe2005}.

\begin{figure}[htbp]
\includegraphics[width=10cm]{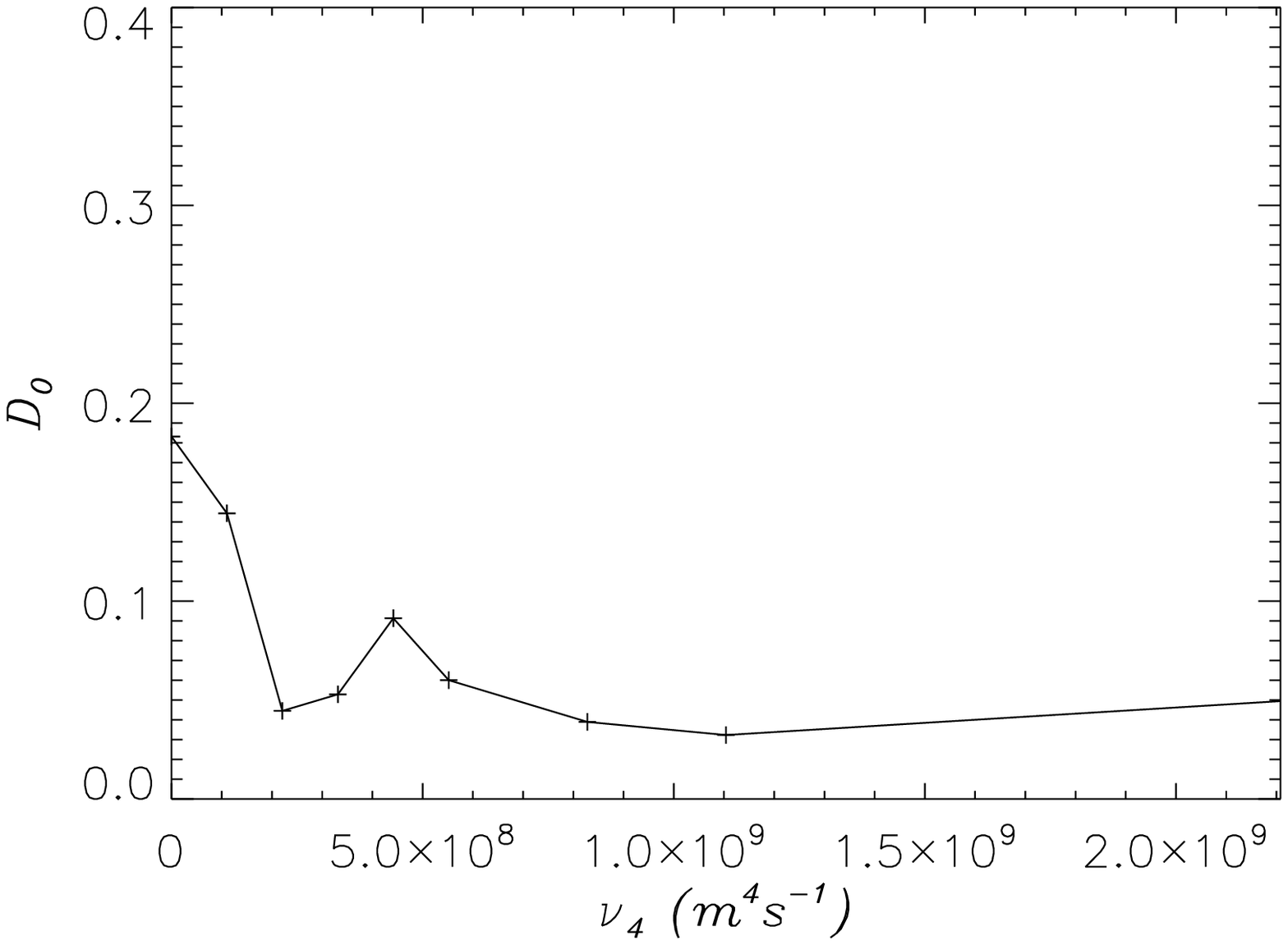}
\caption{Hyper-viscous model: Flux error-landscape {norm $D_0$.}   The optimal value is
  $\nu_4=1.1\times10^9\,$m$^4$s$^{-1}$.}
\label{fig:hypertrue}
\end{figure}

\begin{figure}[htbp]
\includegraphics[width=10cm]{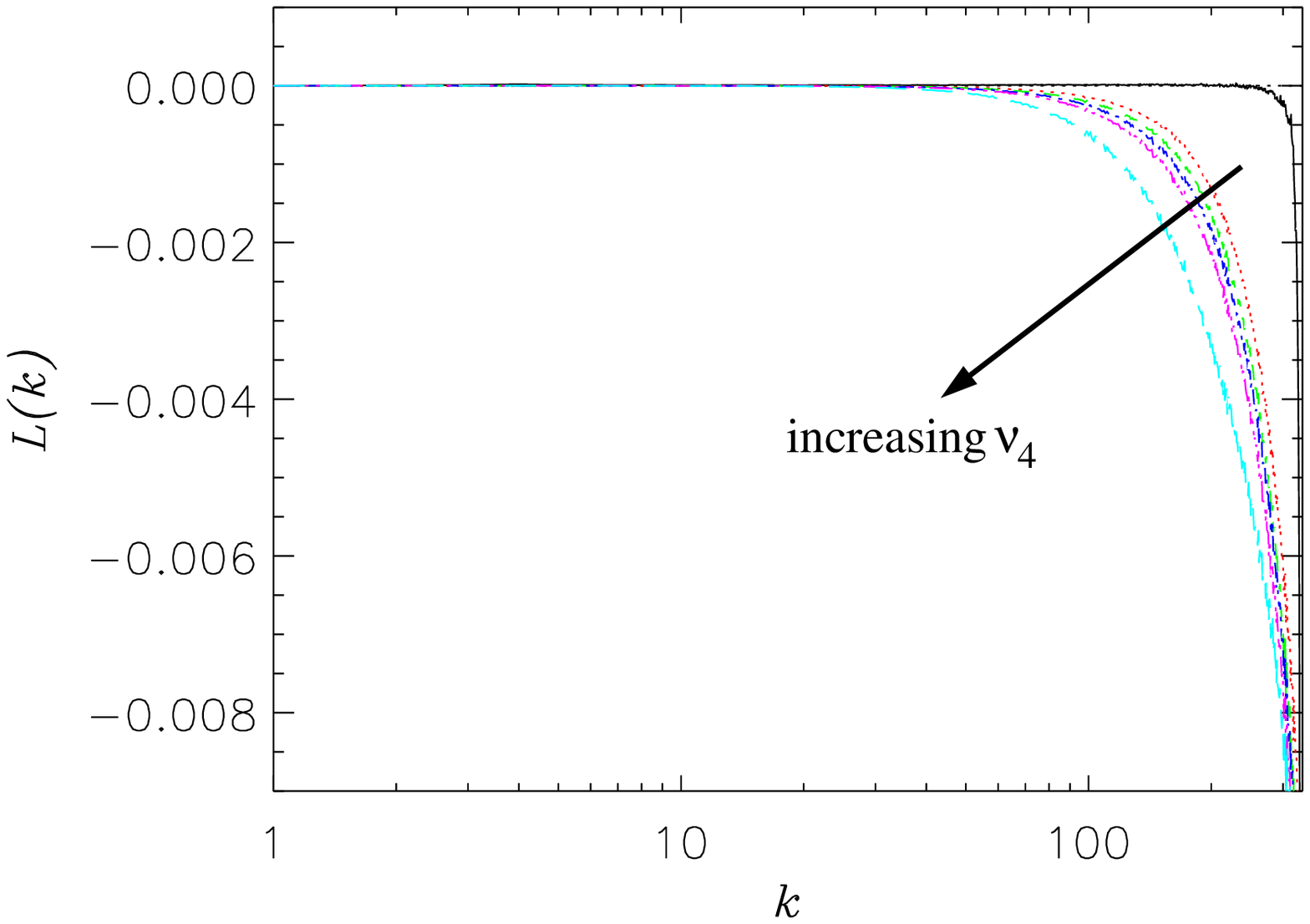}\\
\includegraphics[width=10cm]{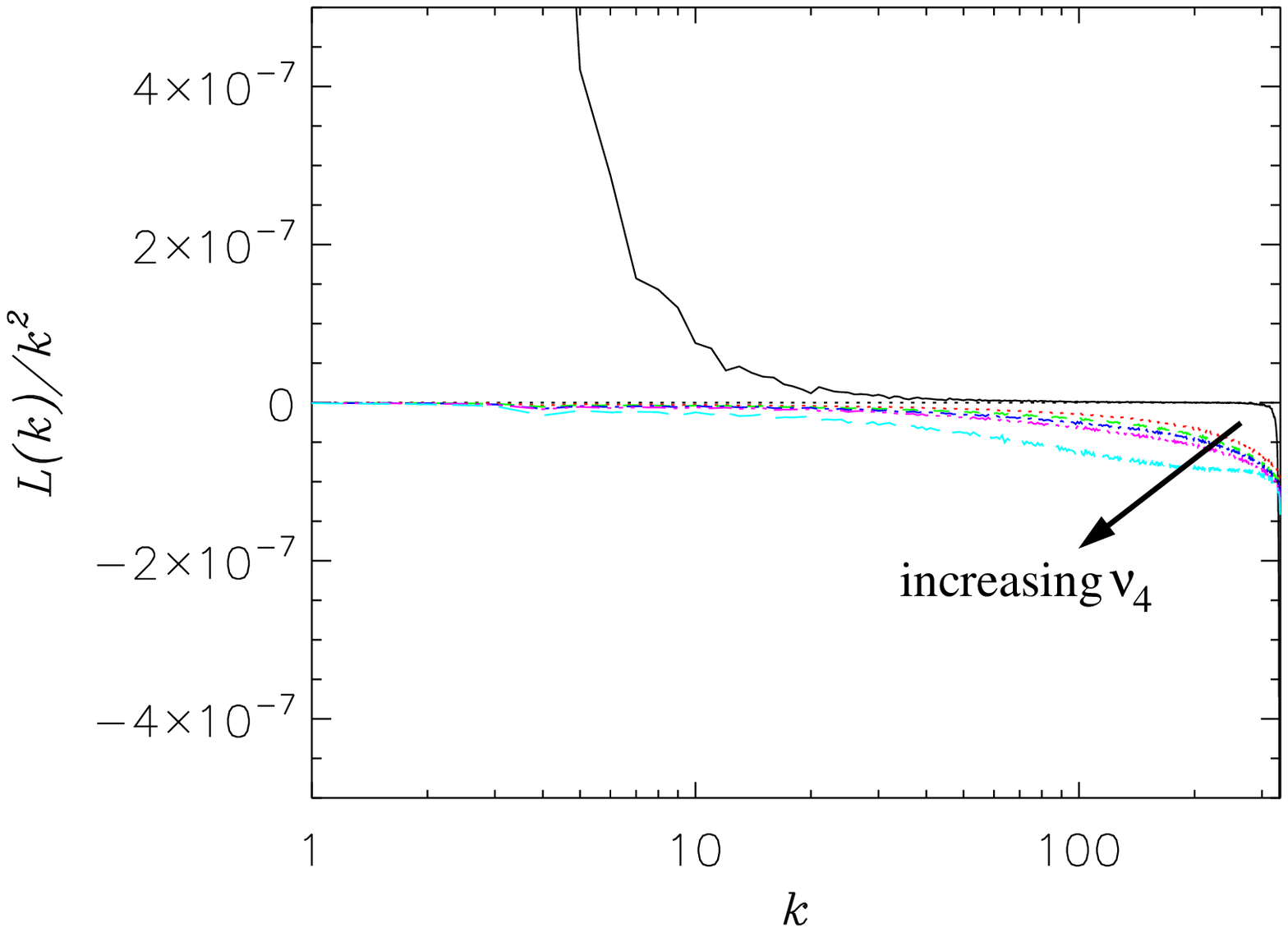}
\caption{Hyper-viscous model: Subgrid transfers for enstrophy ($L(k)$,
  Top) and energy ($L(k)/k^2$, Bottom) for
  $\nu_4=2.2\times10^8\,$m$^4$s$^{-1}$ (red dotted),
  $3.3\times10^8\,$m$^4$s$^{-1}$ (green dashed),
  $4.4\times10^8\,$m$^4$s$^{-1}$ (blue dash-dotted),
  $5.5\times10^8\,$m$^4$s$^{-1}$ (pink dash-triple-dotted),
  $1.1\times10^9\,$m$^4$s$^{-1}$ (cyan long-dashed), and
  $S(k)-\bar{S}(k)$ for benchmark (solid black).}
\label{fig:hyperexp}
\end{figure}

By looking at the hyper-viscous model's flux error-landscape norms
(Fig. \ref{fig:hypertrue}), we identify
$\nu_4=1.1\times10^9\,$m$^4$s$^{-1}$ as the optimal hyper-viscous
model.  The hyper-viscous model much more closely models the
dissipation of enstrophy due to the unresolved scales than the viscous
model, see Fig. \ref{fig:hyperexp}.  Additionally, as the energy
dissipation is $\sim k^2Z\sim k^{0.8}$, \NEU{the rate of energy dissipated at large
scales is insignificant (note} the difference in vertical scales
for energy transfer in Figs. \NEU{\ref{fig:bench2},} \ref{fig:viscexp} and \ref{fig:hyperexp}).  This is a marked improvement, but no
solely-dissipative parameterization will model the {mechanism of upscale energy transfer.} 

\subsection{\label{sec:leith}Leith model}

The Leith model is derived by dimensional analysis \citep{Le1996}.
The local enstrophy dissipation rate is estimated as
\begin{equation}
\eta_\ast = \nu_\ast\big{|}\nabla_\ast\bar{\zeta}\big{|}^2\,,
\label{eq:localeta}
\end{equation}
and an enstrophy cascade spectrum is assumed,
\begin{equation}
Z(k) \propto \eta^{2/3}k^{-1}\,.
\end{equation}
The viscous
range, $k$, is when the viscous enstrophy losses in a given wavenumber
band, $\int\nu k^2 Z(k) dk$, are comparable to the enstrophy
injection, $\eta$, or
\begin{equation}
\eta\sim\nu^3k^6\,.
\label{eq:globaleta}
\end{equation}
Setting the global average dissipation, $\nu$, to the local, grid-scale
dissipation rate, $\nu_\ast$, and equating Eqs. (\ref{eq:localeta}) and
(\ref{eq:globaleta}), we find
\begin{equation}
\nu_\ast \propto \big{|}\nabla\bar{\zeta}\big{|} (\Delta x)^3\,.
\label{eq:leithnu}
\end{equation}
The BVE with Leith model, is \citep{Le1996,FoKeMe2008}
\begin{equation}
\partial_t\bar{\zeta} + \{\bar{\psi},\bar{\zeta}\} =  \nabla\cdot\nu\nabla\bar{\zeta}+ \nabla\cdot\nu_\ast\nabla\bar{\zeta}{+\bar{F}+\bar{Q}}\,,
\label{eq:leith}
\end{equation}
where $\nu=0$ for an infinite Reynolds number flow.
The Leith subgrid term is then
\begin{equation}
\sigma = \nabla\cdot\bigg{[}\bigg{(}\frac{\Lambda\Delta x}{\pi}\bigg{)}^3\big{|}\nabla\bar{\zeta}\big{|}\nabla\bar{\zeta}\bigg{]}\,,
\end{equation}
{where $\Lambda$ is a free parameter.}

\begin{figure}[htbp]
\includegraphics[width=10cm]{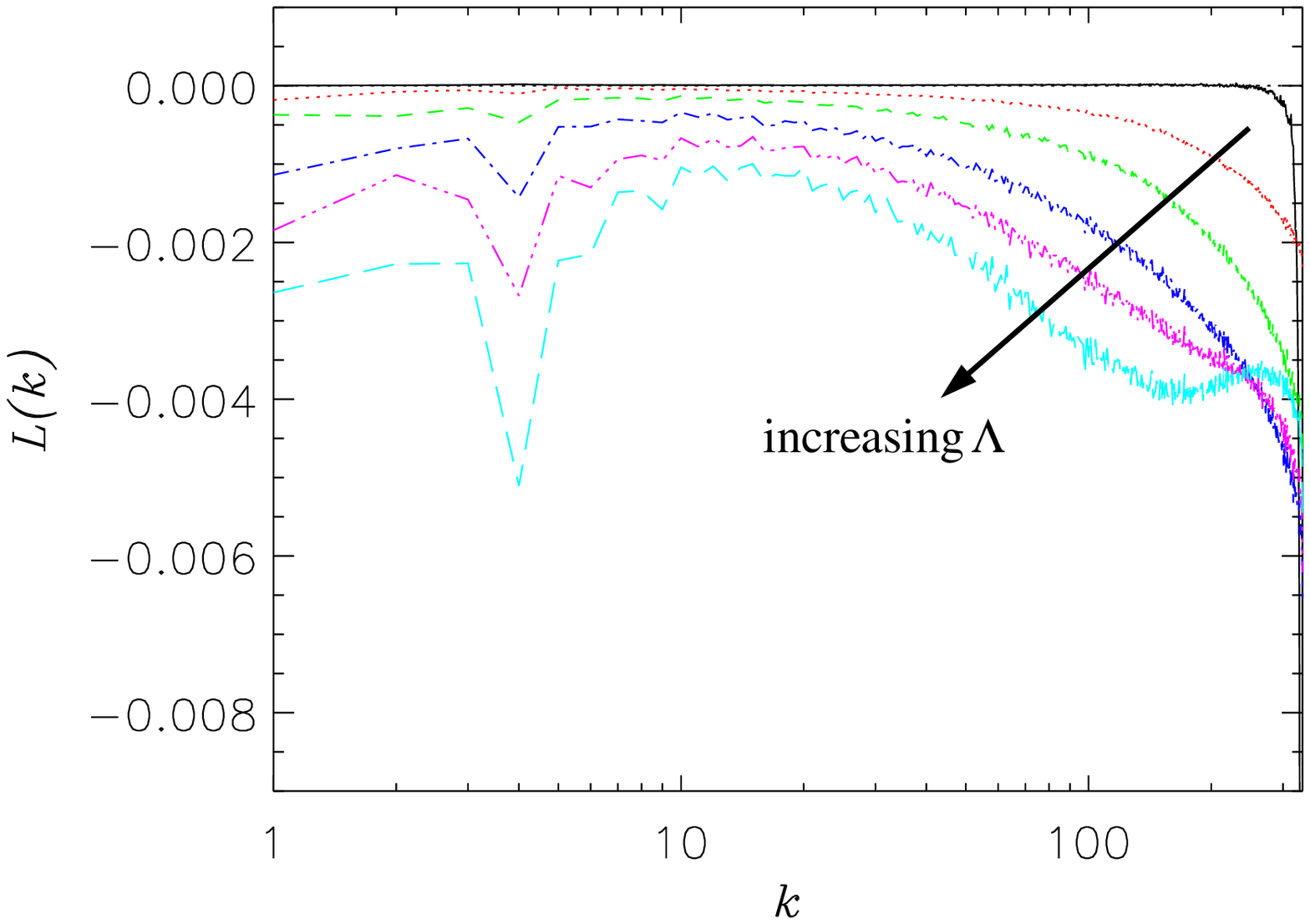}\\
\includegraphics[width=10cm]{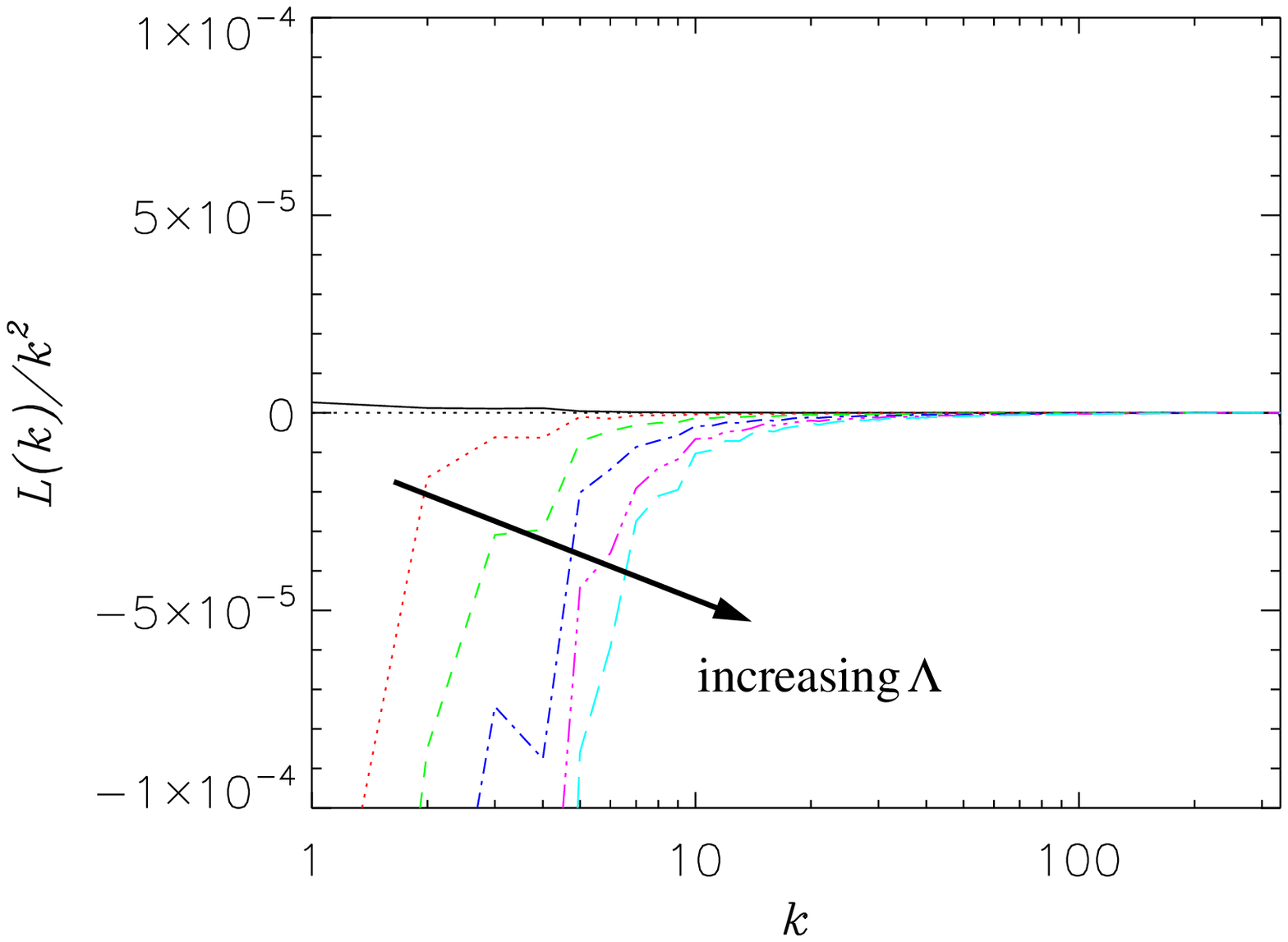}
\caption{Leith model:  Subgrid transfers for enstrophy ($L(k)$, Top)
  and energy ($L(k)/k^2$, Bottom) for {$\Lambda=0.5$ (red dotted),
  $\Lambda=0.75$ (green dashed), $\Lambda=1$ (blue dash-dotted),
  $\Lambda=1.25$ (pink dash-triple-dotted), $\Lambda=1.5$ (cyan
  long-dashed), and benchmark (black solid). The optimal model is $\Lambda=1$.}}
\label{fig:leithexp}
\end{figure}

\begin{figure}[htbp]
\begin{center}
\includegraphics[width=6cm]{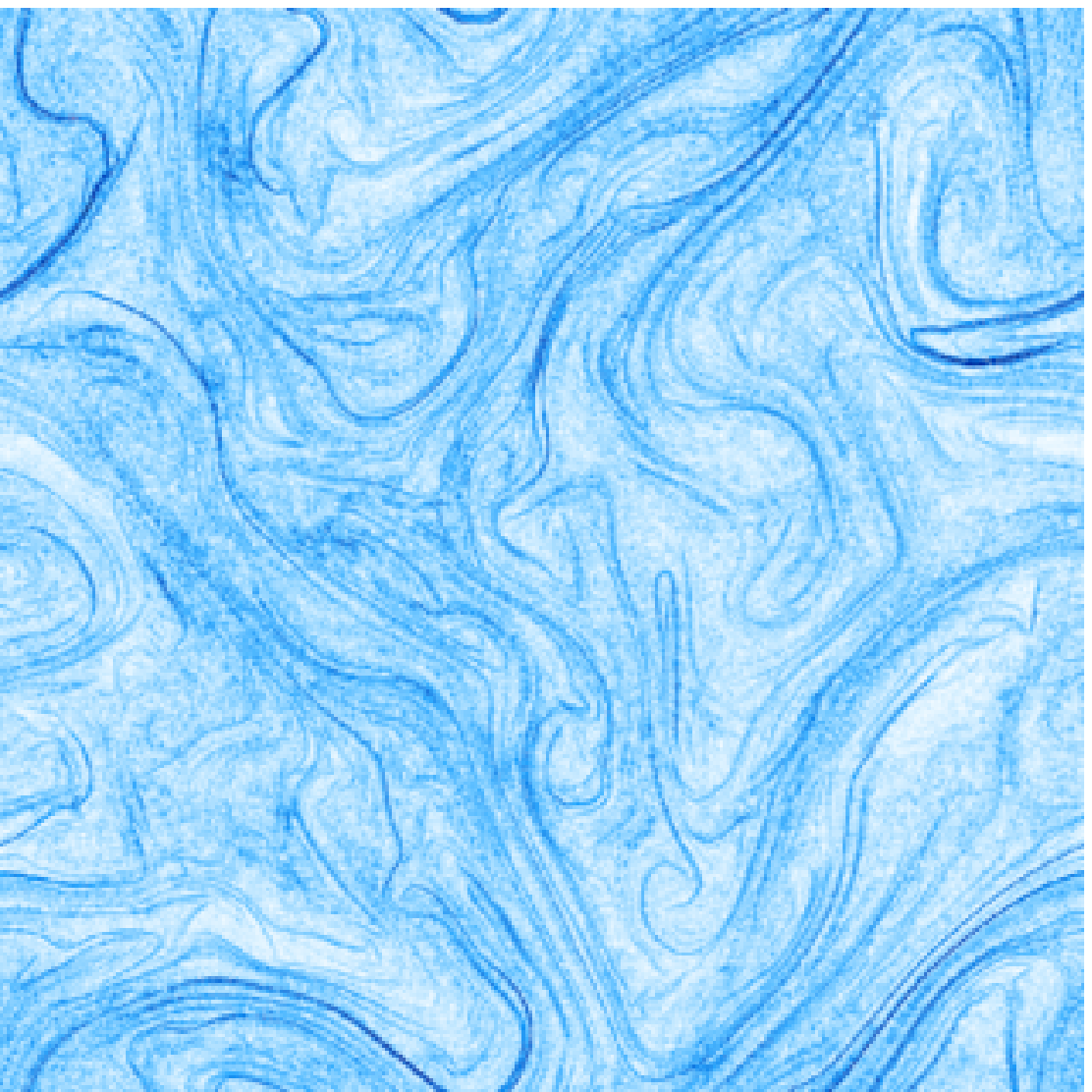}
\includegraphics[width=6cm]{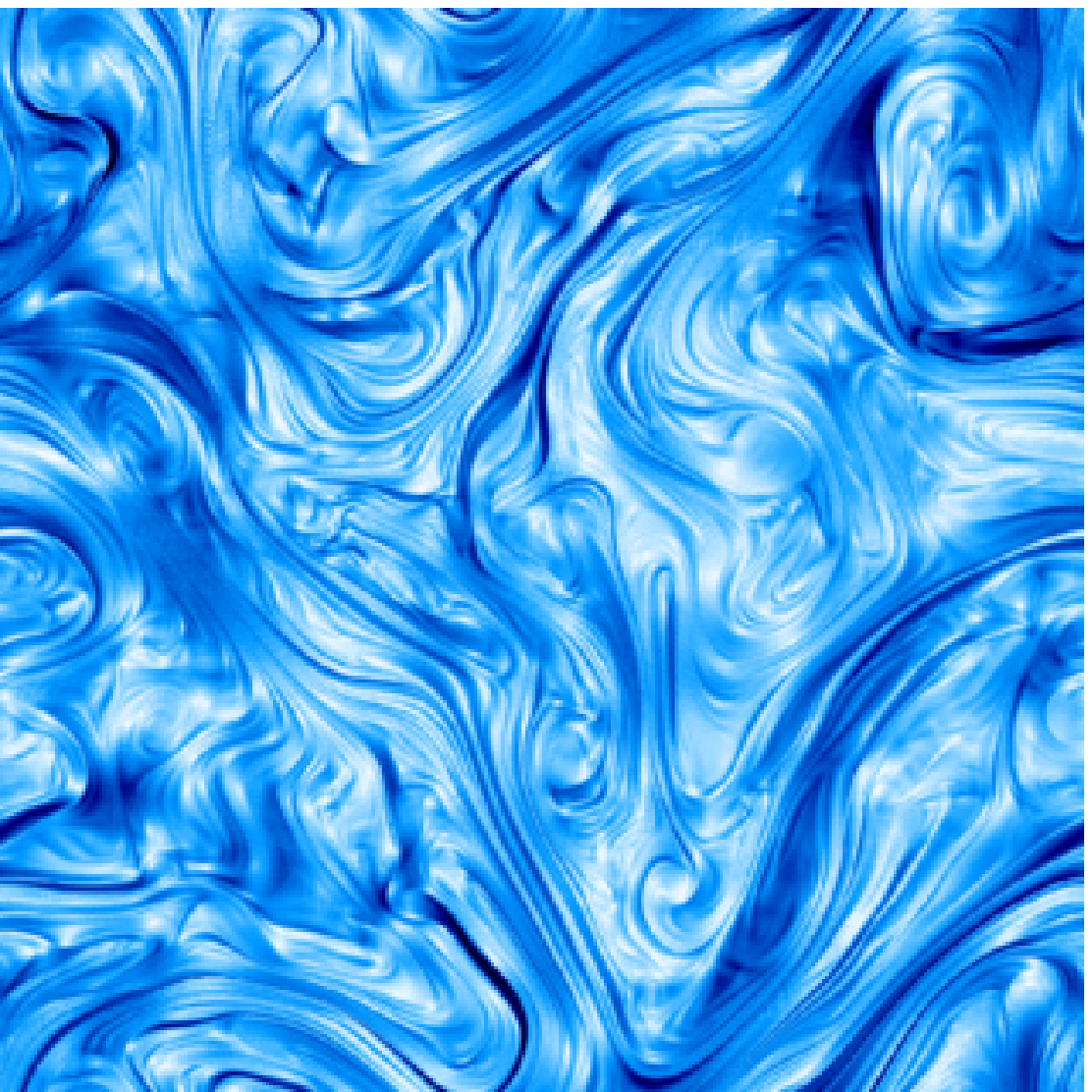}\\
\end{center}
\begin{center}
\includegraphics[width=7cm]{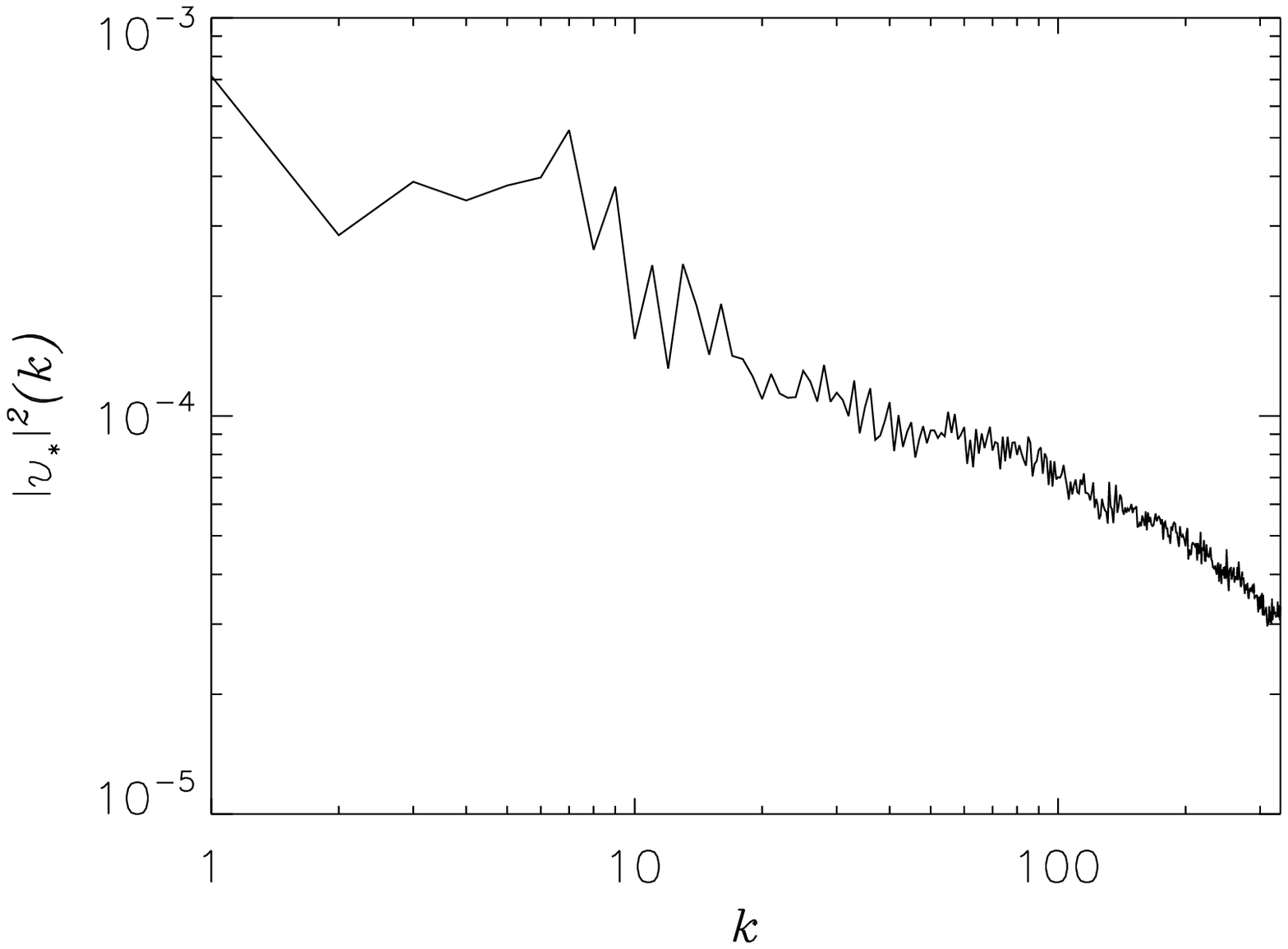}
\includegraphics[width=6cm]{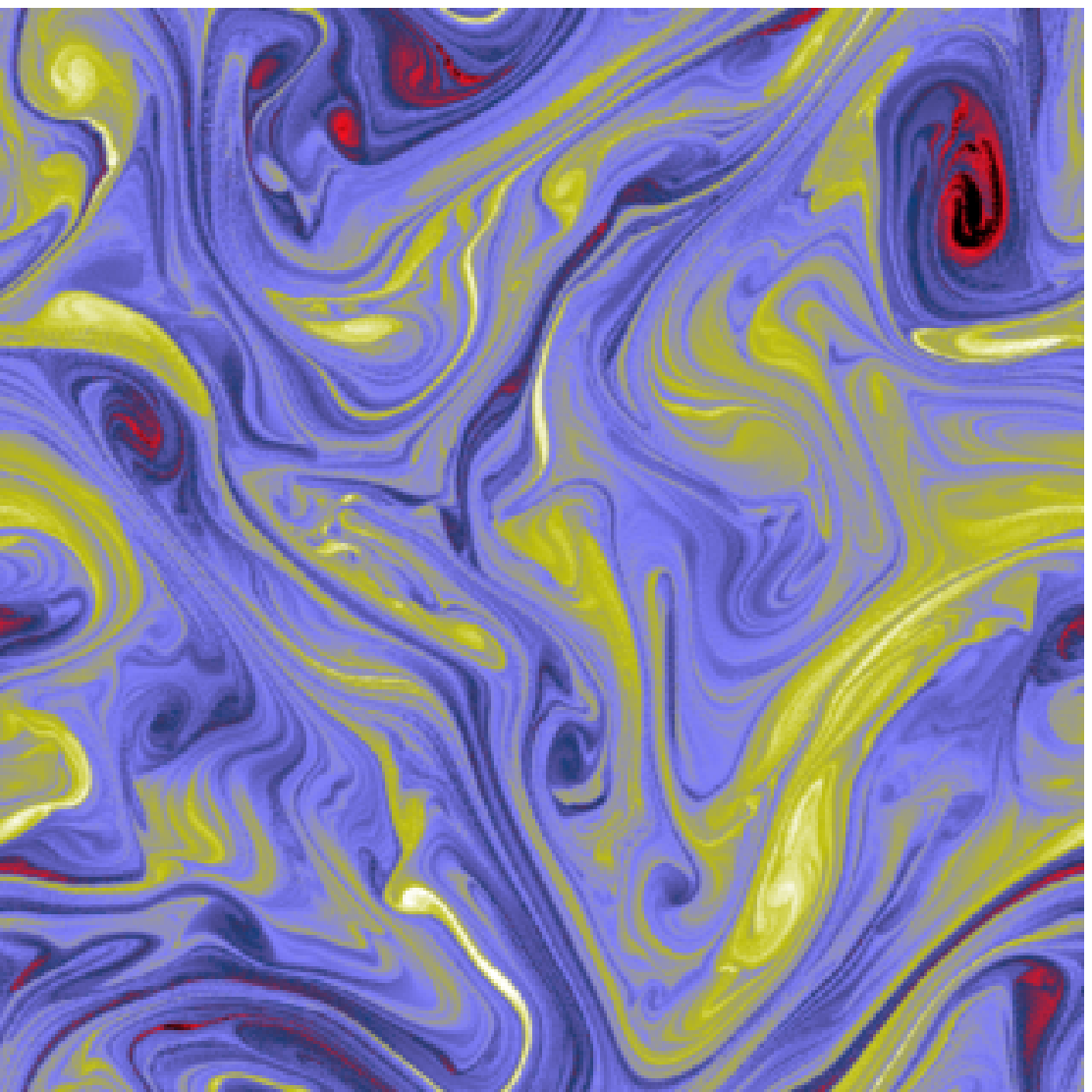}\\
\end{center}
\caption{Snapshots
of $\nu_\ast$ (Top Left) and Fourier power spectrum of
  $\nu_\ast$ (Bottom Left) for Leith model,
  {$\Lambda=1$,} of $\nu_\ast$ (Top Right) for {Smagorinsky, $\Lambda_S=1$,
    } and of  vorticity field (Bottom Right, shown for reference).
All snapshots are at $4\times10^4\,$min.}
\label{fig:leitheye}
\end{figure}

The subgrid transfers for the Leith model are very similar to the
viscous model results (see Fig. \ref{fig:leithexp}).  This is to be
expected as the Leith model is also solely-dissipative.  \NEU{Note} that there is strong enstrophy dissipation at
the forcing scale.  This can be understood by looking at
Fig. \ref{fig:leitheye}.  The Leith viscosity {$\nu_\ast$ is proportional to $|\nabla\bar{\zeta}|$} and, therefore, is concentrated along the
borders between oppositely-signed vortices.  These large-scale
coherent structures of enhanced dissipation then project on the small
wavenumber Fourier-modes {(bottom left panel of
  Fig. \ref{fig:leitheye}).} {Note that the optimal parameter value is found to be $\Lambda=1$.}

\subsection{\label{sec:smag}Smagorinsky model}

The Smagorinsky model \citep{Sm1963,Li67} is the 3D precursor of the
Leith model.  It is derived with a similar dimensional analysis as in
Sec. \ref{sec:leith}, but assuming a 3D direct cascade of energy.
Consequently, the model for eddy-viscosity is
\begin{equation}
\nu_\ast=\bigg{(}\frac{{\Lambda_S}\Delta x}{\pi}\bigg{)}^2|S_{ij}|\,,
\end{equation}
where $S_{ij}=(\partial_jv_i+\partial_iv_j)/2$.  For isotropic,
homogeneous 3D turbulence the Smagorinsky Constant,
$C_S\equiv{\Lambda_S}/\pi\approx0.2$ \citep{MeKa2000}.  {It
  should be noted that Smagorinsky was devised for 3D isotropic flow
  and was not intended for 2D nor geostrophic flows, but has
been employed in global climate models \cite{2000MWRv..128.2935G,2012JCli...25.2755D}.}

\begin{figure}[htbp]
\includegraphics[width=10cm]{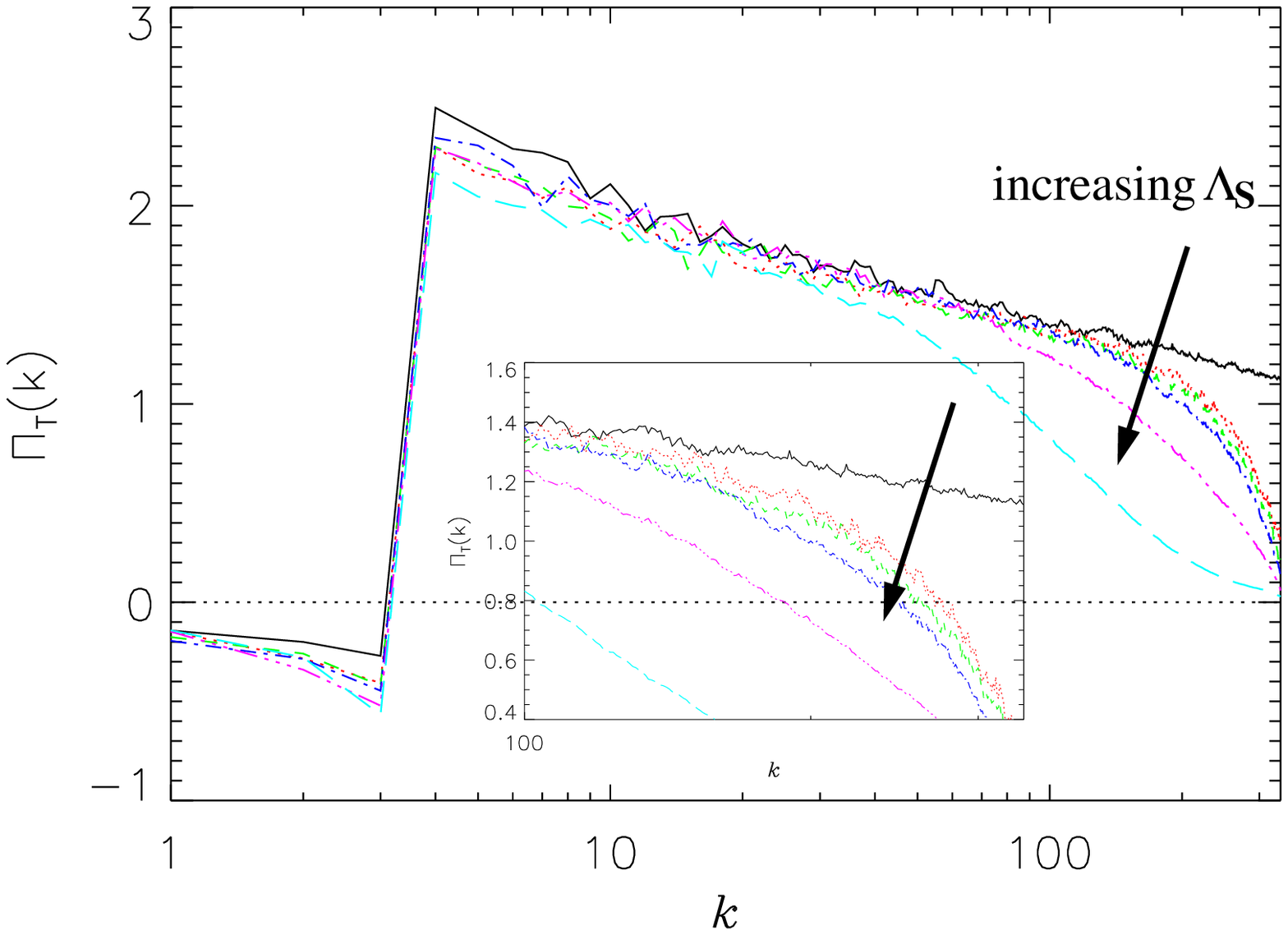}\\
\includegraphics[width=10cm]{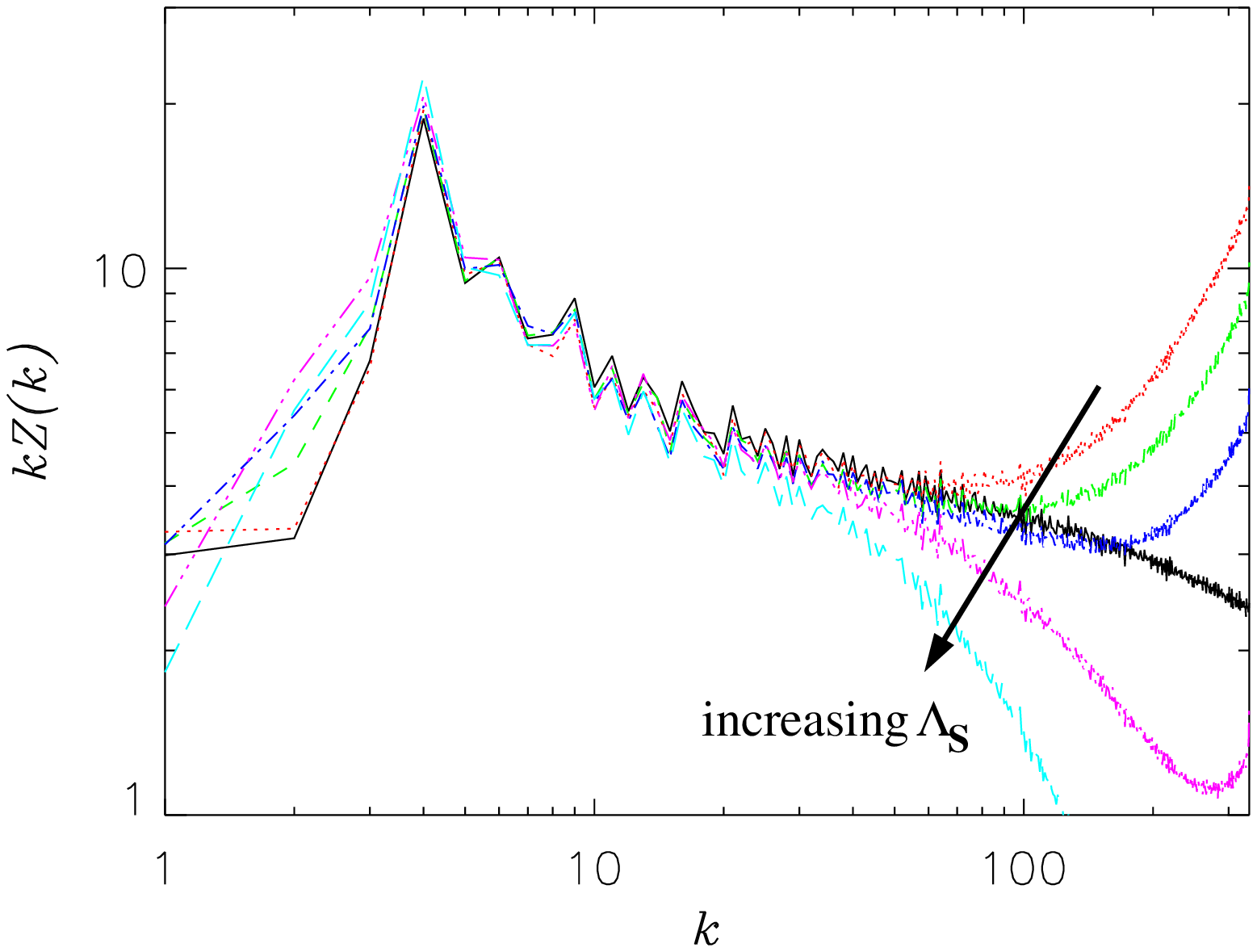}
%
\caption{{Smagorinsky model: (Top) Modeled flux, $\Pi_T(k)$, for
  $\Lambda_S=0.1$ (red dotted), $\Lambda_S=0.3$
  (green dashed), $\Lambda_S=0.5$ (blue dash-dotted),
  $\Lambda_S=1$ (pink dash-triple-dotted), and
  $\Lambda_S=2$ (cyan long-dashed) and $\Pi_S(k)$ for
  $8192^2$ BVE benchmark (solid black).  (Bottom) 
Compensated enstrophy spectrum.}}
\label{fig:smagtrue}
\end{figure}

\begin{figure}[htbp]
\includegraphics[width=10cm]{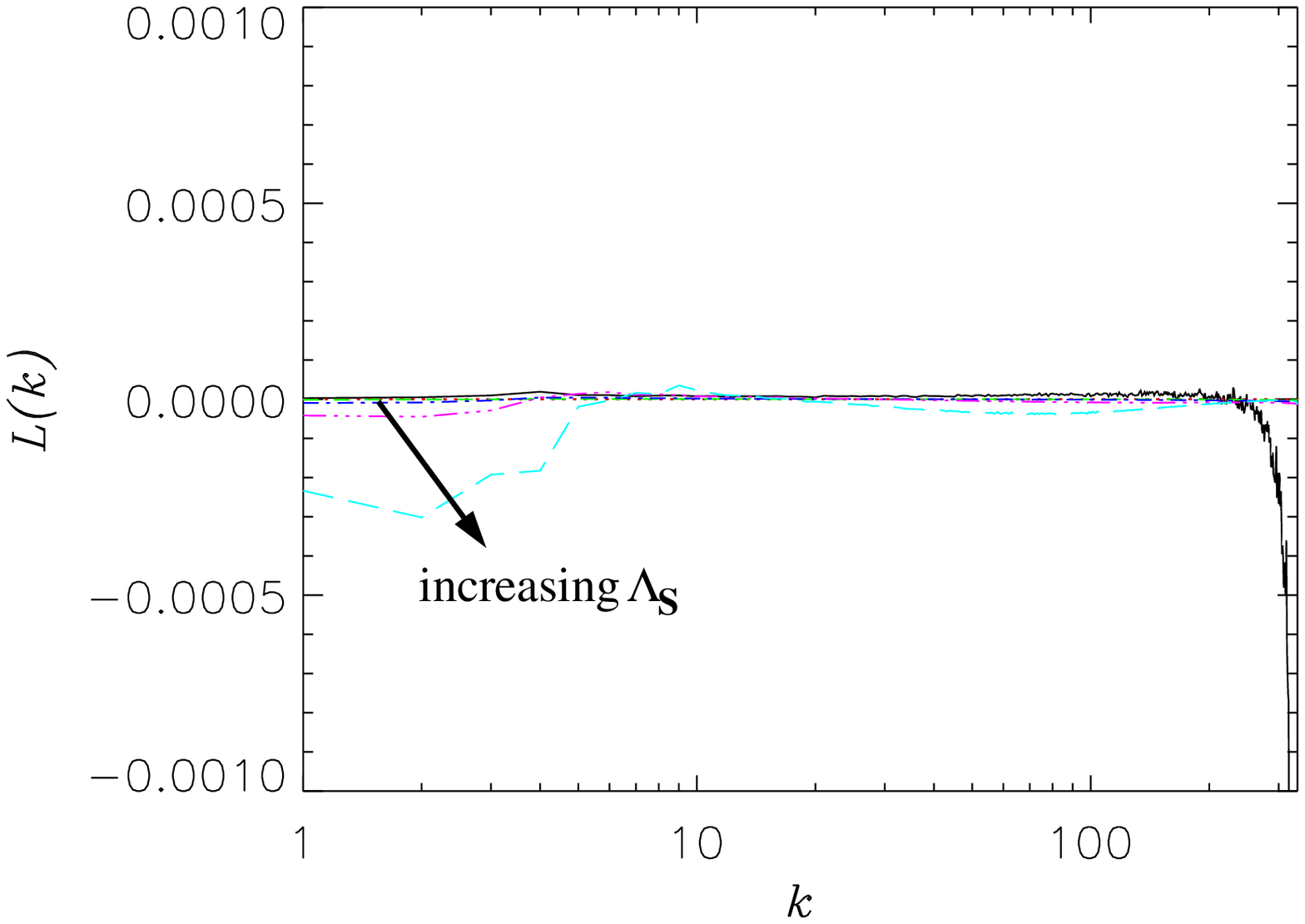}
\includegraphics[width=10cm]{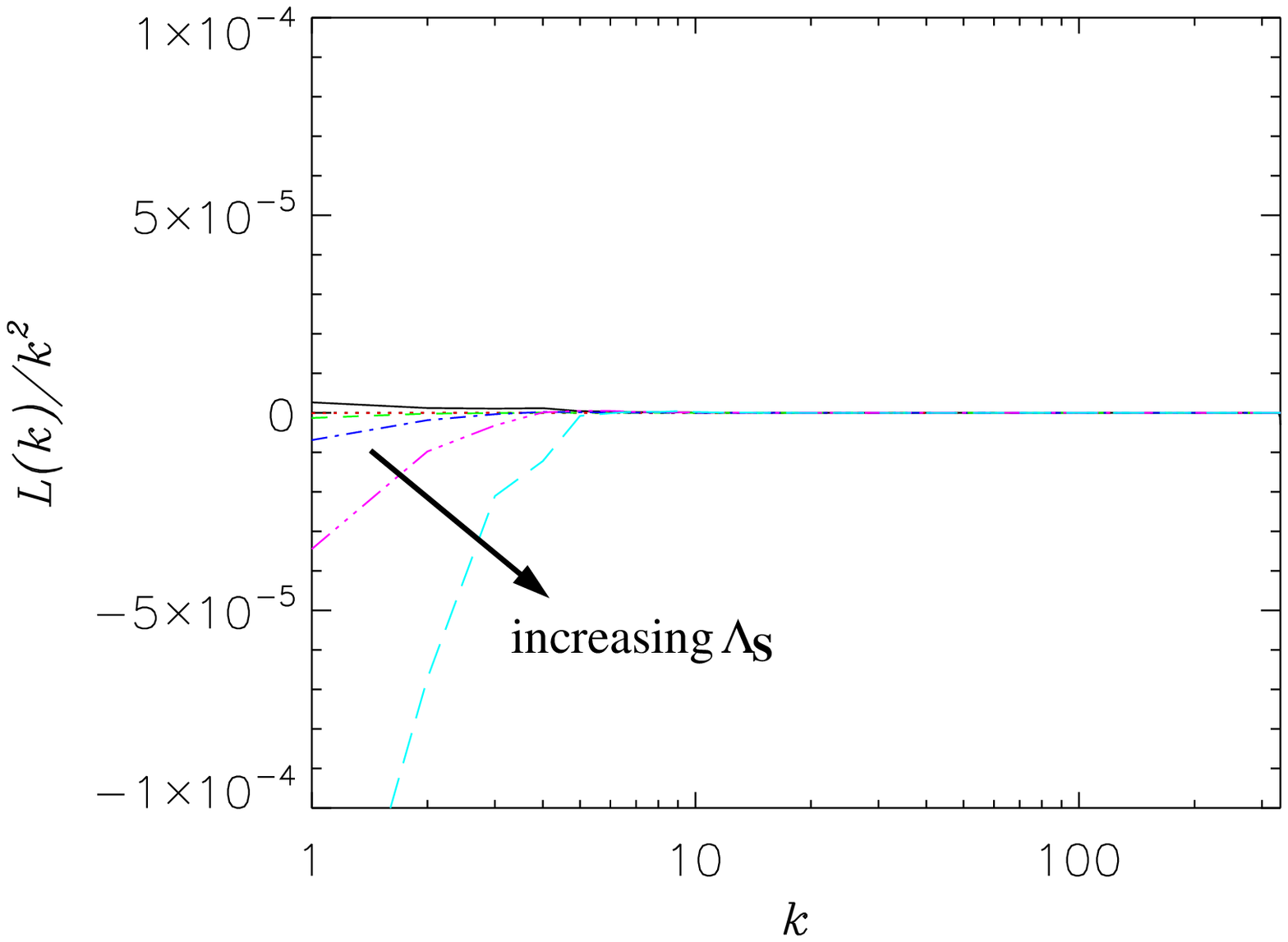}
\caption{{Smagorinsky model: subgrid transfers {of enstrophy ($L(k)$, Top) and energy ($L(k)/k^2$,
  Bottom).}  The model dissipates
  enstrophy and energy unphysically from the large scales.  {Exact viscosities are denoted} in
  Fig. \ref{fig:smagtrue}.}}
\label{fig:smagexp}
\end{figure}

{The enstrophy flux and enstrophy spectrum for Smagorinsky
  (Fig. \ref{fig:smagtrue}), highlight the fact that good spectra can
  be produced without necessarily reproducing the correct dynamics.
  The best spectra are produced for $\Lambda_S=0.5$ (blue dash-dotted)
  while the best flux is produced by $\Lambda_S=0.1$ (red dotted).
  This is opposed to the case for the viscous model where the best
  flux {\sl and} spectrum occur for {the same value of the model's
    one free parameter, $\nu'$.}  The reason for the disparity is that
  the viscous parameterization captures the most important physical
  process, small-scale enstrophy dissipation, while the Smagorinsky
  model unphysically removes enstrophy and energy from the largest
  scales (see Fig. \ref{fig:smagexp} and the real-space visualization
  of $\nu_\ast$ in Fig. \ref{fig:leitheye}).  Therefore, even when the
  combination of modeling and numerical error produces a good
  spectrum, the model is not capturing the correct physical dynamics.}

%
%
\subsection{\label{sec:avm}Anticipated vorticity method (AVM)}

AVM (APVM when applied to potential vorticity, \cite{SaBa1985}) is
so-called because it can be seen as substituting the forward-in-time
vorticity in the BVE,
\begin{eqnarray}
\frac{\bar{\zeta}_{n+1}-\bar{\zeta}_n}{\theta} = - \{\bar{\psi},\bar{\zeta}_n\}\,,
\end{eqnarray}
where $\theta$ is the time step for the anticipation.  Substituting
this anticipated value, $\bar{\zeta}_{n+1}$ {in Eq. (\ref{eq:bve})} results in the
lowest-order AVM,
\begin{eqnarray}
\partial_t\bar{\zeta} = -\{\bar{\psi},\bar{\zeta}_n\} +\theta\{\bar{\psi},\{\bar{\psi},\bar{\zeta}_n\}\}{+\bar{F}+\bar{D}+\bar{Q}}\,.
\end{eqnarray}
In practice, to weight the subgrid model to smaller scales, 
\begin{eqnarray}
\sigma = -\frac{\theta}{k_{max}^{2m}}\{\bar{\psi},\nabla^{2m}\{\bar{\psi},\bar{\zeta}\}\}\,,
\end{eqnarray}
In this study we have used $m=1$ as even this order of diffusive
operator is not practical in finite-volume {and finite-difference
  schemes typically used in global ocean modeling} {because of the
  relationship between high-order derivative accuracy and stencil
  size.}  AVM is not Galilean invariant, i.e., it does not conserve
momentum, but it exactly conserves energy while dissipating enstrophy.
{Note that the subgrid term for the momentum equation is
  $\nabla\cdot\tau=\big{[}(-1)^m\frac{\theta}{k_{max}^{2m}}\nabla^{2m}\big{(}\vec{u}\cdot\nabla(\zeta\hat{\vec{z}}\times\vec{u})\big{)}\big{]}\hat{\vec{z}}\times\vec{u}$
    which is perpendicular to the velocity at every point in space.
    AVM then exactly conserves energy even if $\theta$ varies
    spatially and temporally.}

\begin{figure}[htbp]
\includegraphics[width=10cm]{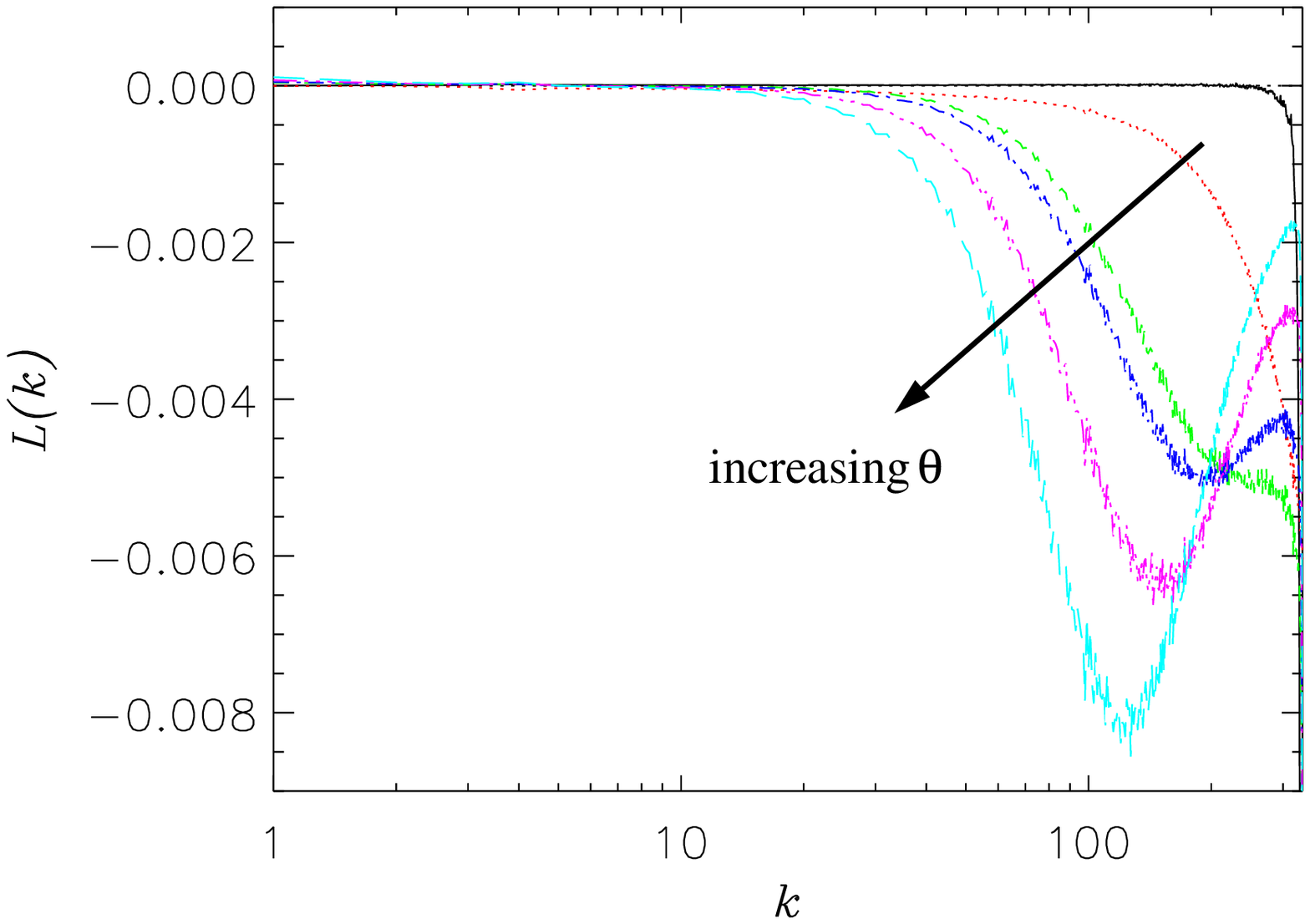}\\
\includegraphics[width=10cm]{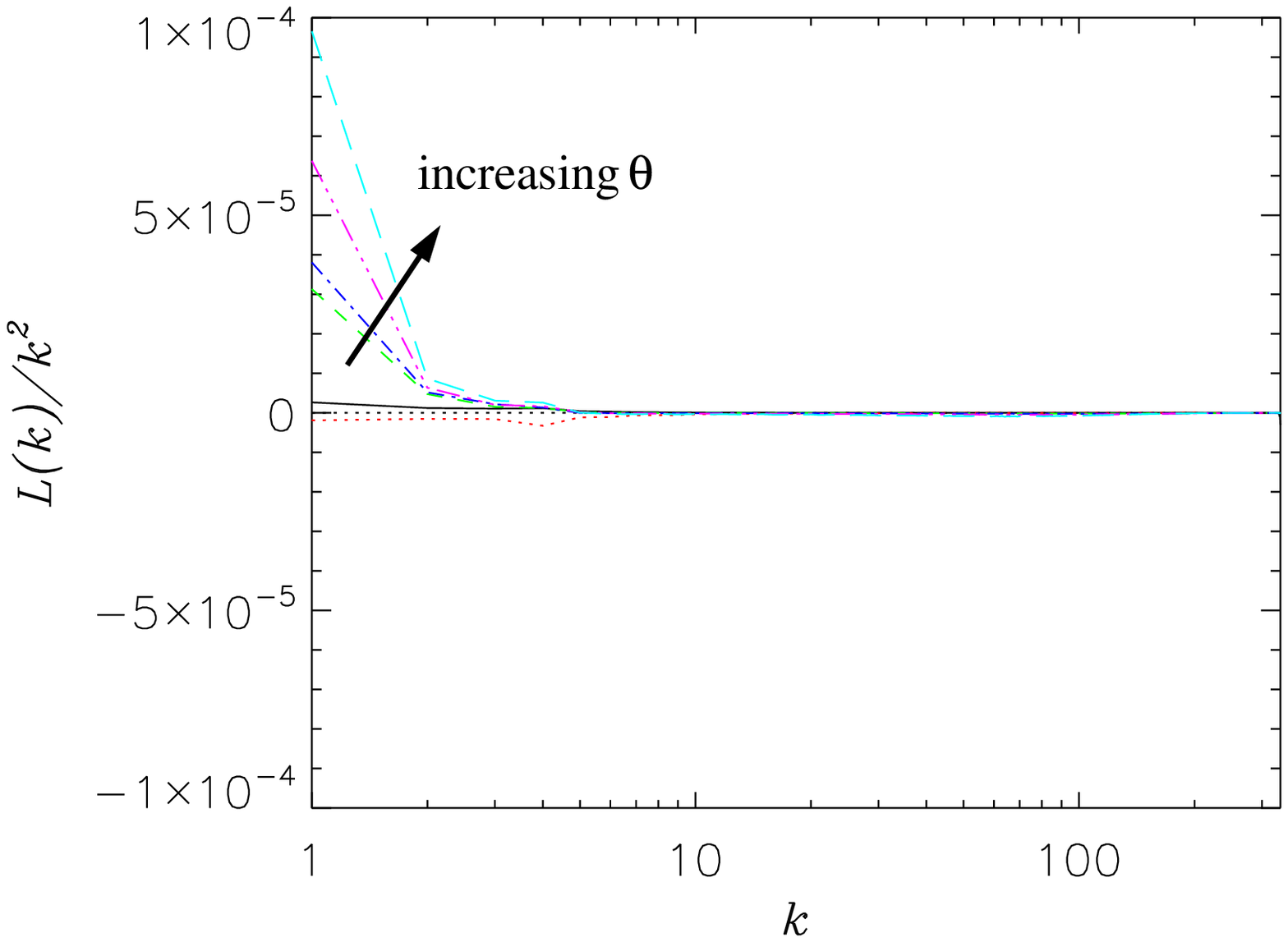}
\caption{AVM: Subgrid transfers of enstrophy ($L(k)$, Top) and energy
  ($L(k)/k^2$, Bottom) {for $8192^2$ BVE benchmark (solid black),}
  for $\theta=0$ (NO MODEL, red dotted), $\theta=0.16$ (green
  dashed), $0.25$ (blue dash-dotted), $0.5$ (pink dash-triple-dotted),
  and $1$ (cyan long-dashed).  The subgrid model transfer in AVM
  changes sign so that the model dissipates no energy, sum of
  $L(k)/k^2$ over all wavenumbers is $o(10^{-12})$, while enstrophy
  dissipation (sum of $L(k)$) is $o(1)$.  The negative energy
  dissipation at large scales mimics the {upscale transfer} from
  unresolved scales, though too strongly.}
\label{fig:avmexp}
\end{figure}
\begin{figure}[htbp]
\includegraphics[width=10cm]{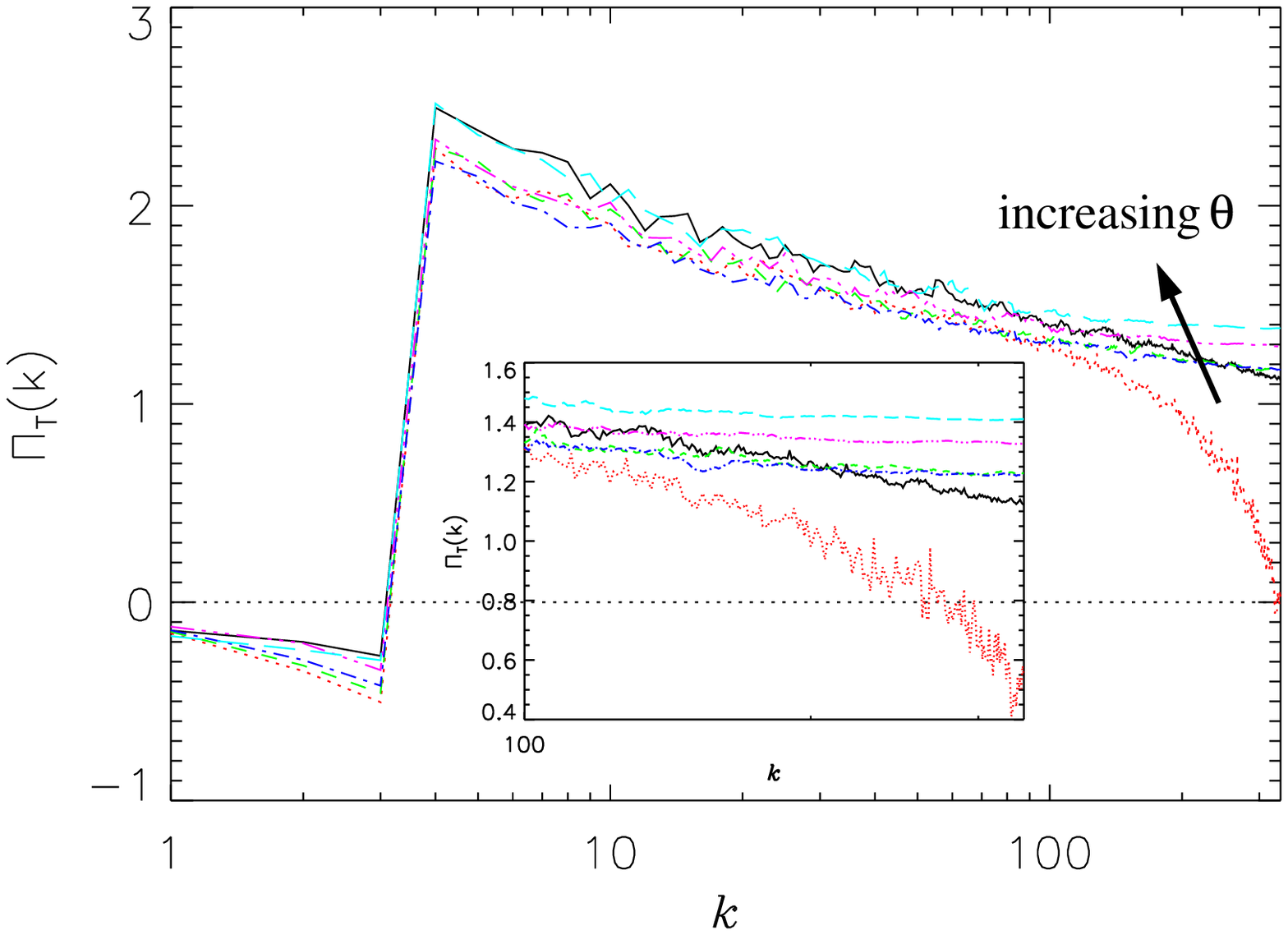}\\
\includegraphics[width=10cm]{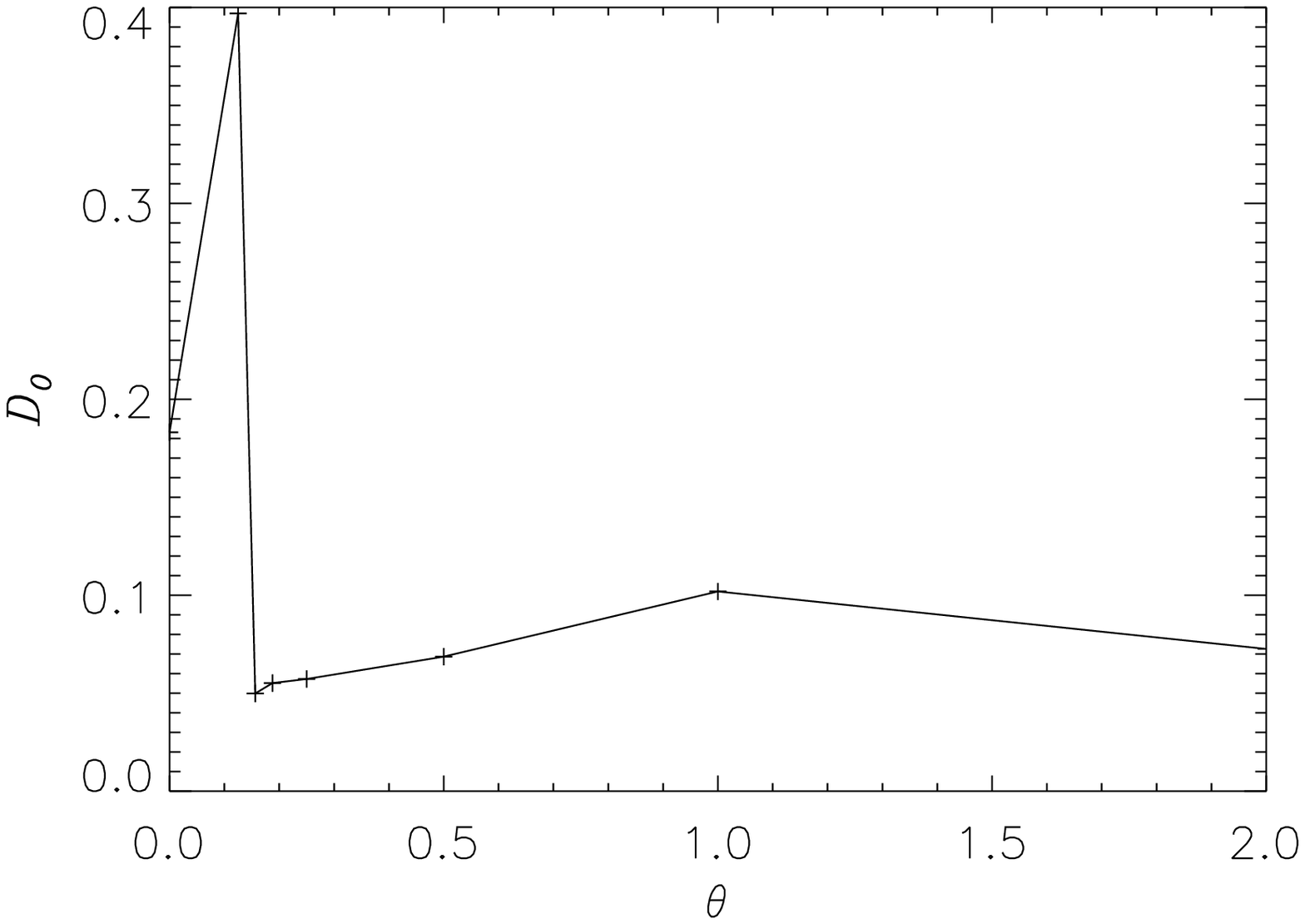}
\caption{AVM: (Top) Modeled flux, $\Pi_T(k)$. {Exact values
    of $\theta$ are given} in Fig. \ref{fig:avmexp}. (Bottom) Flux
  error-landscape {norm.  Optimal value is 
  $\theta=0.16$.} }
\label{fig:avmtrue}
\end{figure}

As AVM dissipates enstrophy at small scales, $L(k)<0$ for large $k$
(see Fig. \ref{fig:avmexp}), it must also remove some small-scale
energy, $k^{-2}L(k)<0$.   Since AVM exactly conserves energy, this
energy shows up at large scales.  AVM is the only parameterization
studied here that reproduces this signature of the correct transfer.
The physical effect, however, is over estimated by at least an order
of magnitude. This can be mitigated by reducing $\theta$.  However,
too small $\theta$ (0.125$dt$ for our flow) results in an excess of
energy at all scales \citep{VaHu1988}.  For $m=1$, as used here, AVM
is unable to mimic that eddy viscosity should only act in a small
range of wavenumbers near $k_{max}$ \citep{VaHu1988}.  {Note that
  setting the anticipation time equal to the time step, $\theta=1$,
  very closely reproduces the low-wavenumber flux
  (Fig. \ref{fig:avmtrue}).  This large value for $\theta$, however,
  makes the eddy viscosity act at even larger scales
  (Fig. \ref{fig:avmexp}).  If larger values of $m$ were practical in
  actual ocean applications, a two parameter optimization might yield
  a very robust model.  Holding constant $m=1$,} the optimal value of
$\theta$ is {$0.16$.}


%
%

\subsection{\label{sec:alpha}$\alpha-$model}

The $\alpha-$model takes a different approach than the other
parameterizations.  It is a non-dissipative, solely dispersive model
-- a mathematical regularization (smooth, and hence computable
solutions are ensured even in the limit $\nu\rightarrow0$) of the fluid equations
{\cite{HoMaRa1998,1998PhRvL..81.5338C,1999PhyD..133...49C,1999PhyD..133...66C,1999PhFl...11.2343C,2001PhyD..152..505F}.}  The result is that the vorticity is advected by a
smoothed velocity,
$\overline{\vec{u_s}}=(1-\alpha^2\nabla^2)^{-1}\bar{\vec{u}}$, with
a filter scale $\sim\alpha$,
\begin{equation}
\partial_t\bar{\zeta} + \nabla\cdot\bigg{(}\overline{\vec{u_s}}\bar{\zeta}\bigg{)}
= \nu\nabla^2\bar{\zeta}+ {\bar{F}+\bar{D}+\bar{Q}}\,,
\label{eq:LANS}
\end{equation}
where
$\nabla\cdot\bigg{(}\overline{\vec{u_s}}\bar{\zeta}\bigg{)}=\{\bar{\psi_s},\bar{\zeta}\}$.
The alpha subgrid term is
\begin{equation}
\sigma = \{\bar{\psi},\bar{\zeta}\} - \{\bar{\psi_s},\bar{\zeta}\}\,.
\label{eq:lanssigma}
\end{equation} 
{Note that the $\alpha-$model has complex conservation properties
  in that the energy balance equation is in the $H^1_\alpha$ norm,
  $\int \overline{\vec{u_s}}\cdot\bar{\vec{u}}dA$, and enstrophy is in
  the $L^2$ norm, $\int \bar{\zeta}^2dA$.  The subgrid energy transfer
  is $L_\alpha(k)/k^2$ is related to the subgrid enstrophy
  transfer by $L_\alpha(k)= L(k)/(1+\alpha^2k^2)$.}

\begin{figure}[htbp]
\includegraphics[width=10cm]{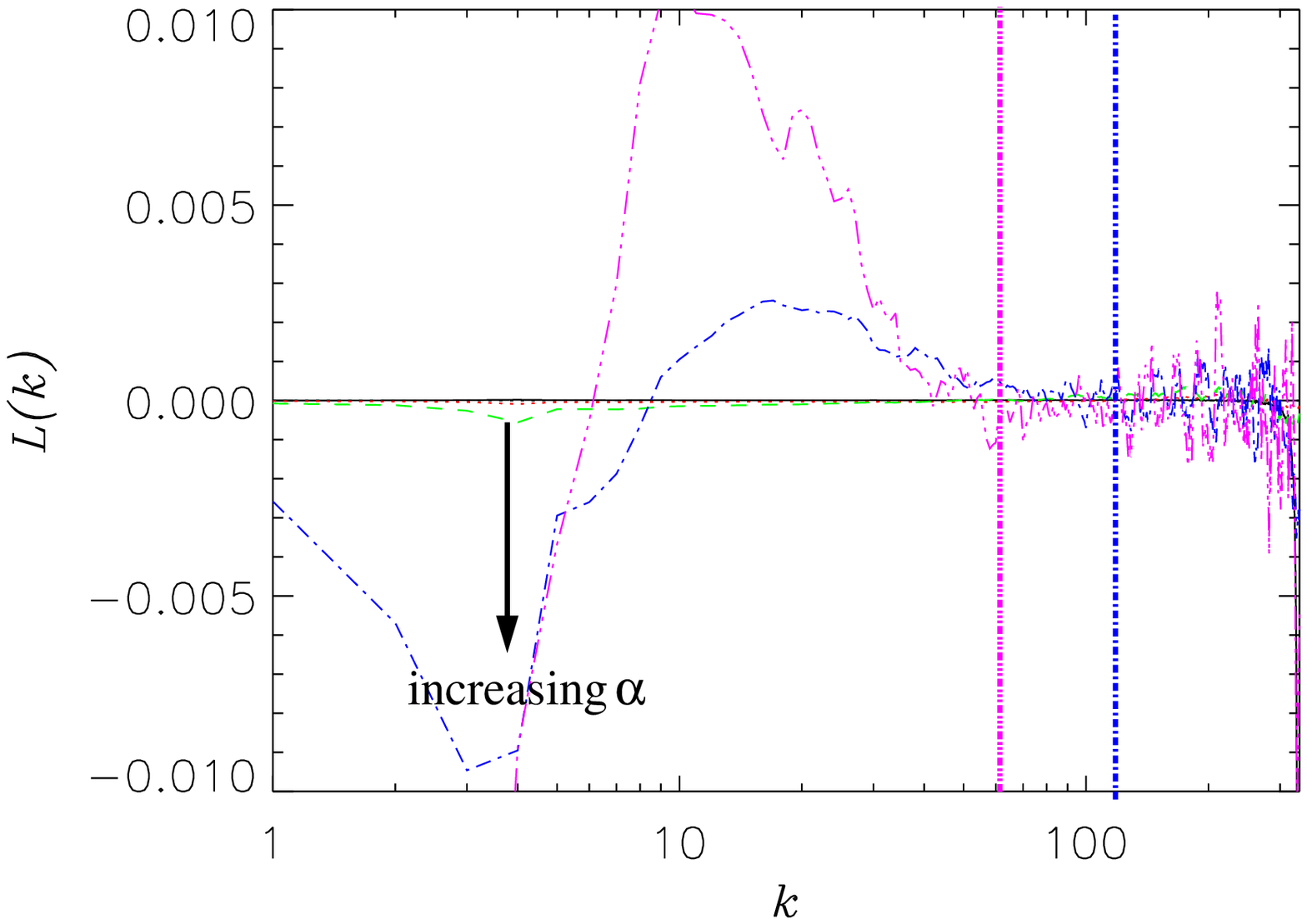}\\
\includegraphics[width=10cm]{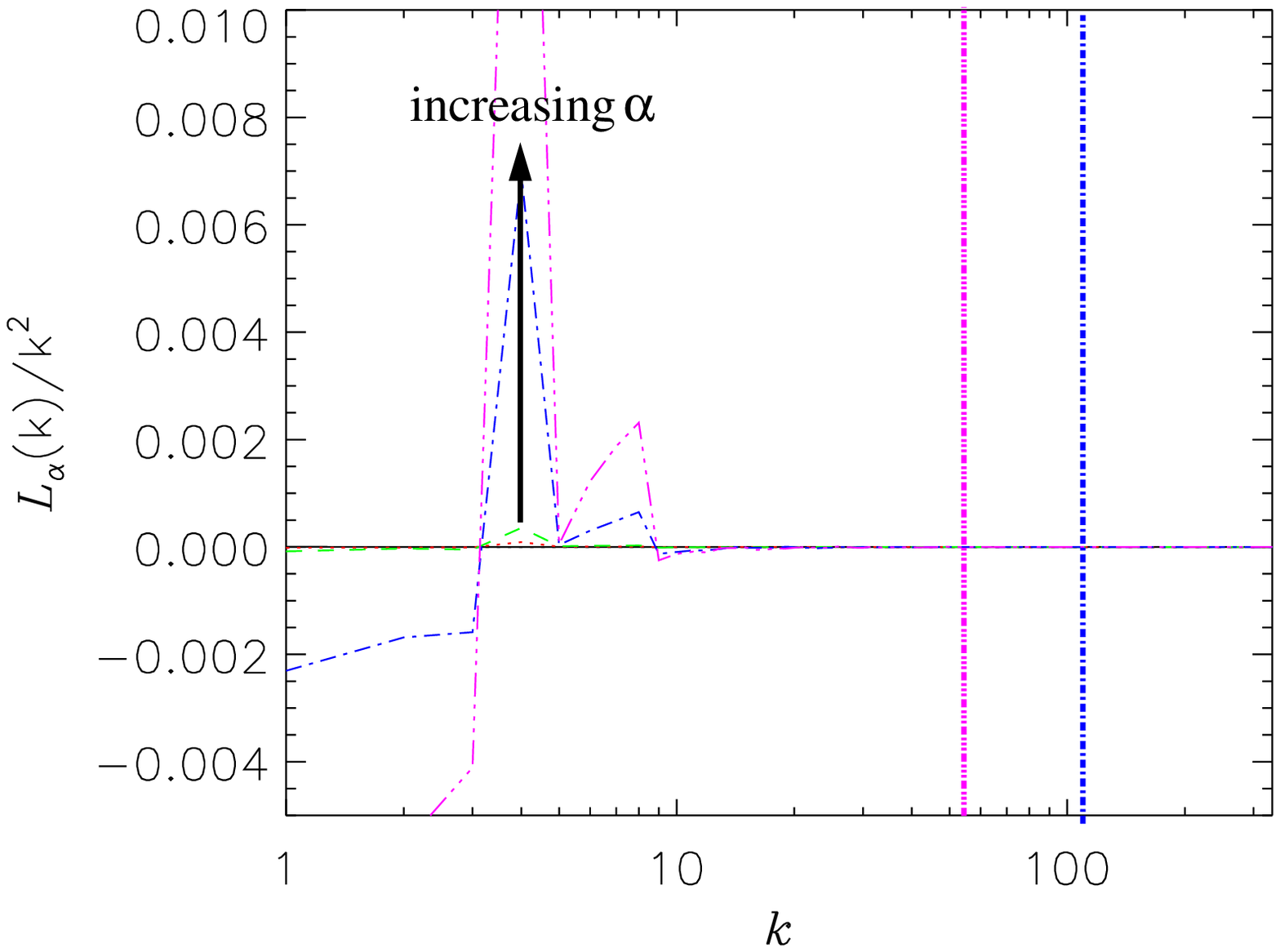}
\caption{$\alpha-$model: subgrid transfers of enstrophy ($L(k)$, Top)
  and energy {($L_\alpha(k)/k^2$,} Bottom) for $\alpha=\Delta x$ (red dotted),
  $2\Delta x$ (green dashed), $9\Delta x$ (blue dash-dotted;
  {vertical line shows wavenumber}), $16\Delta x$ (pink
  dash-triple-dotted; {vertical line shows wavenumber}), and
  benchmark (solid black, {nearly zero except for $k\gtrapprox300$ in $L(k)$).}  Due to numerical cancellation noise in
  Eq. (\ref{eq:lanssigma}), smoothing has been applied to the plots.}
\label{fig:alphaexp}
\end{figure}

The subgrid transfers, Fig. \ref{fig:alphaexp}, for the $\alpha-$model
are very large and in the wrong direction.  As the model dissipates
neither energy nor enstrophy the transfers are conservative; they
remove energy and enstrophy from above the forcing scale and deposit
them below the forcing scale.  As the filter width, $\alpha$, is increased so is the
amount of large-scale energy and enstrophy moved down-scale {to scales larger than $\alpha$ (vertical lines in Fig. \ref{fig:alphaexp}).}  

\begin{figure}[htbp]
\includegraphics[width=4.425cm]{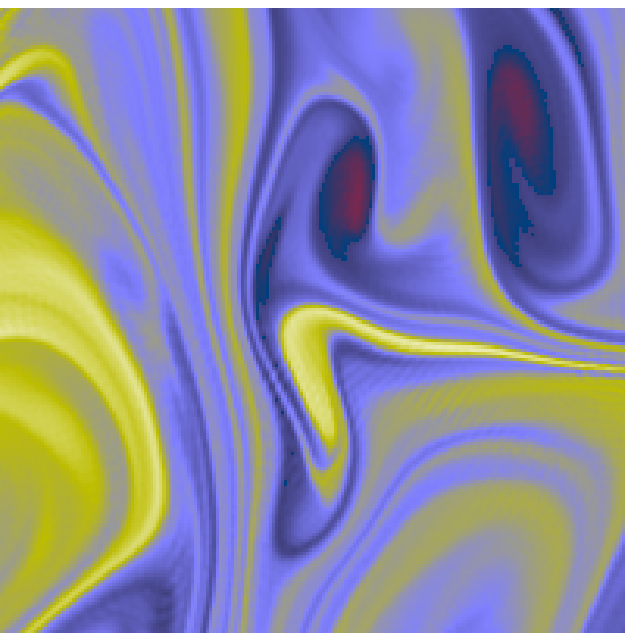}
\includegraphics[width=4.425cm]{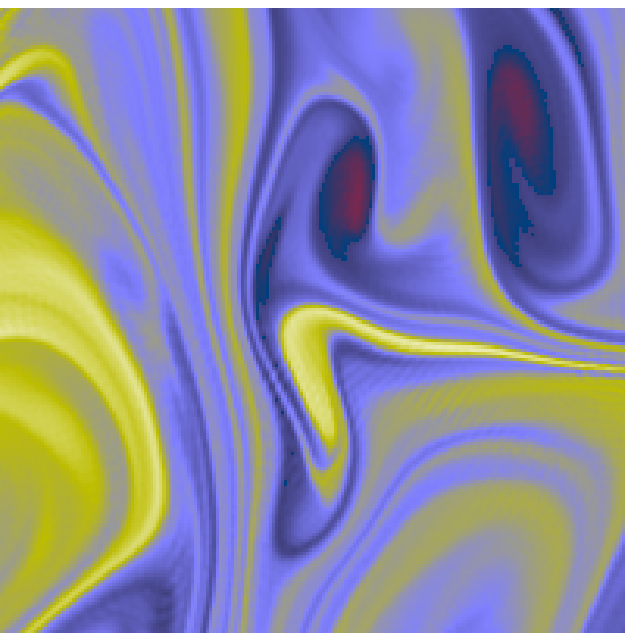}
\includegraphics[width=4.425cm]{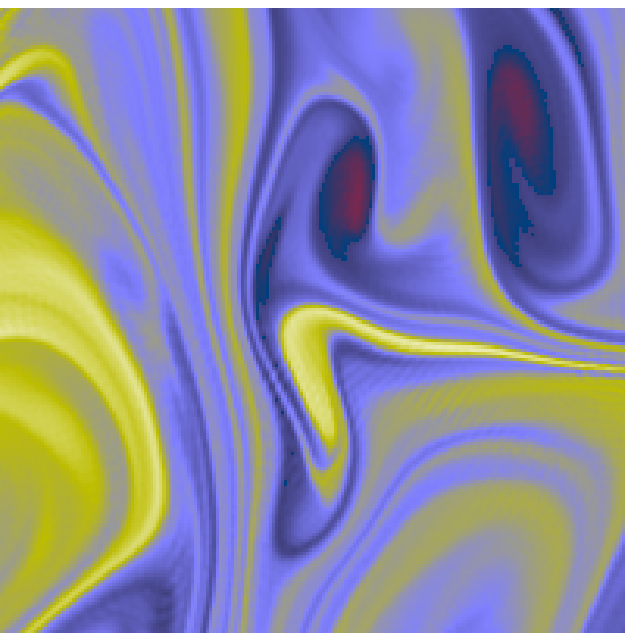}\\
\includegraphics[width=4.425cm]{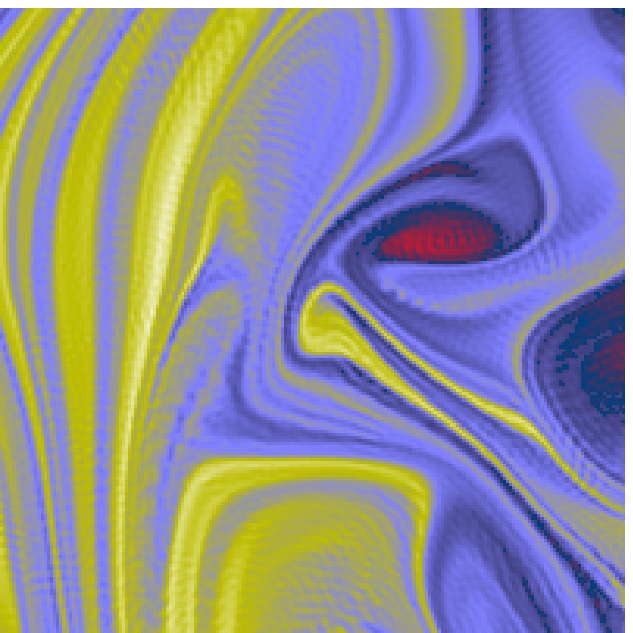}
\includegraphics[width=4.425cm]{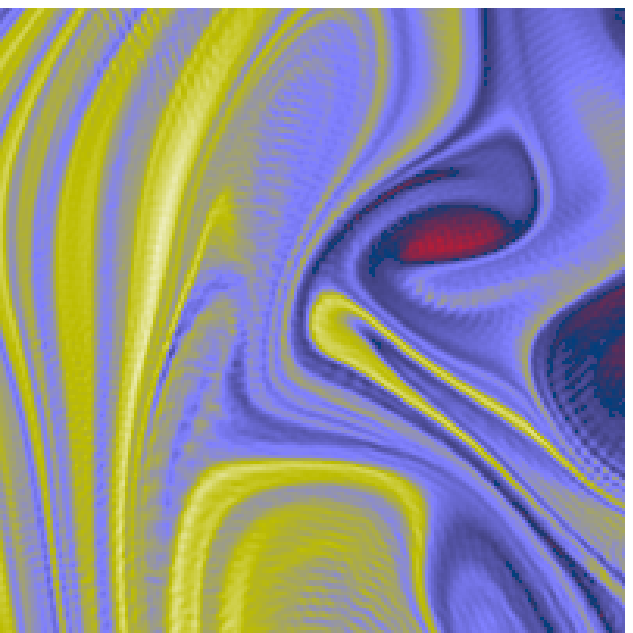}
\includegraphics[width=4.425cm]{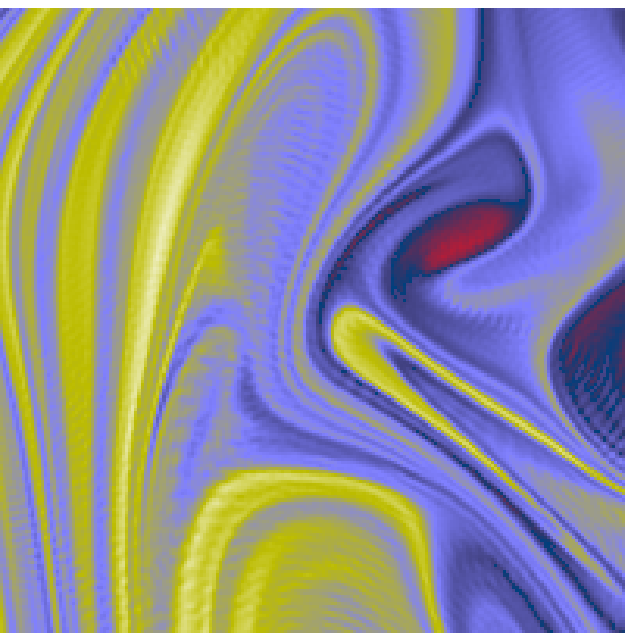}\\
\includegraphics[width=4.425cm]{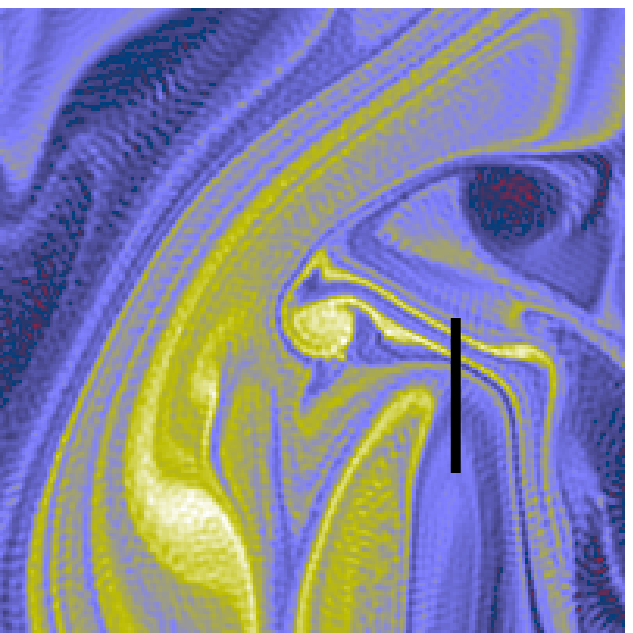}
\includegraphics[width=4.425cm]{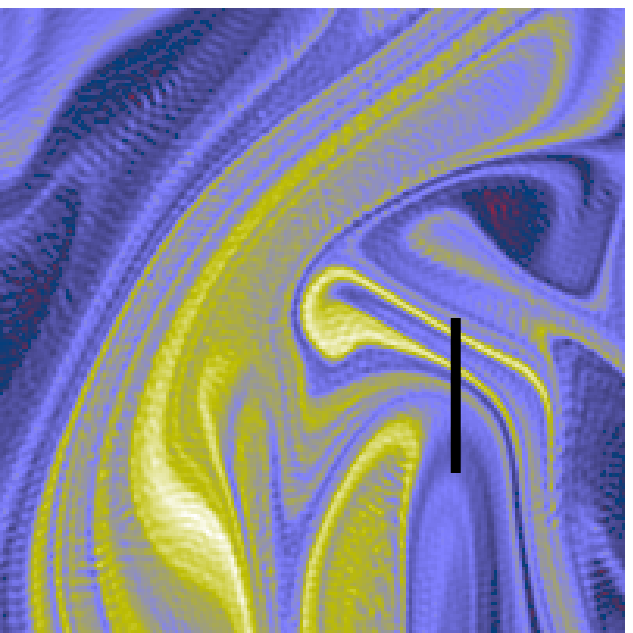}
\includegraphics[width=4.425cm]{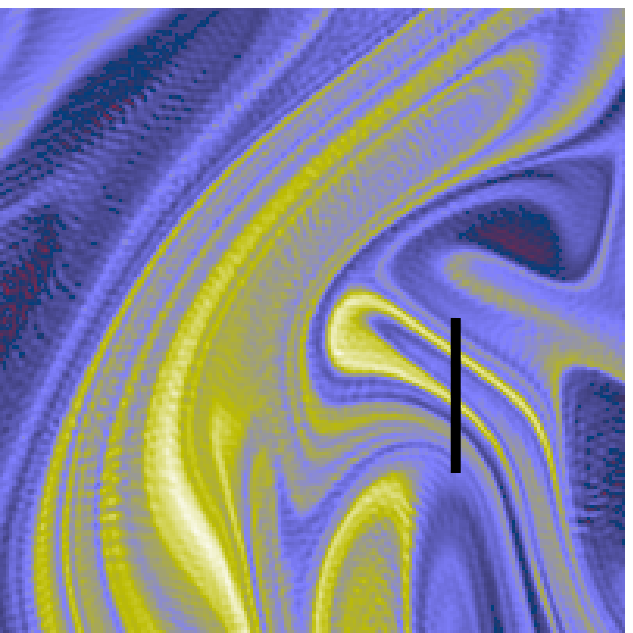}\\
\includegraphics[width=4.425cm]{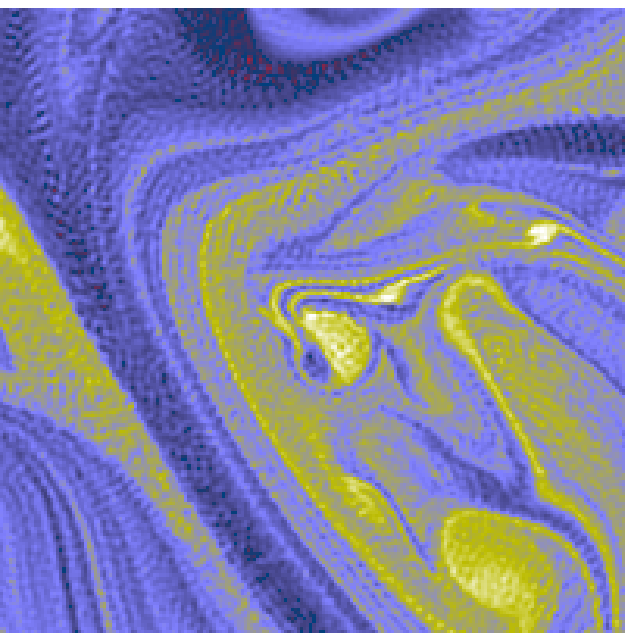}
\includegraphics[width=4.425cm]{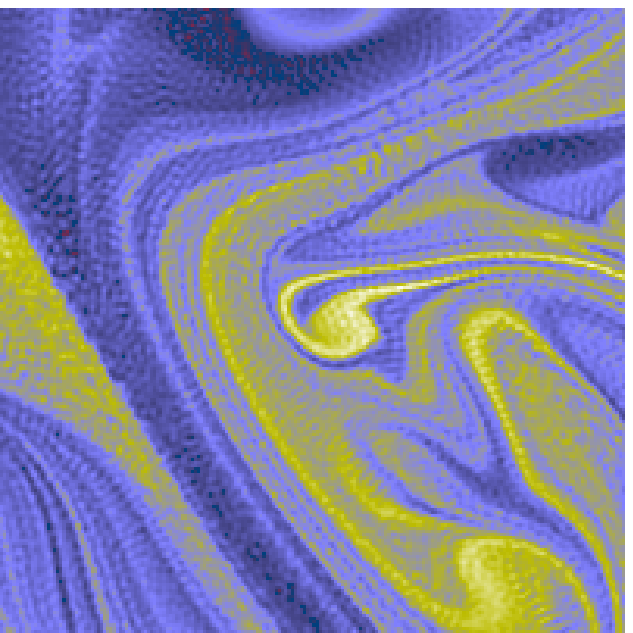}
\includegraphics[width=4.425cm]{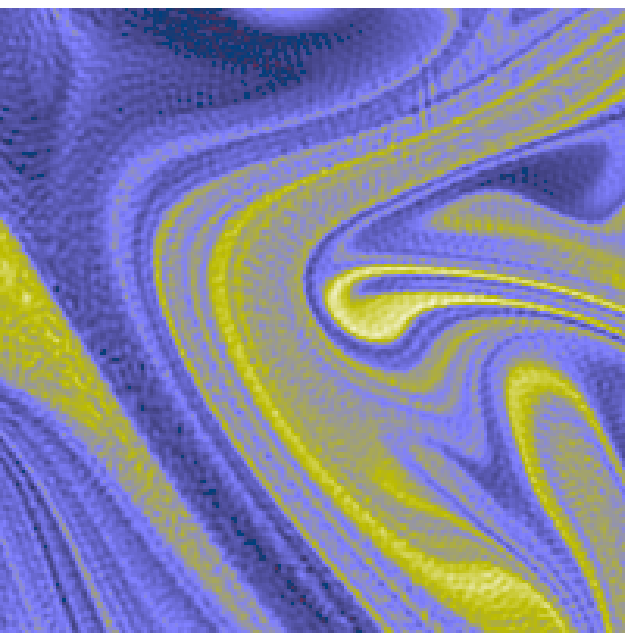}
\caption{Vorticity field, $\zeta$: Vortex merger event {(tracked
    to center of field of view which is $1/5^2$ of the entire domain).}  Time runs from top to bottom
  starting 0min after initialization in steps of $10^4\,$min.  {\sl
    1st column} LANS $\alpha=2\Delta x$, {\sl 2nd column} $9\Delta x$,
  {\sl 3rd column} $16\Delta x$.  $\alpha=2\Delta x$ is the most
  realistic result.  {Cuts from 3rd row (black lines) are plotted in Fig. \ref{fig:darryl}.}}
\label{fig:alphatrace}
\end{figure}

\begin{figure}[htbp]
\includegraphics[width=10cm]{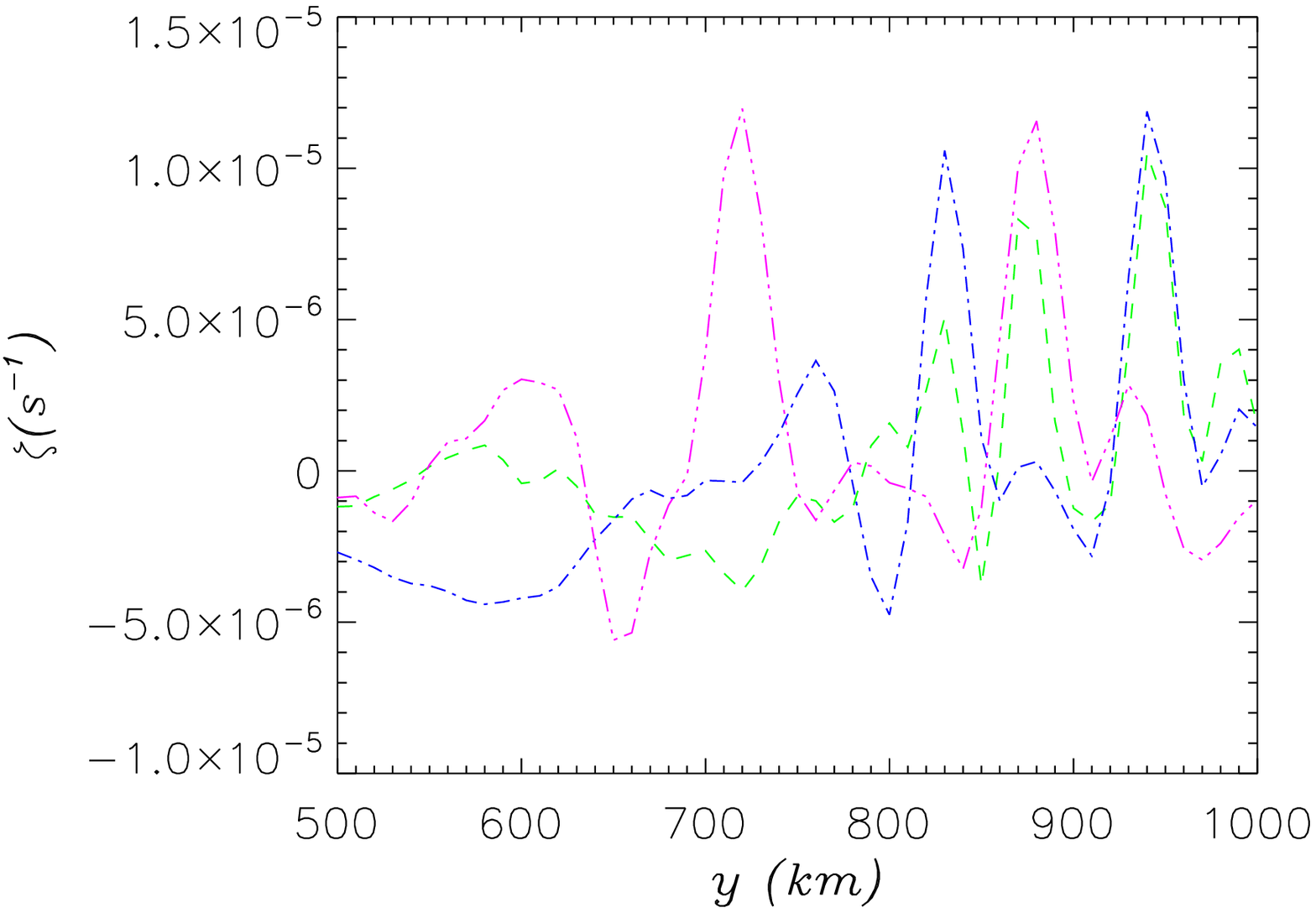}
\includegraphics[width=10cm]{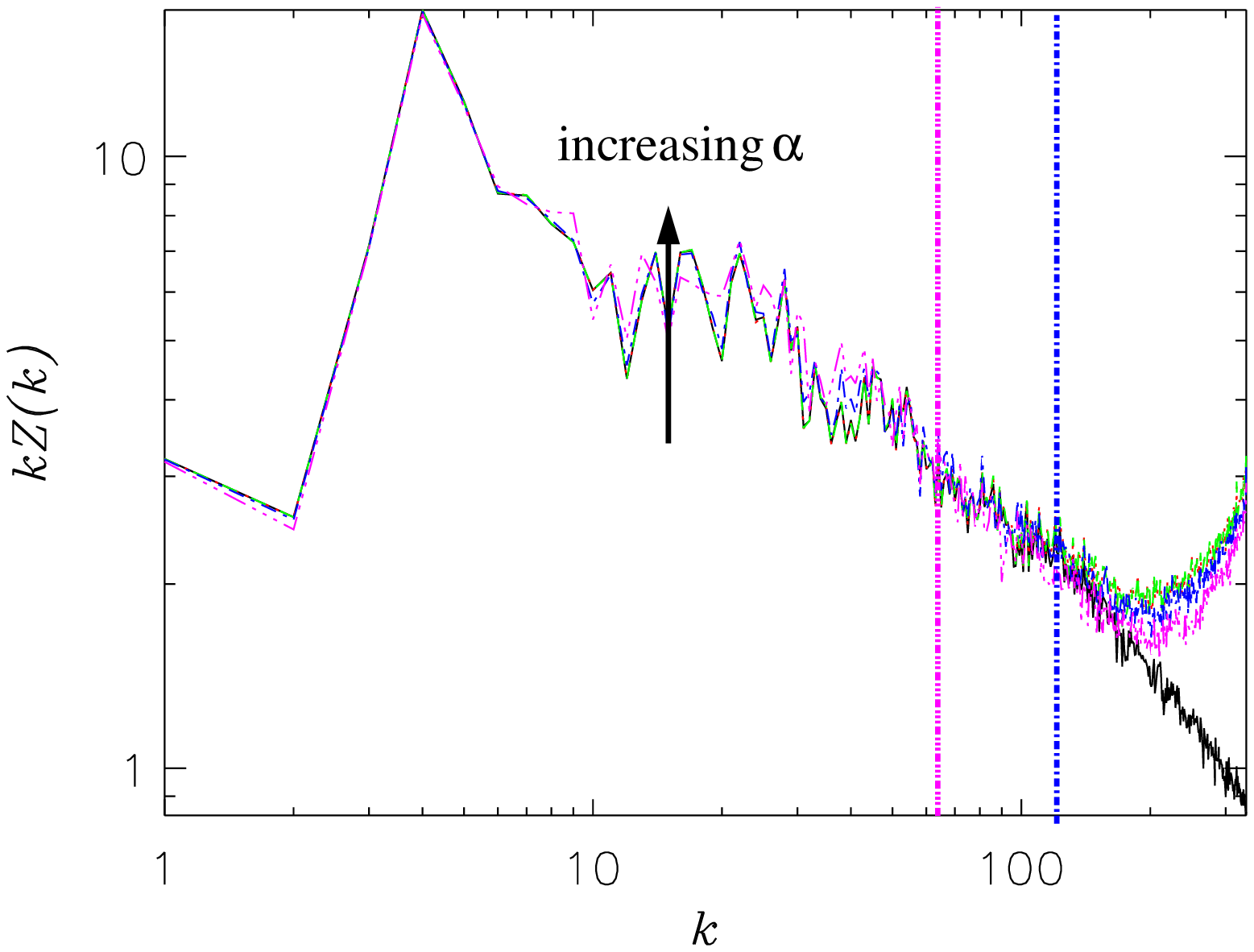}
\caption{{$\alpha-$model: (Top) Cut of vorticity field,
    $2\times10^4\,$min into simulation for $\alpha=2\Delta x$ (green
    dashed), $9\Delta x$ (blue dash-dotted), and $16\Delta x$ (pink
    dash-triple-dotted).  Section of cut is indicated in 3rd row of
    Fig.  \ref{fig:alphatrace}. (Bottom) Compensated enstrophy spectrum for same time.} }
\label{fig:darryl}
\end{figure}

{The physical effects of the $\alpha-$model are visualized in
  Fig. \ref{fig:alphatrace}: small-scale vortical motions are removed
  from the advecting field.  As $\alpha$ is increased the rotation of
  the central, yellow(light) V-shaped, vorticity feature is reduced.
  This can be seen by viewing each row from left to right.  To
  visualize the effect on the vorticity filaments, 1D cuts are taken
  as indicated by the black lines in the third row.  The vorticity values are plotted
  in Fig. \ref{fig:darryl}.  There is a translation due to the removal
  of small-scale vorticity from the advecting field.  Disregarding
  this, it is seen that the filaments are slightly larger as $\alpha$
  is increased.  The vorticity peaks are also taller.  This indicates
  that the dissipation of the filaments is reduced as $\alpha$ is
  increased.  The effect is also seen in the spectra: enstrophy is
  removed from the largest (and smallest) scales and deposited at
  scales bracketed by the forcing scale and $\alpha$.  One
  interpretation could be that the $\alpha-$model reduces both the
  roll-up and the thinning of filamentation.
The reduced roll-up reduces spatially averaged vorticity gradients and,
hence, reduces dissipation.  The reduction in thinning of the filaments
does not appear to be
 large enough to be significant for the dissipation of individual,
  small filaments.  Also due to this, more vorticity and enstrophy
  remains at super$-\alpha$ scales.}

{Note that our $\alpha-$model spectra do not compare to results
  found by \cite{NaSh2001}: their forcing kept $Z_s(10)$ constant
  rather than enstrophy injection constant, dissipated based on
  $\zeta_s$ not $\zeta$, and plotted different quantities than we have
  here.  They studied $|u_s|^2$ and $|\zeta_s|^2$ which are not the
  ideal invariants for the $\alpha-$model.  Finally, unlike for the 3D
  $\alpha-$model \cite{2001PhyD..152..505F}, no change in the scaling
  of the dissipation scale with Reynolds number is expected for the 2D
  $\alpha-$model \cite{LuKuTa+2007}.  This suggests 2D$-\alpha$ will
  not perform as a LES in the same regard as its 3D counterpart and,
  perhaps, explains our results.}


\subsection{\label{sec:comparison}Comparison of parameterizations}

\begin{figure}[htbp]
\includegraphics[width=10cm]{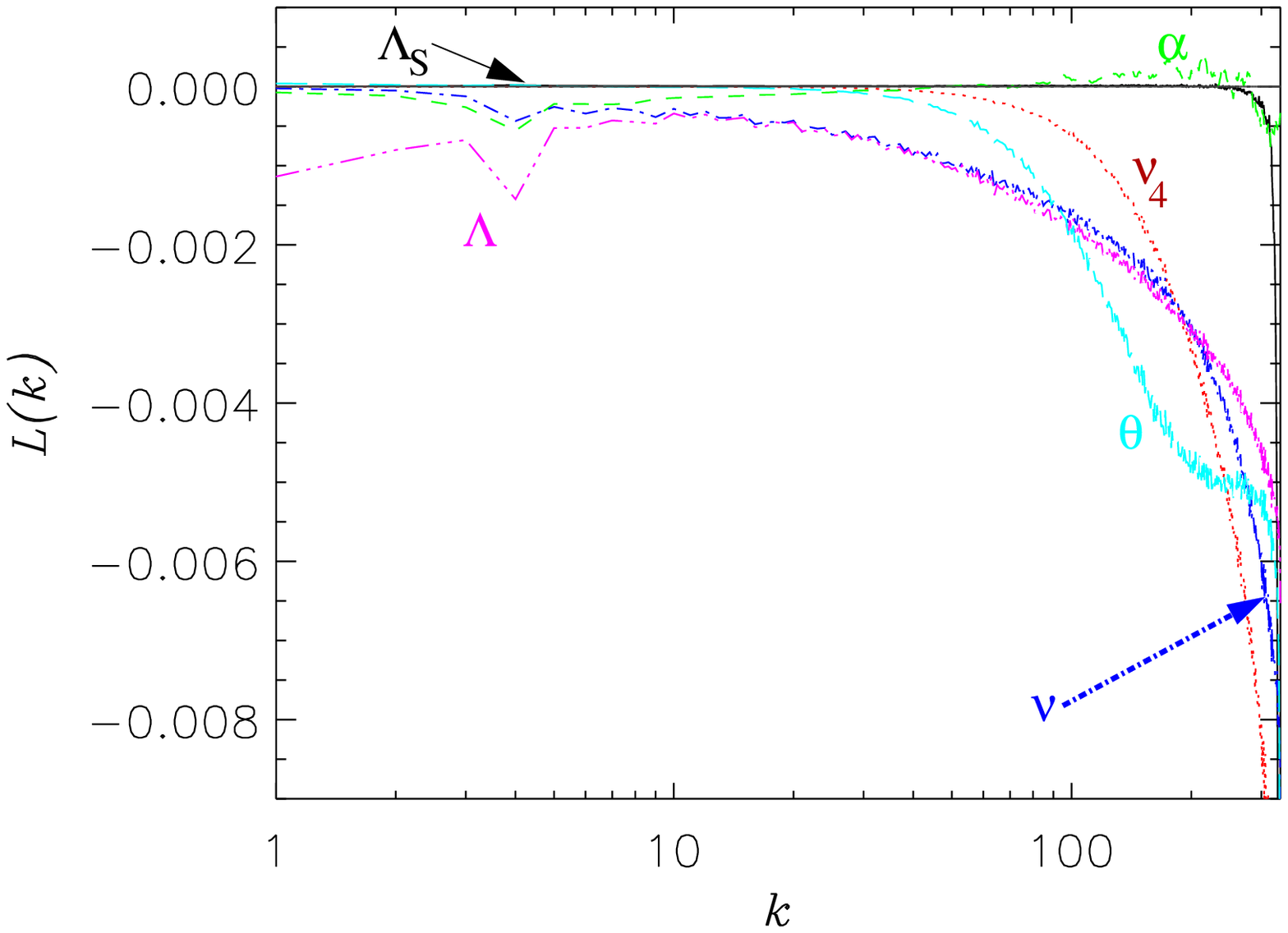}\\
\includegraphics[width=10cm]{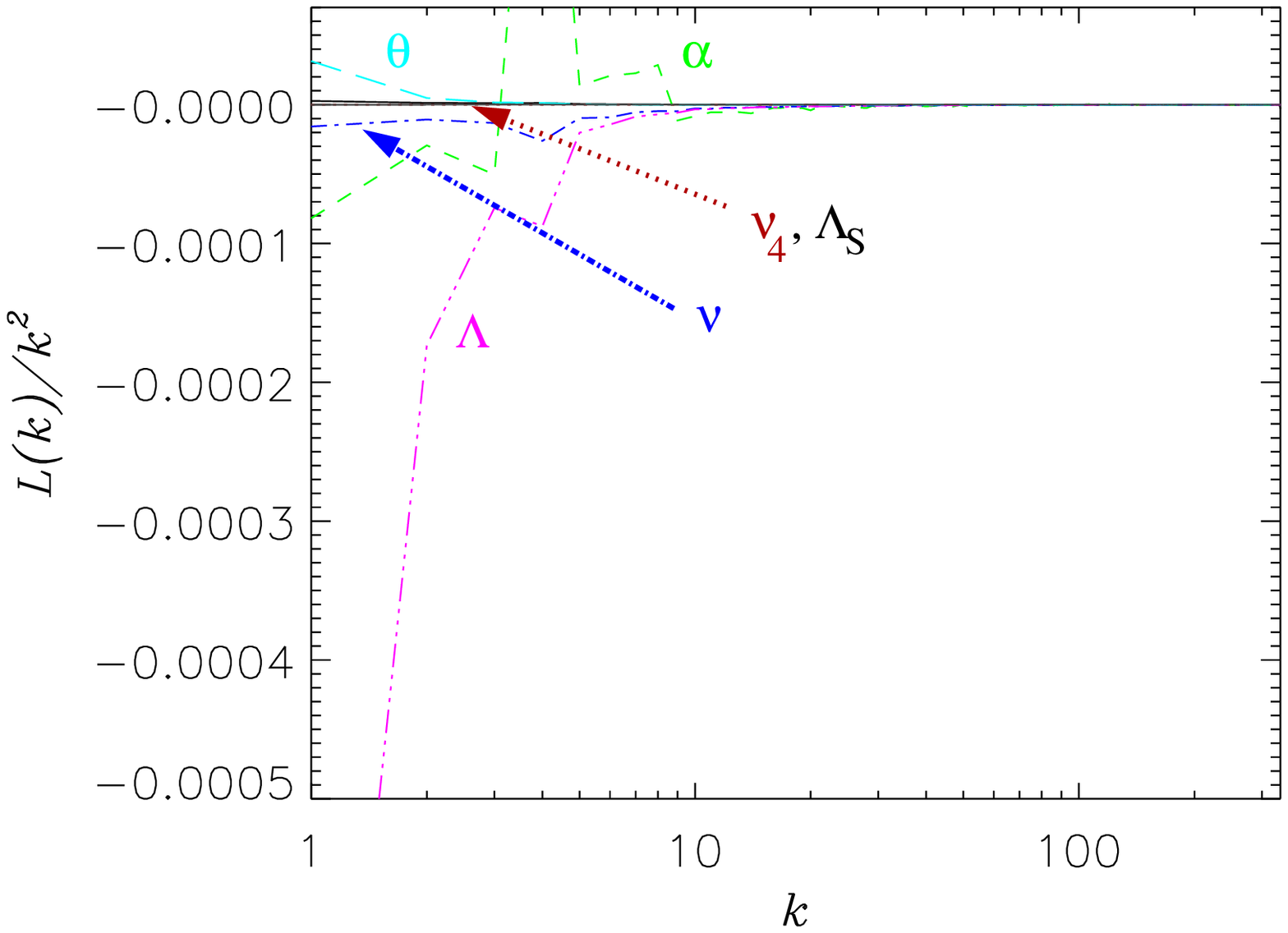}
\caption{Subgrid transfers of enstrophy ($L(k)$, Top) and energy
  ($L(k)/k^2$, Bottom) for benchmark (solid black), hyper-viscous
  $\nu_4=1.1\times10^{-9}m^4s^{-1}$ (red dotted), LANS $\alpha=2\Delta
  x$ (green dashed), viscous $\nu=11m^2s^{-1}$ (blue dash-dotted),
  Leith {$\Lambda=1$} (pink dash-triple-dotted), AVM {$\theta=0.16$}
  (cyan long-dashed), {and Smagorinsky $\Lambda_S=0.1$
    (solid grey).} }
\label{fig:compareexp}
\end{figure}

The subgrid transfers of the six parameterizations are compared in
Fig. \ref{fig:compareexp}.  {Concentrating on the subgrid
  enstrophy transfer, we can eliminate the $\alpha-$model because it
  unphysically generates enstrophy for $100\lessapprox
  k\lessapprox200$ and Smagorinsky can be eliminated because it
  essentially eliminates zero small-scale enstrophy (the grey line is
  flat an indistinguishable from zero on this vertical scale).  Of the
  remaining models, the hyper-viscous is closest to mimicking the true
  subgrid transfers of both energy and enstrophy,} though for the largest
wavenumbers, $k>200$, the viscous and Leith parameterizations perform
similarly.  The AVM is the only method that reproduces the correct
sign of the energy transfer, but it removes enstrophy preferentially
from {intermediate scales} instead of the smallest resolved
scales.  This method would likely perform better for $m>1$
\citep{VaHu1988}.

\begin{figure}[htbp]
\includegraphics[width=10cm]{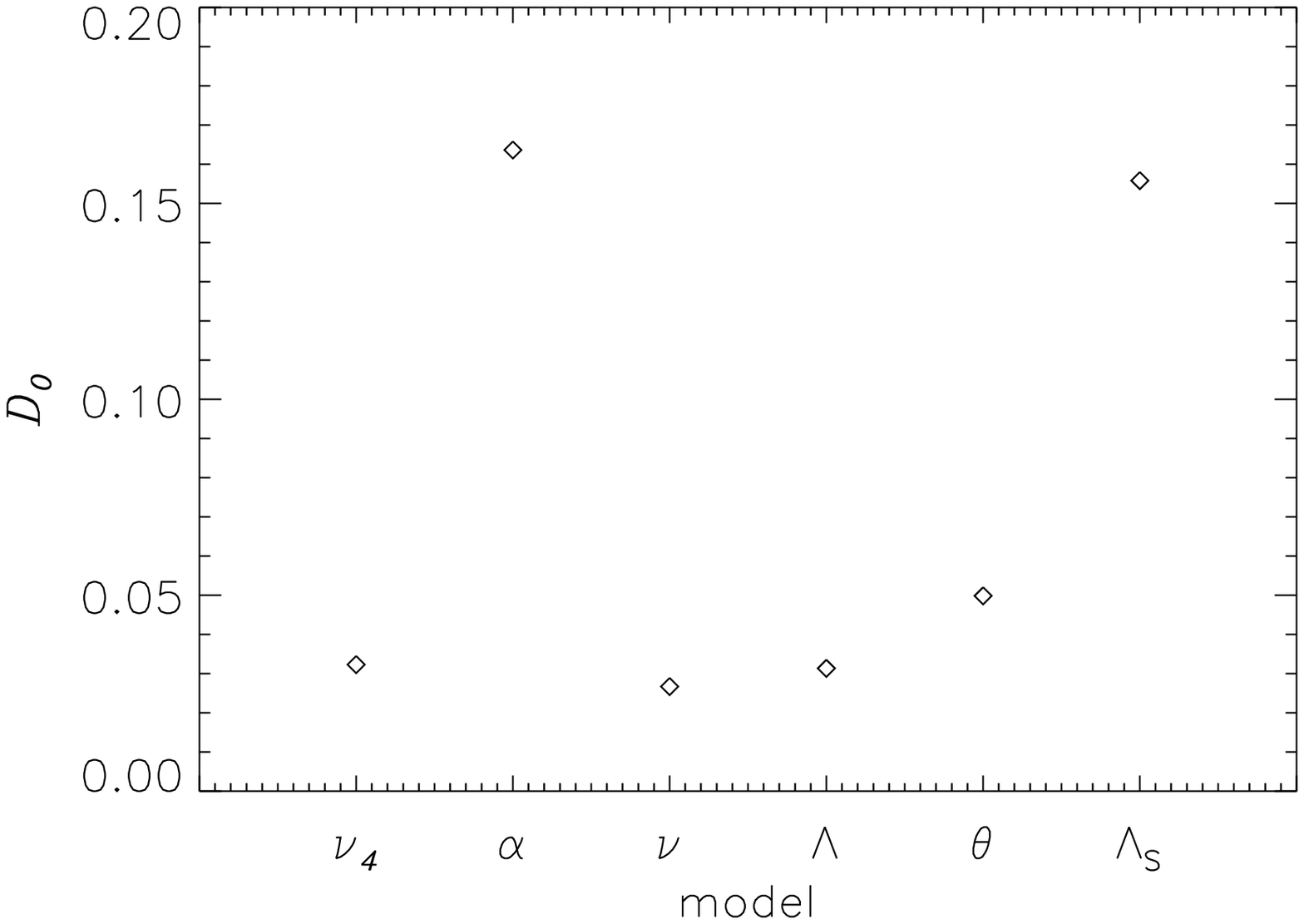}
\includegraphics[width=10cm]{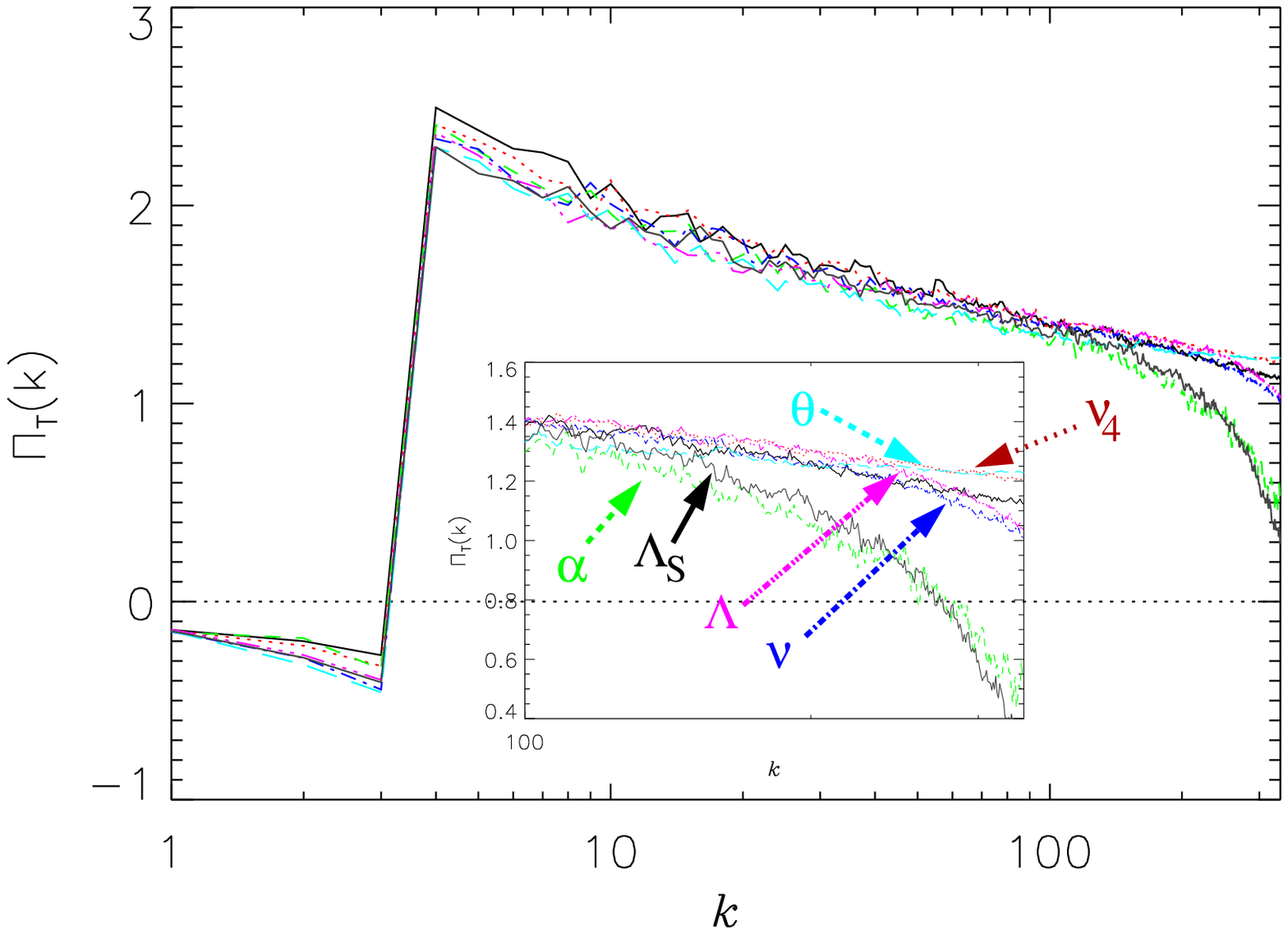}
\caption{{(Top)} Enstrophy flux error-landscape {norm $D_0$} for hyper-viscous
  ($\nu_4$), $\alpha-$model ($\alpha$), $\nabla^2$ viscosity ($\nu$),
  Leith ($\Lambda$), AVM ($\theta$), and Smagorinsky ($\Lambda_S$).
{(Bottom) Modeled enstrophy flux, $\Pi_T$, for for benchmark
    (solid black), hyper-viscous $\nu_4=1.1\times10^{-9}m^4s^{-1}$
    (red dotted), LANS $\alpha=2\Delta x$ (green dashed), viscous
    $\nu=11m^2s^{-1}$ (blue dash-dotted), Leith $\Lambda=1$ (pink
    dash-triple-dotted), AVM {$\theta=0.16$} (cyan long-dashed), and
      Smagorinsky $\Lambda_S=0.1$ (solid grey).  }  }
\label{fig:comparetrue}
\end{figure}

The enstrophy flux error landscape norms are given in
Fig. \ref{fig:comparetrue}.  The $\alpha-$model {and Smagorinsky are} the obvious outliers
with {a factor of five poorer performance.  The viscous, hyper-viscous
and Leith parameterizations have very
similar performance.    The AVM is within a factor of two in performance.
  Again, this could likely
be improved upon by using a larger value of $m$.}

\begin{figure}[htbp]
\includegraphics[width=10cm]{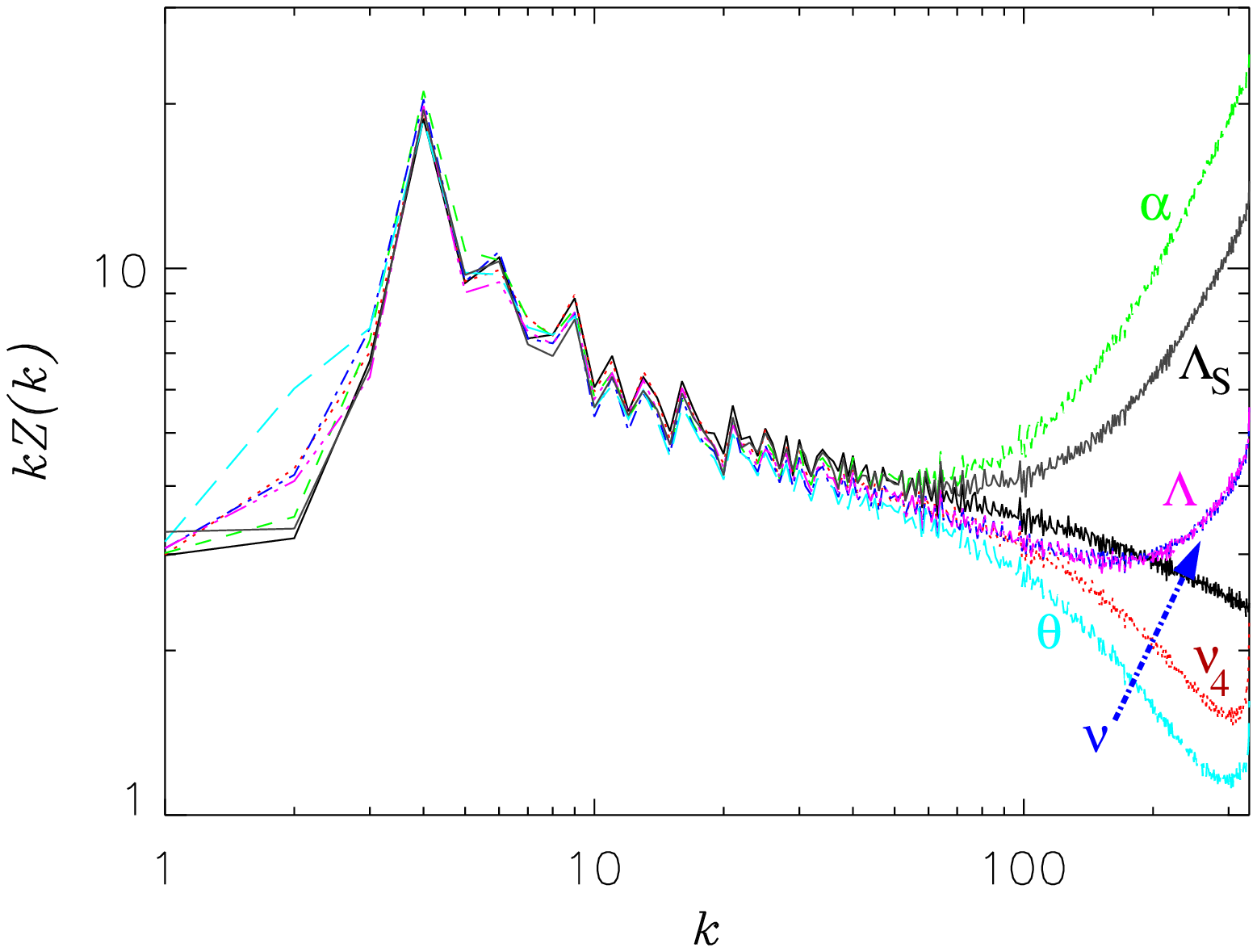}
\caption{Enstrophy spectra for benchmark (solid black), hyper-viscous
    $\nu_4=1.1\times10^{-9}m^4s^{-1}$ (red dotted), LANS
    $\alpha=2\Delta x$ (green dashed), viscous ${\nu'}=11m^2s^{-1}$ (blue
    dash-dotted), Leith $\Lambda=1$ (pink dash-triple-dotted), AVM
    {$\theta=0.16$} (cyan long-dashed), {and Smagorinsky $\Lambda_S=0.1$
    (solid grey).}}
\label{fig:compare}
\end{figure}

This similarity in performance between hyper-viscous, viscous, and
Leith parameterizations can also been seen in the resulting enstrophy
spectra, Fig. \ref{fig:compare}.  All three methods have the same
spectra for $k\lessapprox100$.  {Neither the $\alpha-$model
nor Smagorinsky reduces the pile-up of numerical thermalization 
noise \cite{CiBoDe2005}} in the small scales.  As seen in the previous results, the AVM
method with $m=1$ is dissipative at too large scales to perform as
well as the viscous, hyper-viscous, or Leith parameterizations.
{Smagorinsky performs poorly because it removes enstrophy from the
largest rather than the smallest resolved scales.}

\section{\label{sec:conclusions}Conclusions}

We have compared six popular turbulence parameterizations in the
enstrophy cascade regime of the barotropic vorticity equation on a
$f-$plane (equivalently, 2D Navier-Stokes) in forced-dissipative
simulations.  The hyper-viscous, viscous, and Leith models all perform
well down to about $10\Delta x$.  
The hyper-viscous model
reproduces the largest-resolved-scales ($1\le k\le100$) flux the best of the three and
the viscous model best reproduces the smallest-resolved-scales  ($k\ge200$) flux.
The Leith model, {because its diffusion is anisotropic,} is expected to carry-over its performance to
anisotropic flows {(e.g., the 3D baroclinic ocean system)} which would
be challenging for the viscous and hyper-viscous models.  {The
  Smagorinsky model does not work in the enstrophy cascade regime--it
  removes enstrophy from the largest rather than the smallest resolved
  scales.}  The anticipated vorticity method without a strong enough
weighting to small scales,
{larger values of $m$,}
 does not perform as well as the prior three
parameterizations.  As even this order of diffusive operator is not
practical in the finite-volume {and finite-difference schemes
  typically used in global ocean modeling} (e.g.,
\cite{RiThKl+2010}), we chose not {to} investigate higher-orders.

We have confirmed \cite{LuKuTa+2007}'s suggestion that the
Lagrangian-averaged $\alpha-$model does not perform as a turbulence
model in this system {(see also \cite{NaSh2001}).  Analytically,
  one expects the numerical degrees of freedom to scale with Reynolds
  number the same as unparameterized Navier-Stokes.  The model reduces
  rotation due to small-scale vorticity and, less dramatically, also
  reduces the thinning of vortex filaments due to stretching.  The
  balance of the effect is a net reduction of dissipation of the
  vorticity filaments and a piling of energy and enstrophy to
  sub-forcing/super$-\alpha$ scales (enhancing the flux in this
  spectral region).}

{One possible MOLES closure has not been scoped here,} the use of
monotone transport as the model for LES closure. These closures,
commonly referred to Monotone Implicit Large-Eddy Simulation (MILES),
require the evaluation of the nonlinear transport be carried out in
physical space, something that is not possible within the global
spectral model utilized for this study. Our future work, discussed
briefly below, will utilize a traditional finite-volume approach where
an evaluation of MILES will be possible.  {Combined models have
also not been investigated here due to the enormous parameter space
that would entail.}

Subgrid transfers have been measured before, e.g., for the APVM
\citep{VaHu1988}, and again for the APVM, hyper-viscosity, and
implicit large-eddy simulations \citep{ThKeWo2011}.  Error-landscapes
for LES have been produced for various quantities like spectra
\citep{MeGeBa2003,MeSaGe2006,MeGeSa2007,Me2011}.  By combining these
two techniques, however, we have introduced a method for determining
the optimal turbulence parameterization also in flows different than
those considered here: the error-landscape of the enstrophy flux at
small-scales in a 2D flow can be replaced by the error-landscape of
the modeled {flux in a 3D baroclinic system \NEU{(see \ref{sec:appendix}).}
  {We emphasize that MOLES comparisons based on spectra alone do
    not ensure that the correct dynamics are being reproduced by a
    parameterization.  For example, consider the $\Lambda_S=0.5$
    ($C_S\approx0.16$) result for the Smagorinsky model (blue
    dash-dotted line in Fig. \ref{fig:smagtrue}).
    The spectrum is best approximated by this run, but for the wrong
    reasons as the non-linear flux is more poorly reproduced than for
    $\Lambda_S=0.1$.  For the viscous model, however, which physically
    correctly removes enstrophy from the small scales, both the
    spectrum and the flux are best reproduced for
    ${\nu'}=11\,$m$^2$s$^{-1}$.  In this latter case, the spectrum is
    reproduced because the dynamics are reproduced.}

{For a 3D baroclinic system, at the scales on which the MOLES
  acts \NEU{($5-10\,$km),} the system will be approximately \NEU{QG.  Because of the similarities between QG and 2D \cite{FoKeMe2008}, we have some expectation} that our
  results will hold: Smagorinsky, the $\alpha-$model, and APVM with
  $m=1$ will not perform as well as viscosity, hyper-viscosity, and
  Leith.  In fact, because of the anisotropic diffusion offered by
  Leith, it will likely perform the best.  Our results {may} not extend
  to the 3D system, however, if additional physics comes in to play,
  like vertical mixing over small horizontal scales.}  

Our next step is to move into an idealized, 3D baroclinic system, most
likely a re-entrant zonal channel that can serve as an idealized
Antarctic Circumpolar Current. While the move to three dimensions
allows for the direct simulation of baroclinic instability, it also
necessitates the development of analysis tools that can accurately
account for energy and enstrophy transfers between the disparate
horizontal and vertical scales of motion. Furthermore, the move to a
3D baroclinic system entail\NEU{s} the use of a height-based vertical
coordinate. Such a system requires the transport of one or more
tracers in order to close the system via an equation of state.
\NEU{The theoretical analysis of such a system is outlined in
  \ref{sec:appendix}.}  And finally, as we move to more realistic and,
thus, bounded domains, our ability to simulate the governing
equations, as well as analyze the fluxes, via global spectral
expansions will be increasingly cumbersome. As a result, we plan on
utilizing a traditional finite-volume global ocean model
\cite{Ringler:8qyuvoZk} in the next phase of this study.  \\



{\bf Acknowledgments} We acknowledge discussions with B. Fox Kemper,
{D. Holm,}
and B. Wingate.  This work was supported by the Earth System Modeling
and Regional and Global Climate Modeling programs of the Office of
Biological and Environmental Research within the US Department of
Energy's Office of Science.



\appendix
\section{\NEU{3D baroclinic case}}
\label{sec:appendix}
\NEU{In this section, we apply our methodology  to a 3D baroclinic
system of equations.  The hydrostatic, traditionally shallow, and simple Boussinesq equations, e.g., as solved by \cite{Ringler:8qyuvoZk}, are
\begin{eqnarray}
\label{cvelocity}
\frac{\partial {\bf u}}{\partial t}
+ \frac{1}{2}\nabla_h \left| {\bf u} \right|^2
+ (f + \zeta) \hat{\vec{z}}\times \vec{u}
+ w\frac{\partial {\bf u}}{\partial z}
  = - \frac{1}{\rho_0}\nabla_h p 
   + F + D + Q\\
\nabla_3\cdot\vec{v}=0\\
\partial_t \rho +\nabla_h (\rho\vec{u}) + \partial_z(\rho w) = 0 \label{eq:mass}\\
\partial_t (\rho T) +\nabla_h (\rho T\vec{u}) + \partial_z(\rho Tw) = 0
\end{eqnarray}
where $\vec{v}=\vec{u}+w\hat{\vec{z}}$ is the full 3D velocity field, $\vec{u}$ and $w$ are the horizontal and vertical components,
$\nabla_h$ is the 2D horizontal gradient, $\nabla_3$ is the full 3D gradient,
$f$ is the Coriolis force, $\zeta=\hat{\vec{z}}\cdot\nabla_h\times\vec{u}$ is the vertical
component of vorticity, $\rho_0$ is the background density, $p$ is the
pressure, and $T$ is the temperature.  We have assumed constant salinity and
a linear equation of state for simplicity.}

\NEU{Unlike for the 2D case, for the 3D
system, we must also consider transfer between available potential (APE) and kinetic
energies.  The time evolution of the horizontal kinetic energy, $KE\equiv\vec{u}^2/2$,  is given by
\begin{equation}
\partial_t(KE) + {\nabla_3\cdot\big{[}\vec{v}(KE+\frac{p}{\rho_0})\big{]}}={-\frac{g\rho}{\rho_0}w}
\end{equation}
Where we make use of the hydrostatic condition, $\partial_z{p}=-g{\rho}$,
and have left off the terms $\vec{u}\cdot\vec{F}+\vec{u}\cdot\vec{D}+\vec{u}\cdot\vec{Q}$ for brevity. We define the APE, $\Phi$, implicitly by
$\big{(}\partial_z\Phi\big{)}_{\rho T}\equiv{g\rho}/{\rho_0}$.
From this, we can derive \cite{Va2006}
\begin{eqnarray}
\frac{D\Phi}{Dt}=
\bigg{(}\frac{D\Phi}{D\rho T}\bigg{)}_z\frac{D\rho T}{Dt}
+ \bigg{(}\frac{D\Phi}{Dz}\bigg{)}_{\rho T}\frac{Dz}{Dt}=\frac{g\rho}{\rho_0}w
\nonumber \\
\partial_t\Phi + {\nabla_3\cdot(\Phi\vec{v})}={\frac{g\rho}{\rho_0}w}
\label{eq:APE}
\end{eqnarray}
The sum of horizontal kinetic energy and available potential energy is
ideally conserved (i.e., when $F=D=Q=0$ and no transport occurs across
the system boundaries).  The exchange between the two energy reservoirs is via
the $g\rho w/\rho_0$ term.}

\NEU{The transfer functions for the 3D ocean system determine the time
rate of change of the horizontal kinetic energy ``spectrum,''
\begin{eqnarray}
\partial_tKE_\kappa=TKK_\kappa+TKKP_\kappa+TAK_\kappa\,,
\end{eqnarray}
where $\mathcal{B}{\{}\cdot\}_\kappa$ will represent projection onto a complete orthonormal basis.
For example, in a zonally-reentrant channel this basis could be Fourier modes in
the zonal direction, sines in the meridional  direction, and baroclinic
eigenmodes in the vertical.
 The transfer of $KE$ from other modes to a given
orthonormal mode (equivalent of $T(k)=S(k)/k^2$) is
\begin{equation}
TKK_\kappa = -\mathcal{B}{\{}\vec{u}{\}}_\kappa^\ast\cdot\mathcal{B}\bigg{\{}\vec{v}\cdot\nabla_3\vec{u}\bigg{\}}_\kappa
\end{equation}
where $\vec{v}\cdot\nabla_3\vec{u}=\nabla_h KE
+ \zeta \hat{\vec{z}}\times \vec{u}
+ w\partial_z {\bf u}$,
the transfer of $KE$ from other modes due to the pressure term is
\begin{equation}
TKKP_\kappa = -\mathcal{B}\big{\{}\vec{v}\big{\}}_\kappa^\ast\cdot\mathcal{B}\bigg{\{}\nabla_3\frac{p}{\rho_0}\bigg{\}}_\kappa\,,
\end{equation}
and net transfer rate from APE to $KE_\kappa$ is  
\begin{equation}
TAK_\kappa = -\mathcal{B}\big{\{}w\big{\}}_\kappa^\ast\cdot\mathcal{B}\bigg{\{}\frac{g\rho}{\rho_0}\bigg{\}}_\kappa\,.
\end{equation}
Similarly, expressions for the transfer functions for
Eq. (\ref{eq:APE}) can be written.  As we have shown for the 2D case,
the method to determine the best turbulence closure for an
eddy-permitting ocean model is to compute the transfer functions
derived here (and their integrals, the fluxes).  Then, the $D_0$ norm
error-landscape will be computed for each potential MOLES by
comparison with a high resolution benchmark run.}





\begin{thebibliography}{10}
\expandafter\ifx\csname url\endcsname\relax
  \def\url#1{\texttt{#1}}\fi
\expandafter\ifx\csname urlprefix\endcsname\relax\def\urlprefix{URL }\fi
\expandafter\ifx\csname href\endcsname\relax
  \def\href#1#2{#2} \def\path#1{#1}\fi

\bibitem{FoKeMe2008}
B.~Fox-Kemper, D.~Menemenlis, \href{http://tinyurl.com/FoKeM08}{Can large eddy
  simulation techniques improve mesoscale-rich ocean models?}, in: M.~Hecht,
  H.~Hasumi (Eds.), Ocean Modeling in an Eddying Regime, Vol. 177, AGU
  Geophysical Monograph Series, 2008, pp. 319--338.

\bibitem{2000AnRFM..32....1M}
C.~{Meneveau}, J.~{Katz}, {Scale-Invariance and Turbulence Models for
  Large-Eddy Simulation}, Annual Review of Fluid Mechanics 32 (2000) 1--32.

\bibitem{Sm1963}
J.~{Smagorinsky}, {General Circulation Experiments with the Primitive
  Equations}, Monthly Weather Review 91 (1963) 99.

\bibitem{Le1996}
C.~E. {Leith}, {Stochastic models of chaotic systems}, Physica D 98 (1996)
  481--491.

\bibitem{LeMeCo2005}
M.~{Lesieur}, O.~{M{\'e}tais}, P.~{Comte}, {Large-Eddy Simulations of
  Turbulence}, 2005.

\bibitem{FoKePe2004}
B.~Fox-Kemper, J.~Pedlosky,
  \href{http://dx.doi.org/10.1357/002224004774201681}{Wind-driven barotropic
  gyre {I: C}irculation control by eddy vorticity fluxes to an enhanced removal
  region}, Journal of Marine Research 62~(2) (2004) 169--193.

\bibitem{FoKe2005}
B.~Fox-Kemper, \href{http://dx.doi.org/10.1175/JPO2743.1}{Reevaluating the
  roles of eddies in multiple barotropic wind-driven gyres}, Journal of
  Physical Oceanography 35~(7) (2005) 1263--1278.

\bibitem{SaBa1985}
R.~{Sadourny}, C.~{Basdevant}, {Parameterization of Subgrid Scale Barotropic
  and Baroclinic Eddies in Quasi-geostrophic Models: Anticipated Potential
  Vorticity Method.}, Journal of Atmospheric Sciences 42 (1985) 1353--1363.

\bibitem{VaHu1988}
G.~K. {Vallis}, B.-L. {Hua}, {Eddy viscosity of the anticipated potential
  vorticity method}, Journal of Atmospheric Sciences 45 (1988) 617--627.

\bibitem{ChGuRi2011}
Q.~{Chen}, M.~{Gunzburger}, T.~{Ringler}, {A Scale-Invariant Formulation of the
  Anticipated Potential Vorticity Method}, Monthly Weather Review 139 (2011)
  2614--2629.

\bibitem{2012arXiv1206.2607G}
F.~{Gay-Balmaz}, D.~D. {Holm}, {Parameterizing interaction of disparate scales:
  Selective decay by Casimir dissipation in fluids}, ArXiv e-prints\href
  {http://arxiv.org/abs/1206.2607} {\path{arXiv:1206.2607}}.

\bibitem{HoMaRa1998}
D.~D. {Holm}, J.~E. {Marsden}, T.~S. {Ratiu}, {Euler-Poincar{\'e} Models of
  Ideal Fluids with Nonlinear Dispersion}, Physical Review Letters 80 (1998)
  4173--4176.

\bibitem{1998PhRvL..81.5338C}
S.~{Chen}, C.~{Foias}, D.~D. {Holm}, E.~{Olson}, E.~S. {Titi}, S.~{Wynne},
  {Camassa-Holm Equations as a Closure Model for Turbulent Channel and Pipe
  Flow}, Physical Review Letters 81 (1998) 5338--5341.

\bibitem{1999PhyD..133...49C}
S.~{Chen}, C.~{Foias}, D.~D. {Holm}, E.~{Olson}, E.~S. {Titi}, S.~{Wynne}, {The
  Camassa-Holm equations and turbulence}, Physica D Nonlinear Phenomena 133
  (1999) 49--65.

\bibitem{1999PhyD..133...66C}
S.~{Chen}, D.~D. {Holm}, L.~G. {Margolin}, R.~{Zhang}, {Direct numerical
  simulations of the Navier-Stokes alpha model}, Physica D Nonlinear Phenomena
  133 (1999) 66--83.

\bibitem{1999PhFl...11.2343C}
S.~{Chen}, C.~{Foias}, D.~D. {Holm}, E.~{Olson}, E.~S. {Titi}, S.~{Wynne}, {A
  connection between the Camassa-Holm equations and turbulent flows in channels
  and pipes}, Physics of Fluids 11 (1999) 2343--2353.

\bibitem{2001PhyD..152..505F}
C.~{Foias}, D.~D. {Holm}, E.~S. {Titi}, {The Navier-Stokes-alpha model of fluid
  turbulence}, Physica D Nonlinear Phenomena 152 (2001) 505--519.

\bibitem{PiGrHoMi+2007}
J.~{Pietarila Graham}, D.~D. {Holm}, P.~D. {Mininni}, A.~{Pouquet}, {Highly
  turbulent solutions of the Lagrangian-averaged Navier-Stokes {$\alpha$} model
  and their large-eddy-simulation potential}, Phys. Rev. E. 76~(5) (2007) 056310.

\bibitem{PiGrMiPo2009}
J.~{Pietarila Graham}, P.~D. {Mininni}, A.~{Pouquet}, {Lagrangian-averaged
  model for magnetohydrodynamic turbulence and the absence of bottlenecks},
  Phys. Rev. E 80~(1) (2009) 016313.

\bibitem{PiGrMiPo2011}
J.~{Pietarila Graham}, P.~D. {Mininni}, A.~{Pouquet}, {High Reynolds number
  magnetohydrodynamic turbulence using a Lagrangian model}, Phys. Rev. E 84~(1) (2011)
  016314.

\bibitem{NaMa2001}
B.~T. {Nadiga}, L.~G. {Margolin}, {Dispersive Dissipative Eddy Parameterization
  in a Barotropic Model}, Journal of Physical Oceanography 31 (2001)
  2525--2531.

\bibitem{HoNa2003}
D.~D. {Holm}, B.~T. {Nadiga}, {Modeling Mesoscale Turbulence in the Barotropic
  Double-Gyre Circulation}, Journal of Physical Oceanography 33 (2003) 2355.

\bibitem{Wi2004}
B.~A. {Wingate}, {The Maximum Allowable Time Step for the Shallow Water
  {$\alpha$} Model and Its Relation to Time-Implicit Differencing}, Monthly
  Weather Review 132 (2004) 2719.

\bibitem{HoWi2005}
D.~D. {Holm}, B.~A. {Wingate}, {Baroclinic Instabilities of the Two-Layer
  Quasigeostrophic Alpha Model}, Journal of Physical Oceanography 35 (2005)
  1287.

\bibitem{2008JCoPh.227.5691H}
M.~W. {Hecht}, D.~D. {Holm}, M.~R. {Petersen}, B.~A. {Wingate}, {Implementation
  of the LANS-{$\alpha$} turbulence model in a primitive equation ocean model},
  Journal of Computational Physics 227 (2008) 5691--5716.

\bibitem{2008JPhA...41H4009H}
M.~W. {Hecht}, D.~D. {Holm}, M.~R. {Petersen}, B.~A. {Wingate}, {The
  LANS-{$\alpha$} and Leray turbulence parameterizations in primitive equation
  ocean modeling}, Journal of Physics A Mathematical General 41 (2008) H4009.

\bibitem{NaSh2001}
B.~T. {Nadiga}, S.~{Shkoller}, {Enhancement of the inverse-cascade of energy in
  the two-dimensional Lagrangian-averaged Navier-Stokes equations}, Physics of
  Fluids 13 (2001) 1528--1531.

\bibitem{LuKuTa+2007}
E.~{Lunasin}, S.~{Kurien}, M.~A. {Taylor}, E.~S. {Titi}, {A study of the
  Navier-Stokes-{$\alpha$} model for two-dimensional turbulence}, Journal of
  Turbulence 8 (2007) 30.

\bibitem{Kr1971}
R.~H. {Kraichnan}, {Inertial-range transfer in two- and three-dimensional
  turbulence}, Journal of Fluid Mechanics 47 (1971) 525--535.

\bibitem{MaVa1993}
M.~E. {Maltrud}, G.~K. {Vallis}, {Energy and enstrophy transfer in numerical
  simulations of two-dimensional turbulence}, Physics of Fluids 5 (1993)
  1760--1775.

\bibitem{MiRoRe+2011}
P.~D. {Mininni}, D.~L. {Rosenberg}, R.~{Reddy}, A.~{Pouquet}, {An hybrid
  MPI-OpenMP scheme for scalable parallel pseudospectral computations for fluid
  turbulence}, Parallel Computing 37 (2011) 316.

\bibitem{Va2006}
G.~K. {Vallis}, {Atmospheric and Oceanic Fluid Dynamics}, 2006.

\bibitem{XiWaCh+2008}
Z.~{Xiao}, M.~{Wan}, S.~{Chen}, G.~L. {Eyink}, {Physical mechanism of the
  inverse energy cascade of two-dimensional turbulence: a numerical
  investigation}, Journal of Fluid Mechanics 619 (2008) 1.

\bibitem{MeGeBa2003}
J.~{Meyers}, B.~J. {Geurts}, M.~{Baelmans}, {Database analysis of errors in
  large-eddy simulation}, Physics of Fluids 15 (2003) 2740--2755.

\bibitem{MeSaGe2006}
J.~{Meyers}, P.~{Sagaut}, B.~J. {Geurts}, {Optimal model parameters for
  multi-objective large-eddy simulations}, Physics of Fluids 18~(9) (2006)
  095103.

\bibitem{MeGeSa2007}
J.~{Meyers}, B.~J. {Geurts}, P.~{Sagaut}, {A computational error-assessment of
  central finite-volume discretizations in large-eddy simulation using a
  Smagorinsky model}, Journal of Computational Physics 227 (2007) 156--173.

\bibitem{Me2011}
J.~{Meyers}, {Error-landscape assessment of large-eddy simulations: a review of
  the methodology}, J. Sci. Comp. 49 (2011) 65--77.

\bibitem{DaGu2001}
S.~{Danilov}, D.~{Gurarie}, {Nonuniversal features of forced two-dimensional
  turbulence in the energy range}, Phys. Rev. E 63~(2) (2001) 020203.

\bibitem{CiBoDe2005}
C.~{Cichowlas}, P.~{Bona{\"i}ti}, F.~{Debbasch}, M.~{Brachet}, {Effective
  Dissipation and Turbulence in Spectrally Truncated Euler Flows}, Physical
  Review Letters 95~(26) (2005) 264502.

\bibitem{Li67}
D.~K. {Lilly}, {The representation of small-scale turbulence in numerical
  simulation experiments}, in: Proc. IBM Scientific Computing Symp. Environ.
  Sci., 1967, p. 195.

\bibitem{MeKa2000}
C.~{Meneveau}, J.~{Katz}, {Scale-Invariance and Turbulence Models for
  Large-Eddy Simulation}, Annual Review of Fluid Mechanics 32 (2000) 1--32.

\bibitem{2000MWRv..128.2935G}
S.~M. {Griffies}, R.~W. {Hallberg}, {Biharmonic Friction with a
  Smagorinsky-Like Viscosity for Use in Large-Scale Eddy-Permitting Ocean
  Models}, Monthly Weather Review 128 (2000) 2935.

\bibitem{2012JCli...25.2755D}
T.~L. {Delworth}, A.~{Rosati}, W.~{Anderson}, A.~J. {Adcroft}, V.~{Balaji},
  R.~{Benson}, K.~{Dixon}, S.~M. {Griffies}, H.-C. {Lee}, R.~C. {Pacanowski},
  G.~A. {Vecchi}, A.~T. {Wittenberg}, F.~{Zeng}, R.~{Zhang}, {Simulated Climate
  and Climate Change in the GFDL CM2.5 High-Resolution Coupled Climate Model},
  Journal of Climate 25 (2012) 2755--2781.

\bibitem{RiThKl+2010}
T.~D. {Ringler}, J.~{Thuburn}, J.~B. {Klemp}, W.~C. {Skamarock}, {A unified
  approach to energy conservation and potential vorticity dynamics for
  arbitrarily-structured C-grids}, Journal of Computational Physics 229 (2010)
  3065--3090.

\bibitem{ThKeWo2011}
J.~{Thuburn}, J.~{Kent}, N.~{Wood} (Eds.), {Energy and Enstrophy Cascades in
  Numerical Models}, 2011.

\bibitem{Ringler:8qyuvoZk}
T.~D. Ringler, M.~R. Petersen, D.~Jacobsen, R.~L. Higdon, P.~W. Jones, M.~E.
  Maltrud, {A Multi-Resolution Approach to Global Ocean Modeling}, Ocean
  Modeling, {\sl under revision}.

\end{thebibliography}







\end{document}